\documentstyle[version2,eqsecnum,aps,latexsym]{revtex}
\begin{document}
\draft
\begin{title}
Does backreaction enforce the averaged null energy condition in
semiclassical gravity?
\end{title}
\author{\'Eanna \'E. Flanagan and Robert M.~Wald}
\begin{instit}
Enrico Fermi Institute, 5640 South Ellis Avenue, \\
University of Chicago, Chicago, IL 60637-1433.
\end{instit}

\begin{date}
\newline
\vskip 0.5cm
\centerline{EFI 96-08,  gr-qc/9602052}
\centerline{Submitted to Phys. Rev. D}
\vskip 0.3cm
\centerline{June, 1996}
\end{date}
\input epsf.tex
\def\plotoneNew#1{\begin{center}
    \epsfxsize=0.5\textwidth \epsfbox{#1} \mbox{} \end{center}}

\begin{abstract}

The expectation value, $\langle T_{ab} \rangle$, of the renormalized
stress-energy tensor of quantum fields generically violates the
classical, local positive energy conditions of general relativity.
Nevertheless, it is possible that $\langle T_{ab} \rangle$ may still
satisfy some nonlocal positive energy conditions. Most prominent among
these nonlocal conditions is the averaged null energy condition
(ANEC), which states that $\int \langle T_{ab} \rangle k^a k^b \,
d\lambda \ge 0$ along any complete null geodesic, where $k^a$ denotes
the geodesic tangent, with affine parameter $\lambda$.  If ANEC holds,
then traversable wormholes cannot occur.  However, although ANEC holds
in Minkowski spacetime, it is known that ANEC can be violated in
curved spacetimes if one is allowed to choose the spacetime and
quantum state arbitrarily, without imposition of the semiclassical
Einstein equation, $G_{ab} = 8 \pi \langle T_{ab} \rangle$. In this
paper, we investigate whether ANEC holds for self-consistent solutions
of the semiclassical Einstein equation.  We study a free, linear,
massless scalar
field with arbitrary curvature coupling in the context of perturbation
theory about the flat spacetime/vacuum solution,
and we modify the perturbed semiclassical
equations by the ``reduction of order" procedure to eliminate
spurious solutions.  We also restrict attention to
the limit in which the lengthscales determined by the state and metric
are much
larger than the Planck length.
At first order in the metric and state
perturbations, and for pure states of the scalar field, we find that
the ANEC integral vanishes, as it must for any positivity result to
hold.  For mixed states, the ANEC integral can be negative.  However,
we prove that if we average the ANEC integral
transverse to the geodesic, using a suitable Planck scale smearing
function, a strictly positive result is obtained in all cases except
for the trivial flat spacetime/vacuum solution. Similar results hold
for pure states at
second order in
perturbation theory, when we additionally specialize to the situation
where
incoming classical gravitational radiation does not dominate the first
order metric perturbation.
These results suggest --- in agreement
with conclusions drawn by Ford and Roman from entirely independent
arguments --- that if traversable wormholes do exist as
self-consistent solutions
of the semiclassical equations, they cannot be macroscopic but must be
``Planck scale''. In the course of our analysis, we investigate a
number of more general issues relevant to doing perturbative
expansions of the semiclassical equations off of flat spacetime,
including an analysis of the nature of the semiclassical Einstein
equation and of prescriptions for extracting physically relevant
solutions. A large portion of our paper is devoted to the treatment of
these more general issues.
\end{abstract}

\narrowtext
\twocolumn

\section{Introduction and Summary}

\subsection{Brief overview of the issues addressed here}
\label{overview}

A characteristic feature of general relativity is that it provides a
framework for understanding many objects and phenomena in which
spacetime behaves in ways that are qualitatively completely different
from our everyday experience and intuition.  For example, solutions of
Einstein's equation could in principle exist which describe the
creation of closed timelike curves or whose topologies are nontrivial.
Whether or not such solutions exist depends on the nature of the
matter that inhabits spacetime.

For the types of matter normally considered to be physically realistic,
all observers measure locally non-negative energy
densities in the approximation where matter is treated
classically.  This condition that $T_{ab} u^a u^b \ge 0$ for all
timelike $u^a$ (known as the ``weak energy condition'')
as well as other, similar, local positive energy conditions
are sufficient to strongly constrain the space of
solutions to Einstein's equation.  In particular, macroscopic
traversable wormholes are forbidden when this condition is satisfied
\cite{Morris-Thorne-Yurtsever,Topological-Censorship}.  Moreover, the
positivity of locally measured energy density plays a key role in the
positive energy theorems \cite{posenergy} and the
singularity theorems \cite{Hawking-Ellis,Wald}, which predict that
general relativity must break down at the endpoint of gravitational
collapse.

However, it is well known that quantum fields violate all the
local, pointwise energy conditions \cite{Epstein,Zeldovich}.  For
example, the Casimir vacuum
for the electromagnetic field between two perfectly conducting plates
has a negative local energy density; indirect effects of this have
been observed experimentally \cite{Casimir}.  Squeezed states of light
also violate the energy conditions \cite{squeezed} and also have been
produced experimentally \cite{lab}.  Energy condition violations are
also fundamental to the evaporation of black holes, and also to
particle production in a gravitational field (such as that sometimes
hypothesized to seed galaxy formation in the early Universe)
\cite{particleproduction}.

These ubiquitous violations of energy
conditions have led people to consider the possibility that
the semiclassical equations could admit
solutions that are qualitatively very
different from classical solutions, such as solutions with negative ADM
mass or solutions in which gravitational collapse occurs without the
formation of singularities. In particular, in recent years there
has been considerable speculation
that semiclassical solutions could exist which
contain macroscopic
traversable wormholes, and, perhaps, even describe the creation of
closed
timelike curves in an initially causally well behaved spacetime
\cite{Morris-Thorne-Yurtsever,Visser-book,khatsym}.

Are such objects allowed in semiclassical gravity?  There are three
different types of possibilities \cite{Ford-Roman95c,Yurtsever95}:

\begin{itemize}

\item
The semiclassical equations might forbid traversable wormholes, the
creation of closed timelike curves, and negative mass objects.  The
space of solutions of the semiclassical equations would then not be
very different qualitatively from that of the classical equations.

\item
The semiclassical equations might allow such objects, but only in
such a way that they always lie
outside the domain of validity of the semiclassical theory,
either because the curvature scales are Planckian somewhere in the
corresponding spacetimes, or because the quantum fluctuations in the
stress
tensor are comparable to its expected value.

\item
The semiclassical equations might allow such objects in situations where
the semiclassical theory is a good approximation and the objects are
``macroscopic'' in size (as opposed to Planck-scale).
\end{itemize}

In the last several years, a variety of evidence has accumulated that
indicates against the third of these possibilities, and in the
direction of either the first or second. In particular, the following
evidence has been adduced
against the possibility of creating ``time machines''
via macroscopic,
traversable wormholes: First, it has been argued
that appropriate nonlocal energy conditions may hold, which
prevent traversable wormholes (no less time machines) from being
produced
\cite{Ford-Roman95c,Yurtsever95,Klinkhammar91,Klinkhammar92,WaldYurtsever,Yurtsever94,Ford,Ford91,Ford-Roman90,Ford-Roman92,Ford-Roman95,Ford-Roman93,Ford-Roman95b};
see Sec.~\ref{nonlocalconstraints} below.  Second, it has been argued
\cite{KuoFord} that, for a wide variety of
states in flat spacetime, whenever the expected value of the energy
density is negative, then the fluctuations in the stress tensor are
comparable to the expected value \cite{suppress}.  This suggests that
the 
semiclassical equations should not be trusted in the case of solutions
which depend in a crucial way on energy condition violations, such as
in the case of traversable wormholes. Finally,
it has been argued that even if
traversable wormholes could be produced, quantum field effects near a
chronology horizon will result in a singular $\langle T_{ab} \rangle$,
which could prevent the occurrence of closed timelike curves
\cite{KimThorne,Hawking,KayWald}.

One of the principal purposes of this paper is to present
additional evidence that nonlocal energy conditions which are
sufficiently strong to rule out the occurrence of
macroscopic, traversable wormholes may hold in
semiclassical gravity.
We shall investigate the validity of averaged null energy
condition (ANEC) in perturbation
theory off of Minkowski spacetime. The key
new ingredient in our analysis
is that we will impose the semiclassical Einstein equation
\begin{equation}
G_{ab}[g_{cd}] = 8 \pi  \langle {\hat T}_{ab}[g_{cd}] \rangle,
\label{basic0}
\end{equation}
on the spacetime and quantum state. Although
we shall find that ANEC can be violated even for solutions of
Eq.~(\ref{basic0}), we shall show that in perturbation theory, a
transverse smearing over several Planck lengths of the ANEC integral is
sufficient to
ensure positivity.  Our results thus
suggest that violations of ANEC in semiclassical gravity may be confined
to the Planck scale, where the semiclassical approximation itself is
suspect. In particular, since violations of ANEC are necessary for
producing traversable wormholes \cite{Topological-Censorship},
at the very least, it
should be necessary for traversable wormholes to have a ``Planck scale
structure''.

Our analysis applies only to non-self-interacting quantum fields.
Thus, although the full semiclassical theory we are considering is
nonlinear due to the coupling to the classical metric, the quantum
portion of the theory is linear.  It is possible that semiclassical
solutions for interacting fields could be qualitatively different from
semiclassical solutions for free fields \cite{Kipsmodel}. However,
we are not aware of any evidence which suggests that this is the
case, provided, of course, that the energy
conditions for the interacting fields are satisfied classically.

An additional principal purpose of this paper is to investigate the
nature of the semiclassical
Einstein equation (\ref{basic0}) and its solutions.
In particular, this equation
has a character that is very similar to the radiation reaction
equation for a classical charged point particle. Equation
(\ref{basic0}) contains time
derivatives of order higher than two, and ,correspondingly, there exist,
in effect, ``extra degrees of freedom" in its solutions, including
so-called runaway solutions which grow exponentially
in time.  We build on recent work of Simon
\cite{Simon,Simon1,Simon2,Simon3}, and discuss in detail the
pathologies that arise and possible resolutions.  Our conclusion is that
in the special case of perturbation theory about flat space, it is
possible to resolve the difficulties by a ``reduction of order"
prescription, but
that in general there are still open questions with respect to the
extraction of physical predictions from the semiclassical equations.

\subsection{Nonlocal constraints on the stress-energy tensor}
\label{nonlocalconstraints}

We now
briefly discuss, as background, the status of nonlocal energy
conditions in relativity; see Yurtsever \cite{Yurtsever94} for a
recent review.  Let $(M,g_{ab})$ be a globally hyperbolic spacetime,
let $\phi$ be
a quantum field on this spacetime,
and consider the expected stress tensor, $\langle T_{ab} \rangle$
on all states of this field.  Although at
any given point in the spacetime, we may choose the state so as to
make the energy density be
arbitrarily negative \cite{Yurtsever94}, there can exist {\it nonlocal}
constraints on the stress tensor --- i.e., quantum field theory does
seem to
restrict the amount and nature of energy condition violations.  A
complete understanding of these nonlocal constraints is not yet in
hand, and the search for such an understanding is one of the key,
active areas of research in semiclassical gravity \cite{Yurtsever94}.
Nevertheless, the results that have been obtained to date
\cite{Yurtsever95,Klinkhammar91,Klinkhammar92,WaldYurtsever,Yurtsever94,Ford,Ford91,Ford-Roman90,Ford-Roman92,Ford-Roman95,Ford-Roman93,Ford-Roman95b,Ford-Roman95c}
suggest that nonlocal constraints on stress tensors may play a key role
in restricting the space of solutions of the semiclassical equations.
The present paper will present additional evidence in this direction.

The nonlocal constraints have the following general form
\cite{note9}.  Let $f^{ab}(x)$ be a tensor distribution on the fixed
spacetime, $(M,g_{ab})$, such that the quantity
\begin{equation}
{\cal E} = \int d^4x \sqrt{-g} f^{ab}(x) T_{ab}(x)
\label{generalsmear}
\end{equation}
is classically positive.  Denote by ${\cal E}_{\rm
min}[f^{ab},g_{cd}]$ the minimum over all quantum states of the
expected value of the quantity (\ref{generalsmear}).  There are now
three different possibilities.  First, it is possible that ${\cal
E}_{\rm
min}[f^{ab},g_{cd}]=-\infty$, so that quantum field theory does not
restrict the value of ${\cal E}$.  This will be the case, for example,
when $f^{ab}$ is proportional to a four dimensional delta function, so
that ${\cal E}$ depends only on the value of the stress tensor at one
point.  The second possibility is that ${\cal E}_{\rm
min}[f^{ab},g_{cd}]$ is finite and negative, so that
\begin{equation}
\int d^4x \sqrt{-g} f^{ab}(x) \langle T_{ab}(x) \rangle \, \ge \,
{\cal E}_{\rm min}[f^{ab},g_{cd}]
\label{generalsmear1}
\end{equation}
for all quantum states.  An interesting possibility -- which appears
worthy of further investigation -- is that this
may be the case whenever $f^{ab}$ is smooth and of compact
support.  A specific conjecture of the form (\ref{generalsmear1})
has also been suggested by Yurtsever \cite{Yurtsever95} (see below).

Ford and Roman
\cite{Ford,Ford91,Ford-Roman90,Ford-Roman92,Ford-Roman95,Ford-Roman93,Ford-Roman95b,Ford-Roman95c}
have derived a number of results of the form (\ref{generalsmear1})
in both flat and curved spacetime quantum field
theory, which they call ``quantum inequalities''. For example, Ford
showed that the flux $\Delta E$ of negative energy through some
surface in flat spacetime, when averaged over a time $\Delta t$, must
satisfy $\Delta E \agt - \hbar /
\Delta t$, a result reminiscent of the time-energy uncertainty
relation except for the minus sign \cite{Ford91}.  Similar results can
also be
derived for the spatial average of energy density over a length
$\Delta L$ in two dimensions \cite{EF}.  More recently,
Ford and Roman have derived
constraints on the average over time of the energy density measured at
a particular point by inertial observers \cite{Ford-Roman95} in flat
spacetime, and they have argued that their results can be extrapolated
to curved spacetime so as to
constrain certain types of traversable wormhole spacetimes to be
``Planck scale'' \cite{Ford-Roman95c}.

The third possibility with respect to the quantity ${\cal E}_{\rm
min}[f^{ab},g_{cd}]$ is that it vanish (or be positive), so that
\begin{equation}
\int d^4x \sqrt{-g} f^{ab}(x) \langle T_{ab}(x) \rangle \, \ge \,
0
\label{generalsmear2}
\end{equation}
for all quantum states.  Inequalities of the form
(\ref{generalsmear2}) are usually called ``averaged energy
conditions'' \cite{Tipler}.  An example of a constraint of this type
is the well known fact that in Minkowski spacetime, the integral of
the energy density over a constant time slice (i.e.~the Hamiltonian)
is a positive operator.

A particular averaged energy condition --- upon which
much attention has been focused --- is the averaged null energy
condition (ANEC), which states that
\begin{equation}
\int_\gamma \langle T_{ab} \rangle k^a k^b \, d\lambda \ge 0,
\end{equation}
where the integral is along any complete,
achronal null geodesic $\gamma$,
$k^a$ denotes the geodesic tangent, and $\lambda$ is an affine
parameter \cite{Roman88}.  The reason that this and other similar
conditions (with
null replaced by timelike) are useful is that they dovetail nicely with
the methods used to prove global results about spacetimes in
general relativity.  Many of the standard global results that were originally
proved to hold when pointwise energy conditions are satisfied, can be
shown to also hold under the weaker assumption that the stress tensor
satisfies ANEC.  For example, in spacetimes in which ANEC is satisfied,
the
topological censorship theorem of Friedman, Schleich and Witt
\cite{Topological-Censorship} rules out traversable wormholes.  Under
the same hypotheses, the
Penrose-Sorkin-Wolgar positive mass theorem shows that the
asymptotic mass must be positive \cite{Penrose-Sorkin-Wolgar}.
Finally, the positivity of the ANEC integral along future complete
null geodesics is sufficient to prove some singularity theorems
\cite{Roman88}.

The averaged null energy condition is therefore of considerable
interest.  Is it enforced by quantum field theory?  Early
investigations showed that it holds in Minkowski spacetime for free
scalar fields and electromagnetic fields
\cite{Klinkhammar91,Klinkhammar92,WaldYurtsever}, and also in generic,
curved, 2D spacetimes \cite{WaldYurtsever}.  However, it has
been shown that it can be violated in generic, curved four dimensional
spacetimes \cite{WaldYurtsever,Visser}, even if the spacetime is
nearly flat.

The failure of the ANEC condition in general spacetimes does not,
however, sound a death knell for the program of deriving global
results in semiclassical gravity, since there are some
modifications of the original ANEC conjecture
that may give rise to nontrivial constraints on solutions.
One idea, suggested by Yurtsever \cite{Yurtsever95}, is simply
to weaken the conjecture from being an inequality of
the type (\ref{generalsmear2}) to one of the type
(\ref{generalsmear1}), in analogy with the quantum inequalities of
Roman and Ford.  In other words, a modified ANEC conjecture would be
that the quantity ${\cal E}_{\rm min}[f^{ab},g_{ab}]$ is always finite
and not $-\infty$, when the distribution $f^{ab}$ is chosen such that
the quantity ${\cal E}$ is the ANEC integral along a null geodesic.
Yurtsever shows that if this is true, then reasonable assumptions
about the dependence of ${\cal E}_{\rm min}$ on the spacetime geometry
lead to the conclusion that macroscopic, static wormholes are
excluded; only Planck-scale wormholes are (possibly) allowed.

In this paper, however, we shall follow a different path by
investigating
the
validity of ANEC when the spacetime and quantum state are constrained
by the semiclassical Einstein equation (\ref{basic0}), since any
violations of ANEC occurring when this equation fails to hold would
not be physically relevant. In order to analyze generic solutions to
Eq.~(\ref{basic0}), we will be forced to resort to perturbation theory
about the trivial solution, namely, Minkowski spacetime with the
quantum field in the vacuum state.  We use the ``reduction of order"
procedure to eliminate the unphysical solutions of the perturbative
semiclassical Einstein equation.  We make the additional approximation
that ``wavelengths are large compared to the Planck scale", and for
the portions of our analysis involving second order perturbations, we
also will need to assume that incoming classical gravitational
radiation does not dominate the metric perturbation at first order.
In the ``note in proof'' section of Ref.~\cite{WaldYurtsever},
violations of ANEC for pure states were obtained at first order in
deviation from flatness.  A key result of our analysis is that this
type of counterexample is eliminated by imposing the semiclassical
equation: When Eq.~(\ref{basic0}) holds, the ANEC integral always
vanishes for pure states at first order in deviation from
flatness. This result has the side consequence that we must go to
second order perturbation theory in order to give a complete analysis
of the positivity properties of the ANEC integral for pure states in
nearly flat spacetimes.

As will be described in more detail in the next subsection, we shall
show that ANEC can be violated. However, a suitable {\it transversely
smeared} ANEC integral is always non-negative in the context of our
perturbation expansions.  The condition that a smeared ANEC integral
always be non-negative in general spacetimes is clearly a much weaker
condition than the usual ANEC condition.  Nevertheless, when the width
of the smearing function is of the order of the Planck length as it is
in our analyses, the positivity of a smeared ANEC integral would be
sufficient to derive interesting constraints on the spacetime
geometry.  For example, suppose that a spacetime contains a
macroscpoic traversable wormhole.  Then it must contain one geodesic
$\gamma$ for which the ANEC integral is negative.  However, the
transversely smeared ANEC integral centered on $\gamma$ must be
positive.  There are now two qualitatively different possibilities ---
either the compensating positive contribution to the smeared ANEC
integral comes from within a few Planck lengths, or it comes from a
macroscopic distance away, corresponding to the tail of the
smearing function.  

In the first case the stress tensor and Einstein tensor
must vary significantly over lengthscales of the order of the Planck
length, and therefore the spacetime presumably lies outside the domain
of validity of semiclassical gravity.  In the second case, there can
be macroscopic regions of spacetime which ANEC is violated.  The
existence of violations of ANEC of this type was suggested by some
results of Visser \cite{Visser96}, and in Appendix \ref{visserexample}
we present an explicit example of an approximate self consistent
solution which violates ANEC in this way.  In this second case,
however, the positivity of smeared ANEC would restrict the {\it
amount} of violation to be incredibly small compared to the distant,
positive mass.  We argue in Appendix \ref{visserexample} below that
such violations of ANEC would be far to small to allow macroscopic
traversable wormholes.  Analogous
arguments apply to spacetimes with negative asymptotic mass and with
compactly generated chronology horizons.  This provides evidence in
favor of the second (or first) of the three possibilities discussed in
Sec.~\ref{overview}.  Consequently, if the semiclassical equations
were to enforce the positivity of a transversely smeared ANEC integral
in general spacetimes, with a smearing width of order of the Planck
length, this would provide almost as strong a constraint on physical
possibilities as the positivity of the ANEC integral itself.

Our positivity result for the transversely smeared ANEC integral in
perturbation theory is the first nonlocal constraint on stress tensors
that has been proved in a generic class of four-dimensional, curved
spacetimes.  Our perturbative result suggests that something similar
may be true in general spacetimes, and consequently that the behavior
of solutions in semiclassical gravity --- within the domain of
validity of that theory --- may be qualitatively similar to classical
solutions.

\subsection{Brief summary of our assumptions and results}
\label{summary}

We consider a massless scalar field with arbitrary coupling,
$\xi$, to the scalar curvature.
We wish to consider a one-parameter
family (with parameter denoted
by $\varepsilon$) of spacetimes $\left(M,g_{ab}(\varepsilon)\right)$
and quantum states satisfying
Eq.~(\ref{basic0}), with the spacetime reducing to Minkowski
spacetime and the quantum state reducing to the vacuum state
when $\varepsilon = 0$.
It is somewhat awkward and overly restrictive to attempt to describe the
one-parameter family of quantum states as though they were states
in a single, fixed, Hilbert space, since in a general, curved
spacetime $(M,g_{ab})$, there is no unique preferred Hilbert space of
possible states, and, in general,
there is no unique, preferred way of
identifying the states occurring in Hilbert space constructions in
different spacetimes.  It is much more useful to adopt the algebraic
approach, wherein one
characterizes a state by its $n$-point distributions on spacetime. We
shall
adopt this philosophy here and write the one-parameter
family of states as $\omega(\varepsilon)$. We shall
denote the expected stress-energy
in the state $\omega$ by $\langle T_{ab} \rangle_\omega$.
In fact, since
the expected stress-energy tensor is directly determined from only the
2-point distribution
$G(x,y) \equiv \langle {\hat \Phi}(x) {\hat \Phi}(y)  \rangle_\omega$
of the quantum field, the higher order correlation
functions will play no role in our analysis. Thus,
for the purposes of our analysis
a ``state'' may be viewed as synonymous with a 2-point distribution
on spacetime satisfying the wave equation in each variable,
as well as the positivity and Hadamard conditions (see
subsection \ref{semiclgeneral} below).  However, little harm would be
done in most of our analysis below by pretending that
$\omega(\varepsilon)$ corresponds to a one-parameter
family of density matrices, ${\hat \rho}(\varepsilon)$
in some fixed Hilbert space,  with
$\langle T_{ab} \rangle_\omega = {\rm tr} [{\hat \rho} \, {\hat
T}_{ab}]$.

It is useful to characterize the state $\omega(\varepsilon)$ by the
behavior of its correlation functions in the asymptotic past. Assuming
that suitable asymptotic conditions hold on the spacetime
$\left(M,g_{ab}(\varepsilon)\right)$, and state, $\omega(\varepsilon)$,
we may associate with $\omega(\varepsilon)$ a state
$\omega_{\rm in}(\varepsilon)$ on Minkowski spacetime which
agrees with $\omega(\varepsilon)$ in the asymptotic past under an
appropriate identification of $\left(M,g_{ab}(\varepsilon)\right)$
with Minkowski spacetime. In particular, the 2-point distribution,
$G(x,y;\varepsilon)$,
of $\omega(\varepsilon)$ can then be characterized by the
function $F_{\rm in}(x,y;\varepsilon) = G_{\rm in}(x,y;\varepsilon) -
G_{\rm in,0}(x,y)$ on Minkowski spacetime,
where $G_{\rm in}(x,y;\varepsilon)$ is the 2-point function of
$\omega_{\rm in}(\varepsilon)$ and $G_{\rm in,0}(x,y)$ is the
2-point function of the Minkowski vacuum state, $\omega_{\rm in,0}$.
For Hadamard states, $F_{\rm in}(x,y;\varepsilon)$ will be a smooth
bisolution of the wave equation in Minkowski spacetime, whose initial
data at past null infinity, ${\cal J}^-$,
may be viewed as the freely specifiable initial data for the state
(which, however, is subject to the positivity constraints
discussed in Section \ref{semiclgeneral} below).

It is well known that the semiclassical Einstein equation is of a
``higher derivative" character than the corresponding classical
equation, and that consequently there exist new --- presumably
spurious --- ``degrees of freedom" in semiclassical gravity, closely
analogous to the ``run-away" solutions which occur for the dynamics of
a point charge in classical electrodynamics when radiation reaction
effects are included. Thus, in order to extract any physical
predictions from the semiclassical equations, we need a prescription
either for extracting the ``physical solutions" to these equations or
for modifying the equations themselves so that the spurious solutions
no longer arise. We investigate this issue in depth in
Sec.~\ref{simonstuff}. We conclude that --- at least in the context of
perturbation theory about flat spacetime --- the ``reduction of order"
algorithm for modifying the equations \cite{Simon3} provides a
satisfactory means of eliminating the spurious solutions without
(significantly) sacrificing accuracy at ``long wavelengths", i.e., in
the regime where the dominant lengthscales in the solution are much
larger than the Planck scale. The validity of ANEC is investigated in
the context of solutions to the
reduced order perturbative semiclassical Einstein equation.

We expand the one-parameter family of ``in"-states as
\begin{equation}
\omega_{\rm in}(\varepsilon) = \omega_{\rm in,0} + \varepsilon
\,\omega_{\rm in}^{(1)}
+ \varepsilon^2 \omega_{\rm in}^{(2)} + O(\varepsilon^3)
\label{rhoexpand0}
\end{equation}
and we expand the metric as
\begin{equation}
g_{ab}(\varepsilon) = \eta_{ab} + \varepsilon h^{(1)}_{ab} + \varepsilon^2
h^{(2)}_{ab} + O(\varepsilon^3),
\label{metricexpand}
\end{equation}
where $\eta_{ab}$ is the flat, Minkowski metric.  The metric
perturbation $h_{ab}^{(1)}$ can be written as
\begin{equation}
h_{ab}^{(1)} = h_{ab}^{(1) \, ,{\rm in}} + \Delta h_{ab}^{(1)},
\label{Delh}
\end{equation}
where $h_{ab}^{(1)\, ,{\rm in}}$ satisfies the linearized Einstein
equation in
vacuum and represents incoming classical gravitational waves at
${\cal J}^-$.  The remaining portion $\Delta h_{ab}^{(1)}$ is determined from
the first order state perturbation $\omega_{\rm in}^{(1)}$ via the reduced
order, perturbed
semiclassical equations.  There is a
decomposition similar to Eq.~(\ref{Delh}) for the second order metric
perturbation $h_{ab}^{(2)}$.  Note that to second order, solutions to the
reduced order, perturbative
semiclassical equations can be characterized by their ``initial
data''  at past null infinity ${\cal J}^-$, consisting of the incoming
gravitational
radiation $h_{ab}^{(1) \, ,{\rm in}}$
and $h_{ab}^{(2) \, ,{\rm in}}$, as well as the
incoming, freely propagating (with respect to $\eta_{ab}$)
pieces $F_{\rm in}^{(1)}$ and $F_{\rm in}^{(2)}$ of the two-point functions
associated with $\omega_{\rm in}^{(1)}$ and $\omega_{\rm in}^{(2)}$
(see Figure \ref{scatterpicture}).

{\vskip -1.2cm}
{\plotoneNew{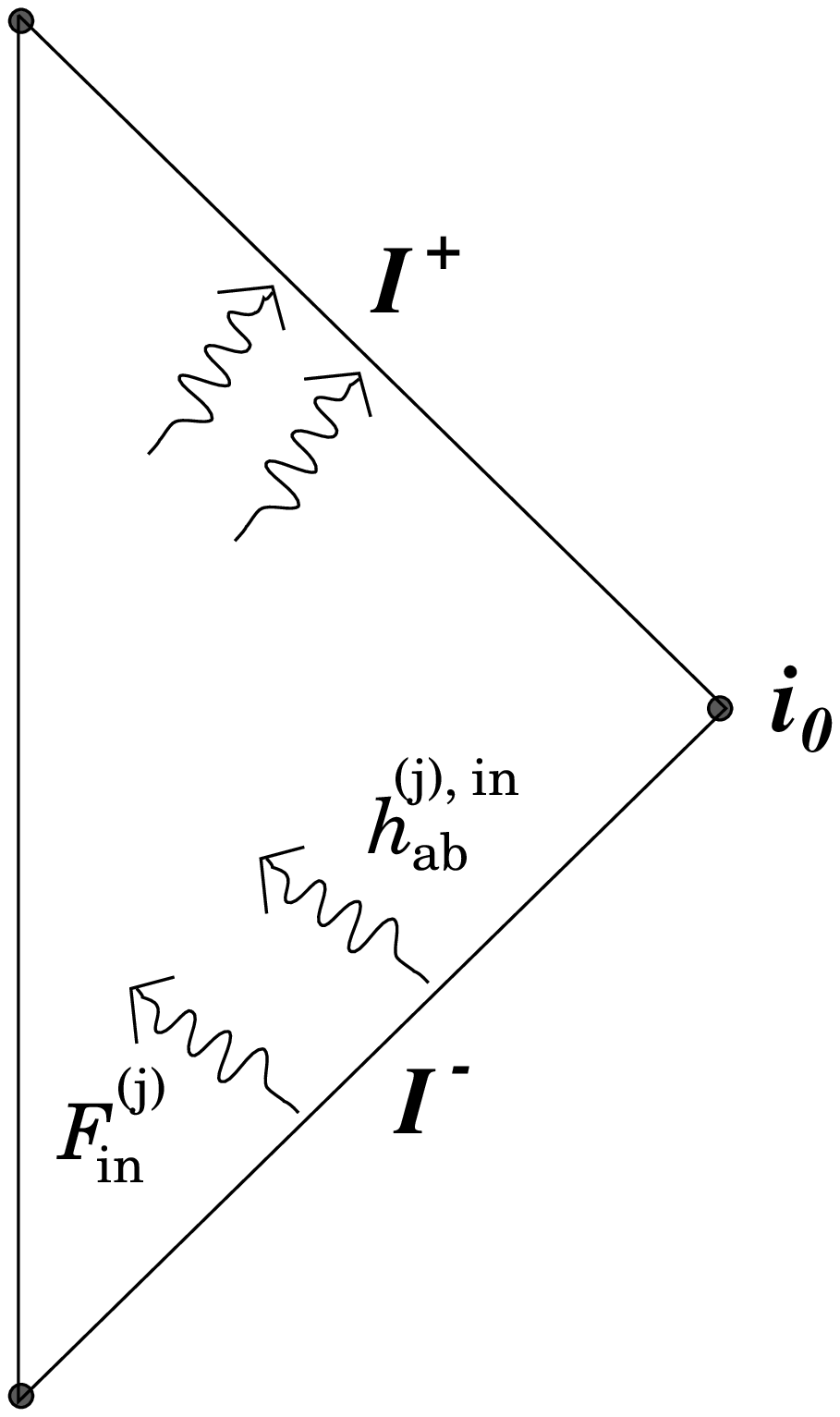}}
{\vskip -0.2truein}
\figure{
The ``scattering picture'' for solutions of the semiclassical
equations.  The spacetime metric is determined by (i) the incoming
first and second order metric perturbations $h_{ab}^{(1) \, ,{\rm
in}}$ and $h_{ab}^{(2) \, ,{\rm in}}$ at ${\cal J}^-$, which describe
freely propagating gravitational radiation, and (ii) the two-point
functions $F_{\rm in}^{(1)}$ and $F_{\rm in}^{(2)}$ at ${\cal J}^-$ of the
first
and second order perturbations $\omega^{(1)}_{\rm in}$ and $\omega^{(2)}_{\rm
in}$ to the incoming state $\omega_{\rm in}(\varepsilon)$ of the scalar
field.
These determine, via the semiclassical equations, corresponding
outgoing quantities at ${\cal J}^+$.  However, we choose to parameterize
the solutions in terms of the incoming quantities at ${\cal J}^-$.
\label{scatterpicture}}
{\vskip 0.6cm}

The ``transverse smearing'' of the ANEC integral referred to in the
previous subsection is defined in Minkowski spacetime as follows.
Let $\gamma$ be a null geodesic, let $\lambda$,
$\zeta$, $x^1$, $x^2$ be coordinates such that the geodesic is given
by $\zeta = x^A = 0$, $A=1,2$, and such that the metric is
\begin{equation}
ds^2 = - 2 d\lambda d\zeta + (d x^1)^2 + (d x^2)^2.
\end{equation}
Now let
\begin{equation}
I_s = \int d\lambda \int d^2 x^A \, T_{\lambda
\lambda}(\lambda,\zeta=0,x^1,x^2) \,  S(x^1,x^2),
\label{smearedanec}
\end{equation}
where $S$ is some positive function that is peaked at $x^A=0$ and that
falls off at large $x^A$.  This is essentially an ANEC type integral,
with an additional averaging in two spatial directions transverse to
the geodesic.  The definition (\ref{smearedanec})
can be generalized
in a natural way to general, curved spacetimes by using Fermi-Walker
type coordinates [see Sec.~\ref{generalizedanec} below for details],
provided, of course, that the smearing function, $S$, vanishes outside
the region where such coordinates are well defined.

Let $\gamma(\varepsilon)$ be a one-parameter family of null geodesics
in $(M,g_{ab}(\varepsilon))$, and let
$I_s(\varepsilon)$ denote the smeared ANEC integral along
$\gamma(\varepsilon)$. We expand $I_s$
as $I_s = \varepsilon I_s^{(1)} + \varepsilon^2 I_s^{(2)} +
O(\varepsilon^3)$.  Using the reduced-order semiclassical equations we
calculate
the dependence of the perturbations of the ANEC integral on the
initial data:
\begin{eqnarray}
I_s^{(1)} &=& I_s^{(1)}[F_{\rm in}^{(1)}] \nonumber \\
\mbox{} I_s^{(2)} &=& I_s^{(2)}[F_{\rm in}^{(1)},F_{\rm in}^{(2)},h_{ab}^{(1)
\, ,{\rm
in}}].
\label{answers}
\end{eqnarray}
We restrict attention to the case where the incoming two-point
functions $F_{\rm in}^{(1)}$ and $F_{\rm in}^{(2)}$ satisfy the regularity
conditions discussed in Appendix \ref{FTstress}.  We choose the
transverse
smearing
function $S({\bf x})$ that
enters into the definition (\ref{smearedanec}) of $I_s$ to be
\begin{equation}
S({\bf x}) \propto  {1 \over 1 + {\bf x}^4 / \Lambda_T^4},
\label{smear2}
\end{equation}
where ${\bf x} = (x^1,x^2)$, and $\Lambda_T$ is greater than a certain
critical length, $\Lambda_{\rm T,crit}$, of the order of the Planck
length [see
Sec.~\ref{firstorderexact} below].  We also
specialize to the limit in which the lengthscales determined by the
incoming state are much larger than the Planck length.
With these assumptions, our results may be summarized as follows (see
table \ref{table1} below):

At first order in perturbation theory, we find that (i) $I_s^{(1)}$
vanishes for pure incoming states, (ii) $I_s^{(1)}$ is positive for mixed
incoming states if the transverse smearing length $\Lambda_T$ is
chosen to be greater than $\Lambda_{\rm T,crit}$, and (iii) $I_s^{(1)}$ can
be made
negative if $\Lambda_T$ is chosen to be sufficiently small, and in
particular, if there were no transverse smearing at all.  Thus, the
transverse smearing is a crucial ingredient in our analysis.  Note
also that result (i) is a necessary condition for $I_s$ to be always
nonnegative; if $I_s^{(1)}$ were nonvanishing for some incoming pure
state, it could be made to have either sign by choosing the sign of
the first order perturbations appropriately.  [See
Sec.~\ref{pureversusmixed} below for an explanation of why pure states
and mixed states behave differently in this regard.]  As was already
mentioned in the previous subsection, the analog of result (i) fails
to hold when the semiclassical Einstein equation is not imposed.

Because the smeared ANEC integral vanishes at first order for pure
states, it is necessary to go to second order in perturbation theory
to see if the positivity of this integral can be violated for pure
states.  A complete calculation of all second order effects would have
required us to derive a formula for the complete corrections to the
stress-energy tensor of the quantum field valid to second order in
deviation from flatness. This would have been a major undertaking in
its own right, and we did not attempt to do this.  Instead, we
specialized to a limit in which not only are
the lengthscales determined by the
incoming state much larger than the Planck length, but,
in addition, the incoming classical gravitational radiation does not
dominate the first order metric perturbation.  Under these conditions,
the unknown term in the second order stress energy tensor is
negligible.  In Sec.~\ref{secondorderanalysis}, we calculate the three
remaining terms in $I_s^{(2)}[F_{\rm in}^{(1)},F_{\rm in}^{(2)},h_{ab}^{(1) \,
,{\rm in}}]$, and we show that in this limit $I_s^{(2)}$ is always
positive.  As before, the positivity only holds with transverse
smearing.

One unsatisfactory feature of our analysis is the following.
The positivity of the transversely smeared ANEC integral holds
in our perturbation expansion only
when the transverse smearing function $S$ falls off like
$x^{-4}$ or more slowly.  In particular, the smeared ANEC
integral with transverse smearing of compact support is
{\it not} always positive. However, in a curved spacetime, the
Fermi-Walker type coordinates needed for the
generalization of the definition (\ref{smearedanec}) will be well
defined only in a
neighborhood of the null geodesic in question. Thus, although we
prove the positivity of a smeared ANEC integral in the context of
perturbation theory, we do not even have an obvious candidate for a
``smeared ANEC conjecture'' outside of this context. Nevertheless,
we interpret our results as having the physical implications described
in the previous subsection.

Finally, it is worth mentioning why we need to consider general, mixed
incoming states instead of just pure states.  In most situations,
whatever is true for pure states will generalize trivially to mixed
states.  Here, however, it turns out that pure states and mixed states
behave qualitatively very differently.  One reason for this (which may
be viewed as an artifact of semiclassical theory) is that the
linearity of the evolution law for states breaks down as a result of
the coupling to the classical metric.  This arises only at second
order in perturbation theory.  However, even at first order in
perturbation theory where $I_s^{(1)}[\omega_{\rm in}^{(1)}]$ depends linearly
on $\omega^{(1)}_{\rm in}$, it is not true that positivity of $I_s^{(1)}$ for
all pure states would imply positivity for mixed states.  To see this,
let ${\cal D}$ denote the space of states, and let ${\cal D}_P$ denote
the space of pure states.  Then the space of allowed state
perturbations $\omega_{\rm in}^{(1)}$ is not ${\cal D}$ but is the tangent
space $T_{\rm vac}({\cal D})$ to ${\cal D}$ at $\omega_0 =
\left|0\right>\left<0\right|$.  Similarly, the linear space of allowed
pure state perturbations is the tangent space $T_{\rm vac}({\cal
D}_P)$ to ${\cal D}_P$ at $\omega_0$.  Now the key point is that,
although ${\cal D}$ is the convex hull of ${\cal D}_P$, $T_{\rm
vac}({\cal D})$ is {\it not} the convex hull of $T_{\rm vac}({\cal
D}_P)$ but is larger than it.  [The convex hull of $T_{\rm vac}({\cal
D}_P)$ is just itself since it is a linear space.]  Therefore, results
for pure states do not generalize to mixed states.  Thus, the
differences between pure and mixed states arise in our analysis
because of our working in perturbation theory.  Roughly speaking, we
find that mixed states at first order in perturbation theory behave
very similarly to pure states at second order.

\subsection{Conventions}

We use the metric signature $(-,+,+,+)$, and the sign conventions of
Refs.~\cite{MTW} and \cite{Wald}, as well as
the abstract index notation explained in Ref.~\cite{Wald}.  The
following is our
convention for Fourier transforms.  If $F= F(x)$ is a function on
Minkowski spacetime, then we define the Fourier transform to be
\begin{equation}
{\tilde F}(k) = \int d^4x \, e^{- i k \cdot x} \, F(x).
\label{FTdef}
\end{equation}
Similarly if $f = f({\bf x})$ is a function on three
dimensional Euclidean space, we define its Fourier transform to be
\begin{equation}
{\tilde f}({\bf k}) = \int d^3x \, e^{- i {\bf k} \cdot {\bf x}} \, f({\bf
x}).
\end{equation}
We use gravitational units in which the speed of light $c$ and
Newton's gravitational constant $G$ are unity, so that $\hbar =
L_p^2$, where $L_p$ is the Planck length.  An index of notation is
given at the end of the paper in table \ref{table2}.

\section{semiclassical gravity with backreaction}

Difficulties arise in the calculation of backreaction effects in
semiclassical gravity for several reasons.  First, as we
discuss further in section \ref{simonstuff}, there are problems associated with
spurious, runaway solutions to the equations.  Second, the dependence
of the renormalized stress tensor on the spacetime geometry is
nonlocal and complicated, making calculations outside of linear
perturbation theory prohibitively difficult.  These difficulties do
not appear in two dimensional models, for the reasons explained in
Ref.~\cite{Wald94}, and in recent years there have been a variety of
calculations of backreaction effects on two dimensional black hole
backgrounds \cite{refs}.  However, in four dimensions, most
backreaction calculations have been restricted to linear order
perturbation theory off some fixed background (except for analyses of
various conformally flat spacetimes \cite{Simon3,Wald7}).  Our
backreaction analysis is apparently the first to go beyond linear
order perturbation theory for a generic class of four dimensional
spacetimes.  We shall build upon and extend the work of Horowitz
\cite{Horowitz}, who considered semiclassical gravity in perturbation
theory about flat spacetime to linear order, without allowing
perturbations to the incoming quantum state.

\subsection{The classical equations}

We now describe in more detail the model of gravity coupled to a
scalar field that we will treat in our analysis.
We consider a scalar field $\Phi$ of
arbitrary curvature coupling $\xi$ and arbitrary mass $m$, so that the
Lagrangian is
\begin{eqnarray}
L = {1\over2} \int d^4x \sqrt{-g} \{ \kappa R - g^{ab} \nabla_a \Phi
\nabla_b \Phi \nonumber \\ \mbox{} - m^2 \Phi^2 - \xi R \Phi^2 \},
\label{lagrangian}
\end{eqnarray}
where $\kappa = 1/(8 \pi G)$.  We will specialize later to the massless
case.  The classical equations of motion are
\begin{equation}
\kappa \, G_{ab} = T_{ab},
\label{Einsteineq}
\end{equation}
and
\begin{equation}
\left[ \Box - m^2 - \xi R \right] \Phi =0,
\label{waveeqn}
\end{equation}
where the stress tensor is
\begin{eqnarray}
T_{ab} = \nabla_a \Phi \nabla_b \Phi - {1 \over 2} g_{ab} (\nabla
\Phi)^2 - {1\over2} g_{ab} m^2 \Phi^2 \nonumber \\ \mbox{} + \xi
\left[ G_{ab} \Phi^2 - 2 \nabla_a ( \Phi \nabla_b \Phi) + 2 g_{ab}
\nabla^c (\Phi \nabla_c \Phi) \right].
\label{stclassical}
\end{eqnarray}
In the classical theory, this stress tensor violates the pointwise
null energy condition for $\xi \ne 0$, but satisfies the averaged null
energy condition for all negative
values of $\xi$, provided that
suitable asymptotic fall-off conditions hold for $\Phi$. More precisely,
for any null geodesic $x^a = x^a(\lambda)$ with
affine parameter $\lambda$ and tangent vector $\lambda^a = (\partial /
\partial \lambda)^a$, it follows from Eqs.~(\ref{Einsteineq}) and
(\ref{stclassical}) that
\FL
\begin{eqnarray}
\int G_{ab} \lambda^a \lambda^b \, d \lambda &=& \int d \lambda \,
{\Phi^{\prime \, 2} \over \kappa - \xi \Phi^2} \,\nonumber \\ \mbox{}
&& + 4 \xi^2 \int d \lambda \, {(\Phi \Phi^\prime)^2 \over (\kappa -
\xi \Phi^2)^2},
\label{xineg}
\end{eqnarray}
provided that $\xi < 0$ and $(\Phi \Phi^\prime)/(\kappa -\xi \Phi^2) \to
0$
as $|\lambda| \to \infty$,
where primes denote derivatives with respect to $\lambda$.  However, for
at least some positive values of $\xi$,  it seems likely that ANEC can
be
violated \cite{note4}.

The failure of the classical stress-energy tensor (\ref{stclassical})
to satisfy pointwise energy conditions when $\xi \ne 0$, and its
possible failure to satisfy the classical ANEC condition for positive
values of $\xi$ indicates that this class of field theories may not be
suitable for a general study of the validity of ANEC in quantum field
theory in curved spacetime.  However, in this paper, we shall be
concerned only with a perturbative analysis of solutions about
Minkowski spacetime.  In this case, it follows
immediately from Eq.~(\ref{xineg}) that, for all values of
$\xi$, the classical ANEC condition holds
for all solutions near the trivial solution of Minkowski spacetime
with $\Phi=0$.
Thus, this class of models should provide a good
testing ground as to whether, in the context of perturbation theory,
quantum fields can attain violations of ANEC which are classically
forbidden.

\subsection{The status of the semiclassical equations and the local
curvature ambiguity}
\label{status}

It is natural to postulate the semiclassical equation
\begin{equation}
\kappa G_{ab}[g_{cd}] = \langle {\hat T}_{ab}[g_{cd}] \rangle_\omega,
\label{basic}
\end{equation}
to describe the backreaction of the quantized matter degrees of freedom
on spacetime.  However, as is well known, the exact status and
domain of applicability of this equation is far from clear,
due mainly to the fact that we do not have a complete, more
fundamental theory of quantum gravity from which Eq.~(\ref{basic})
could be derived. In addition there
exist pathological runaway-type solutions to
these equations, as analyzed in detail by Horowitz
\cite{Horowitz} for the linearized
equations.  Proposals for dealing with these unphysical
solutions have been put forward by Simon
\cite{Simon,Simon1,Simon2,Simon3}, and we will discuss
Simon's suggestions in detail in Sec.~\ref{simonstuff} below.  In this
subsection, we critically examine
the origin, status and uniqueness of the semiclassical equations.
Specifically, we consider the following two basic issues:

\begin{itemize}

\item
How is the ``semiclassical approximation'' derived and what is its
domain of applicability, i.e.,~what is the class of
states $\omega$ and classical metrics $g_{ab}$ for which the
semiclassical approximation is good?

\item
How unique is the semiclassical Einstein equation
(\ref{basic}) itself, i.e.,~how unique is the
prescription to obtain the expected stress energy tensor
$\langle {\hat T}_{ab} \rangle$?
\end{itemize}

With regard to the first issue, it should first be noted that
the classical metric, $g_{ab}$, appearing
in the semiclassical equations presumably must correspond --- from the
vantage point of a complete quantum theory of gravity --- to the
expected value of a quantum metric operator in some
state. However, we wish to
emphasize here that the ``average value'' of a metric which has nonzero
amplitudes to correspond to different spacetime geometries is an
intrinsically gauge non-covariant concept; it cannot even be
defined unless an algorithm is given which completely fixes all gauge
freedom, i.e.,
which rigidly fixes a coordinate system for each spacetime
geometry. This phenomenon can be seen
even in the simple case of a classical
probability distribution which assigns probability $1/2$ to the
spacetime
$(M, g^{(1)}_{ab})$ and probability $1/2$ to the spacetime
$(M, g^{(2)}_{ab})$.  We could say that the expected metric on $M$ is
$\langle g_{ab} \rangle = (g^{(1)}_{ab} + g^{(2)}_{ab})/2$. However, we
could equally well have represented the second spacetime as
$(M, \psi^*g^{(2)}_{ab})$, where $\psi: M \to M$ is any diffeomorphism.
The expected metric would then be computed to be
$\langle {g'}_{ab} \rangle = (g^{(1)}_{ab} + \psi^* g^{(2)}_{ab})/2$.
However,
$\langle {g}_{ab} \rangle$ and $\langle {g'}_{ab} \rangle$ will
{\em not}, in general,
be gauge equivalent (nor is there any guarantee that either
of them will even define Lorentz metrics!). Thus, unless one eliminates
all gauge freedom, the notion of an ``expected metric'' will make sense
only in a limit where the fluctuations of the spacetime geometry about its mean
value
are negligibly small. Such a limit undoubtedly would have to be
contemplated in any case in order to justify a semiclassical
approximation,
but it is worth bearing in mind that the quantity one is trying to
calculate
in the semiclassical theory is ill defined (without complete gauge
fixing) except in this limit. Note that this same phenomenon would occur
in the calculation of the expected value of any gauge dependent
quantity in any non-abelian gauge theory.

There are at least two different approximation schemes by which the
semiclassical equation (\ref{basic}) can be formally derived, and
these give rise to distinct viewpoints on the range of validity of
this equation \cite{others}. In the first scheme one formally expands
a quantum metric and quantum scalar field about Minkowski spacetime
(or, more generally, any classical vacuum solution
\cite{vacuumnote}), and derives the equation of motion (\ref{basic})
for the expected metric by keeping only the ``tree diagrams'' for
gravitons and the ``tree diagrams'' and ``one-loop'' terms in the
scalar field \cite{Jordan,oneloop}.
[As is well known, predictions can be obtained
at one loop order despite the fact that the theory is
non-renormalizable, at the expense of having to introduce two new,
undetermined, coupling constants into the theory; see
Eq.~(\ref{STgeneral}) below.]  This loop expansion formally
corresponds to an expansion in powers of $\hbar$, so keeping only the
``one-loop'' terms corresponds to keeping only the lowest order
correction in $\hbar$ to the equations of motion.  There is no obvious
mathematical or physical justification for dropping the one-loop
``graviton'' terms, since the quantum effects of the metric field are,
a priori, just as important as those of a scalar field
\cite{gravitonST}.  It is possible that the neglect of the one-loop
graviton terms could be justified for certain choices of the state of
the scalar and gravitational fields --- wherein the expected scalar
stress-energy tensor dominates the corresponding effective graviton
contribution, and the fluctuations in the scalar stress-energy tensor
are suitably small --- but we are not aware of any analysis
demonstrating this.  However, in any case, the effects of the graviton
loops would be expected to be qualitatively similar to the effects of
the scalar field loops, so, at the very least, Eq.~(\ref{basic}) can
be justified as a simplified model of the exact equations resulting
from keeping all one-loop terms.  Note that the one-loop graviton
terms could be handled within this approximation by treating the
metric perturbation as a linear field propagating in a background
classical spacetime.  However, the phenomenon described in the
previous paragraph would then manifest itself by the fact that, at
second order in the perturbed metric, the difference between the
Einstein tensor of the expected metric and the expected value of the
Einstein tensor (which effectively acts as a ``graviton stress energy
tensor'') would be gauge non-covariant.

Thus, in this first scheme, the semiclassical Einstein equation
(\ref{basic}) would be, at best, an approximate equation valid for
certain states of the scalar and gravitational fields, and, at worst,
be a ``model equation'' whose properties should be qualitatively
similar to the equation resulting from a complete one-loop
approximation. In either case, higher loop contributions would modify
Eq.~(\ref{basic}) by terms proportional to quadratic and higher powers
of $\hbar$.  In
the context of perturbation theory off of Minkowski spacetime, a
formal expansion in powers of $\hbar$ is equivalent to an expansion in
powers of $1/{\cal L}^2$ (keeping constants of nature such as $\hbar$
fixed), where ${\cal L}$ denotes a typical lengthscale associated with
a solution.  Thus, from the viewpoint of this first scheme we would
expect there to be modifications to Eq.~(\ref{basic}) which are of
order $\sim L_p^2 / {\cal L}^2$, and consequently the domain of
validity of the equation is restricted to ${\cal L} \gg L_p$, where
$L_p$ is the Planck length.

In the second scheme the semiclassical Einstein equation (\ref{basic})
is formally derived from a fully quantum treatment by imagining that
there are $N$ decoupled scalar fields present --- all of which are in
the same quantum state --- and then taking the limit $N \to \infty$,
with $G N = ({\rm constant})$ (see, e.g., Ref.~\cite{HartleHorowitz}).
In this ``$1/N$ expansion'', the graviton loops are suppressed
relative to the matter loops simply because there are $N$ scalar
fields but only one graviton field.  Note that since the scalar fields
are free (i.e., non-self-interacting), only one-loop terms in the
scalar fields arise.  Furthermore, in the $N \to \infty$ limit, the
fluctuations in the expected total stress-energy tensor of the scalar
fields becomes negligible. Thus, one formally obtains
Eq.~(\ref{basic}) exactly in the $N \to \infty$ limit. Corrections to
this equation should be of order $1/N$ or higher.  Thus, one might
expect that if $N$ is sufficiently large, Eq.~(\ref{basic}) would be a
good approximation up to the curvature scale corresponding to the
effective Planck length in the rescaled theory, ${\cal L} \sim L_{\rm
p,eff} = \sqrt{N} L_p$.  Indeed, estimates of the order of magnitude of
successive terms in the loop expansion support this viewpoint.
However, it is far from clear that the loop expansion would provide a
good approximation in this regime, so we do not believe that there are
solid grounds for believing that ``graviton effects" can be neglected
for any finite value of $N$ when ${\cal L} \sim L_{\rm p,eff}$. In
other words, we would expect effects which are 
non-perturbative in $\hbar$ could be important at these scales.  One
piece of evidence that this is the case is that this seems to be the
only way to escape the conclusion that flat spacetime is unstable
\cite{HartleHorowitz}.  As we shall see in detail in Appendix
\ref{firstordersolutions} below, pathological solutions of the
linearized version of Eq.~(\ref{basic}) exist on scales ${\cal L} \sim
L_{\rm p,eff}$, and we will not regard these solutions as being
physical.

We comment that the existence of these two different
points of view on the status of the semiclassical equation has given
rise to some controversies in the literature.  Some implications of
the first point of view have been discussed at length by Simon
\cite{Simon,Simon1,Simon2,Simon3} (see Sec.~\ref{simonstuff} below for
further discussion).  On the other hand,
Suen \cite{Suen} disagreed with Simon's analysis and argued that there
should not be any corrections to Eq.~(\ref{basic}).  In our view, the
existence of higher order corrections to Eq.~(\ref{basic}) depends on
whether or not one justifies that equation in terms of a one-loop
approximation or by the invocation of a $1/N$ limit. Certainly,
corrections to Eq.~(\ref{basic}) {\it will} appear in the physically
realistic case of finite $N$.

The viewpoint will shall adopt in this paper is the following: We will
appeal to the $1/N$ limit to give a mathematically clean justification
for ignoring graviton contributions to Eq.~(\ref{basic}). Thus, in an
$\hbar$ (or, equivalently, a ``long wavelength") expansion, we shall
regard Eq.~(\ref{basic}) as valid to all orders in $\hbar$ (or,
equivalently, to all orders in inverse wavelength), although in most of
our analyses we will not make use of its validity beyond order
$\hbar^2$.  However, as already indicated above, we will not view
Eq.~(\ref{basic}) (or any other known equation) as adequate for
describing phenomena where wavelengths comparable to $L_p$ occur. This
will justify our modification of Eq.~(\ref{basic}) in
Sec.~\ref{simonstuff} below using the reduction of order procedure,
so that it still describes ``long wavelength" behavior accurately, but
no longer predicts pathological behavior at the Planck scale.

We now turn to a discussion of the second issue, i.e.,
the uniqueness of the right hand side of
Eq.~(\ref{basic}). If we assume
that the prescription for obtaining the
expected stress tensor satisfies the axioms discussed in
Ref.~\cite{Wald94}, then the expected stress tensor is
unique up the addition of local, conserved curvature tensors.  These
are tensors which are functionals of the metric, and whose value at a
point depend only on the geometry in an arbitrarily small
neighborhood of that point.  This ambiguity is assumed to
be only a two parameter ambiguity \cite{ambiguity,note3}, because
in general spacetimes there are only two independent conserved local
curvature tensors of dimensions $({\rm{length}})^{-4}$, which are
explicitly given by \cite{localHs}
\FL
\begin{eqnarray}
A_{ab} &=& {1 \over \sqrt{-g}} {\delta \over \delta g^{ab}} \int d^4 x
\sqrt{-g} \, C_{cdef} C^{cdef} \nonumber \\
\mbox{} &=& -2 \Box R_{ab} + {2\over3} \nabla_a \nabla_b R +
{1\over3} \Box R g_{ab} -{1\over3} R^2 g_{ab} \nonumber \\
&& + {4\over3} R \, R_{ab} + (R_{cd} R^{cd})
g_{ab} - 4 R_{acbd} R^{cd}
\label{Adefgeneral}
\end{eqnarray}
and
\FL
\begin{eqnarray}
B_{ab} &=& {1 \over \sqrt{-g}} {\delta \over \delta g^{ab}} \int d^4 x
\sqrt{-g} \, R^2 \nonumber \\
\mbox{} &=& 2 \nabla_a \nabla_b R -2 \Box R \, g_{ab}
 +{1\over2} R^2 \,g_{ab} -2 R \, R_{ab}.
\label{Bdefgeneral}
\end{eqnarray}
Thus, if we denote by
$\langle {\hat T}_{ab} \rangle_{\rm point\ split}$ the
stress tensor given by the point splitting algorithm (briefly
reviewed in the next
subsection), then the appropriate right hand side for
Eq.~(\ref{basic}) must be of the form
\begin{equation}
\langle {\hat T}_{ab} \rangle = \langle {\hat T}_{ab}
\rangle_{\rm point\ split} + \alpha A_{ab} + \beta B_{ab}.
\label{STgeneral}
\end{equation}
Here $\alpha$ and $\beta$ are dimensionless coefficients which can be
regarded as free parameters.  Different renormalization schemes
predict different values of $\alpha$ and $\beta$.
Indeed, $\alpha$ and $\beta$ may be
viewed as new ``coupling constants'' which must be introduced into the
theory as a result of the non-renormalizability of quantum gravity
(coupled to matter) at one-loop order. In quantum gravity,
at higher (graviton)
loop orders, additional new ``coupling constants'' would
have to be introduced as coefficients of the conserved local
curvature terms of the appropriate dimension for that order; these would
constitute a portion of the $O(\hbar^2)$ and higher corrections to the
semiclassical equation referred to above.

An important feature of $\langle {\hat T}_{ab} \rangle$ is that it has
an anomalous behavior under a scaling of the spacetime metric and
the corresponding scaling transformation of the state.
To see this more explicitly, consider, for simplicity, the case of a
massless field.  Under the scaling transformation $g_{ab} \to \mu^2
g_{ab}$,
we must scale the 2-point function, $G$, of the quantum field
as $G(x,y) \to \mu^{-2} G(x,y)$ in order to preserve Eq.~(\ref{Gform})
below. Physically, this scaling of the quantum state can be interpreted
as
preserving the ``particle content'' of that state, so that the new state
in the
new metric $\mu^2 g_{ab}$ corresponds simply to increasing the
wavelength of all of the particles by the factor $\mu$. Note that this
required scaling of $G$ contrasts sharply with the situation in the
classical theory, where the amplitude,
$\Phi$, of the scalar field may be scaled in an arbitrary
manner independently of the scaling of the metric tensor
\cite{scalingnote}. In the
classical theory, the stress-energy tensor also scales in a
straightforward
manner under any combined scaling transformation of $g_{ab}$
and $\Phi$. However, the situation is quite different in
the semiclassical theory because one is forced to introduce a
lengthscale,
$\lambda_0$, in the prescription for defining $\langle {\hat T}_{ab}
\rangle$. In particular, in the point-splitting algorithm (reviewed
briefly in the next subsection), a
lengthscale implicitly enters into the logarithmic term in the local
Hadamard subtraction term $G^H(x,y)$ \cite{Wald94}; a lengthscale
similarly enters all other regularization prescriptions as a
``renormalization point'' or ``cutoff''.
Under a change of this lengthscale
$\lambda_0 \to \mu \lambda_0$, we have \cite{Page,Visser}
\FL
\begin{eqnarray}
\langle T_{ab} \rangle \left[ g_{cd}, \mu \lambda_0 \right] &=&
\langle T_{ab} \rangle \left[ g_{cd}, \lambda_0 \right] \nonumber \\
\mbox{} && + 4 \pi \, \log \mu
\, \left[ a A_{ab} + b B_{ab}\right],
\label{logscalegeneral}
\end{eqnarray}
where $a$, $b$ are specific numerical coefficients that depend on the
curvature coupling $\xi$ [see Eq.~(\ref{abdef}) below]. Equivalently, if
we keep
$\lambda_0$ fixed but scale the metric and state via
\begin{eqnarray}
g_{ab} &\to& \mu^2 g_{ab}  \nonumber \\
\mbox{} G(x,y) &\to& \mu^{-2} G(x,y).
\label{quantumscale}
\end{eqnarray}
we obtain
\FL
\begin{eqnarray}
\langle T_{ab} \rangle \left[ \mu^2 g_{cd}, \lambda_0 \right] &=&
\mu^{-2} \langle T_{ab} \rangle \left[ g_{cd}, \lambda_0 \right]
\nonumber \\
\mbox{} &&- 4 \pi \,
\mu^{-2} \log \mu \, \left[ a A_{ab} + b B_{ab}\right].
\label{logscalegeneral1}
\end{eqnarray}
Note that the ambiguity (\ref{logscalegeneral}) in $\langle {\hat
T}_{ab} \rangle$ resulting from the need to introduce a lengthscale is
subsumed by the more general ambiguity given by
Eq.~(\ref{STgeneral}). Indeed, one way of describing the above
anomalous scaling behavior (\ref{logscalegeneral}) is to say that a
particular linear combination of the two new dimensionless ``coupling
constants'' $\alpha$ and $\beta$ is a ``running coupling constant'',
i.e., in effect, its value depends upon the scale one is considering.

The scaling behavior given by Eq.~(\ref{logscalegeneral}) has two
important consequences.  First, it shows that (at least part of) the
ambiguity occurring in Eq.~(\ref{STgeneral}) for a massless field
cannot be eliminated by any criterion arising from the study of the
quantum field theory of that field propagating in a fixed classical
background spacetime, since that theory does not have a preferred
lengthscale, whereas any prescription fixing $\alpha$ and $\beta$
would have the effect of determining a lengthscale. (Although a
massive field does have an associated lengthscale -- namely $1/m$ --
using this lengthscale to fix $\alpha$ and $\beta$ would give rise to
singular behavior in the $m \to 0$ limit.)  The second consequence is
that the scaling behavior (\ref{logscalegeneral}) will affect the
nature of solutions in the ``long wavelength'' limit. As we will
discuss in detail in Sec.~\ref{scaling1} below, when we perform an
expansion in a
``wavelength parameter'' $L_p / {\cal L}$, we will need to introduce
terms which vary as powers of $\ln [L_p / {\cal L}] (L_p^2/{\cal L}^2)
$ as well as powers of $L_p^2/{\cal L}^2$.

Although the numerical values of the dimensionless
coefficients $\alpha$ and $\beta$
are not known -- and, indeed, are not determinable without a more complete
theory -- we shall assume below that their values are of order unity
``at the Planck scale''. More precisely, if we define
$\langle {\hat T}_{ab} \rangle_{\rm point\ split}$
by choosing the lengthscale $\lambda_0$
arising in that prescription to be $L_p$, then we shall assume that
the correct formula for $\langle {\hat T}_{ab} \rangle$ is given by
Eq.~(\ref{STgeneral}) with $\alpha$ and $\beta$ of order unity.
However, our results concerning ANEC will
be valid for all values of $\alpha$ and $\beta$.

\subsection{The semiclassical equations}
\label{semiclgeneral}

Consider now the semiclassical theory where the metric $g_{ab}$ is
treated classically, but where the scalar field $\Phi$ is treated as a
quantum field. The metric and quantum state, $\omega$ are required to
satisfy the semiclassical Einstein equation (\ref{basic}).
The coupled, evolving degrees of freedom in the semiclassical theory
consist of (i) the metric, and (ii) all the observables associated
with the scalar field. These field observables may be taken to be the
$n$-point correlation functions
$\langle {\hat \Phi}(x_1)...{\hat \Phi}(x_n)  \rangle_\omega$ in the
given quantum state, $\omega$.
However, as we discuss further below, the expected value of the
stress tensor in any state is determined
via the point splitting procedure from a knowledge of only the two-point
function,
\begin{equation}
G(x,y) = \langle {\hat \Phi}(x) {\hat \Phi}(y)  \rangle_\omega.
\label{Gdef}
\end{equation}
Moreover, in free field
theory (which we are considering here) the evolution of the two-point
function is decoupled from that of the higher $n$-point functions.
Thus, if we are only interested in the metric and not in other
observables depending on the state of the scalar field, we can regard
the semiclassical equations as a set of coupled equations for the
metric $g_{ab}$ and the distributional bi-solution $G(x,y)$
to the scalar wave equation (\ref{waveeqn}).
This is a key feature which simplifies our analysis.  From this point
of view, states which differ only in their $n$ point functions for $n
\ne 2$ are effectively identical as far as semiclassical gravity is
concerned.

The appropriate set of bi-distributional solutions for a given, fixed,
globally hyperbolic spacetime $(M,g_{ab})$ can be characterized as
follows \cite{Wald91,Wald94}: Let $S(M)$ denote the space of smooth
solutions of the Klein-Gordon equation (\ref{waveeqn}) with initial
data of compact
support on Cauchy surfaces, and let $C_0^\infty(M)$ be the space of
smooth test functions of compact support  on spacetime.  Define the usual
Klein-Gordon-like symplectic product on pairs $F, G \in S(M)$
\begin{equation}
\Omega(F,G) = -\int_\Sigma \, F
{\mathop{\nabla}\limits^{\leftrightarrow}}_a  G \, d \Sigma^a,
\label{KGprod}
\end{equation}
where $\Sigma$ is any Cauchy surface.  Let $G(f,g)$ denote the
two-point bidistribution evaluated on (``integrated against'') test
functions $f,g \in C_0^\infty(M)$.  Then $G(f,g)$ must be of the form
\begin{equation}
G(f,g) = {1\over2} G^{(1)}(f,g) + {i \over 2} \Omega(Ef,Eg),
\label{Gform}
\end{equation}
where $Ef$ denotes the advanced minus retarded solution with
source $f$.  Furthermore,
the symmetric part $G^{(1)}/2$ of $G$ must satisfy the
positivity conditions $G^{(1)}(f,f) \ge0$ and
\begin{equation}
G^{(1)}(f,f) G^{(1)}(g,g) \ge \Omega(Ef,Eg)^2.
\label{positivity}
\end{equation}
Moreover, in order that the stress tensor be well defined, the
two-point function $G(x,y)$ must be of so-called Hadamard form
\cite{Wald91}.

To specify more explicitly the coupled evolution equations for the
metric
$g_{ab}(x)$ and the two-point function $G(x,y)$, we need to discuss
the point splitting prescription for calculating the stress tensor.  Let
$G^H(x,y)$ denote the locally constructed Hadamard bidistribution
given by the algorithm described in
Refs.~\cite{Wald91,Wald94,Garabedian},
as specified/modified in the following
way.  Use the differential operator
\begin{equation}
{\cal D} \equiv \Box - m^2 - \xi R.
\label{diffop}
\end{equation}
Choose the lengthscale, $\lambda_0$, implicitly appearing in the
logarithmic term in $G^H(x,y)$ to be the Planck length $L_p$,
choose $w_0=0$ in the notation of Ref.~\cite{Wald94}, and
truncate the series expansion after three terms.  Then the
regulated two-point function
\begin{equation}
f(x,y) \equiv G(x,y) - G^H(x,y)
\end{equation}
will be well defined in a neighborhood of the ``diagonal'' $x = y$
of $M \times M$ and will be at least $C^2$ for
Hadamard states \cite{Wald91}.  The
expected value of the stress tensor will be given by
\FL
\begin{equation}
\langle T_{ab}(x)\rangle_{\rm point\ split} =
\lim_{y \to x} {\cal D}_{ab} f(x,y) + Q(x) g_{ab}(x),
\label{Tabformal}
\end{equation}
where ${\cal D}_{ab}$ is a
particular second order differential operator and $Q$
is a particular local curvature invariant  \cite{Wald94}.  Note that
if we modify the
prescription for calculating $G^H$ by truncating the series after say
four terms instead of three terms, then the regularized two-point
function $f(x,y)$ will be altered, but the value (\ref{Tabformal}) of
$\langle T_{ab}(x)\rangle_{\rm point\ split}$ will not be changed.
As explained in the previous subsection, we shall assume that the
correct value of $\langle T_{ab}(x)\rangle$ is given by
Eq.~(\ref{STgeneral}), with $\alpha$ and $\beta$ of order unity.

In summary, the independent variables in the semiclassical
evolution equations consist of a smooth metric $g_{ab}(x)$ and
a bidistribution $G(x,y)$.
$G(x,y)$ is required to satisfy the wave equation
(\ref{waveeqn}) in each variable, as well as Eqs.~(\ref{Gform}),
(\ref{positivity}) and the Hadamard condition. Finally, $g_{ab}$
and $G$ are required to satisfy the semiclassical Einstein equation
(\ref{basic}), with $\langle T_{ab} \rangle$
given by Eqs.~(\ref{STgeneral}) and (\ref{Tabformal}).
The gauge freedom in this formulation of semiclassical
gravity simply consists of
the diffeomorphisms $\varphi: M\to M$,
under which $g_{ab}$ and $G(x,y)$ get transformed by the natural
action of $\varphi$.

Although the above formulation of semiclassical gravity is fully
satisfactory mathematically, the unknown variable $G(x,y)$ has a
distributional character, and it would be more convenient to specify
the independent degrees of freedom in terms of a smooth function. This
can be done as follows in the physically relevant case of a spacetime
$(M, g_{ab})$ which becomes flat in the asymptotic past. First, we
assume that the state is sufficiently regular and that
the approach to flatness of the spacetime occurs at a sufficiently rapid
rate that $G(x,y)$ asymptotically approaches the two-point function
$G_{\rm in}(x,y)$ of a state, $\omega_{\rm in}$,
in Minkowski spacetime $(M,\eta_{ab})$.
For sufficiently regular states of a massless field, asymptotic
flatness of $(M, g_{ab})$ at null
infinity should suffice for this to hold, with $G_{\rm in}(x,y)$ being
the two-point function of the state in Minkowski spacetime with the
same initial data on past null infinity ${\cal J}^-$ as $G(x,y)$ [under a
suitable identification of ${\cal J}^-$ for $(M, g_{ab})$ with ${\cal J}^-$
for $(M,\eta_{ab})$].  For a massive field, it is less clear precisely
what asymptotic conditions on $(M, g_{ab})$ would suffice, but the
necessary conditions presumably would be qualitatively similar to
those for a massless field.  Note that the positivity condition will
hold for $G(x,y)$ if and only if it holds for $G_{\rm in}(x,y)$.  This
is because, in general, the positivity condition (\ref{positivity}) can
be expressed as a condition on initial data on $\Sigma \times \Sigma$,
where $\Sigma$ is any Cauchy surface \cite{Wald91}, and will be
preserved under evolution.  The two-point functions $G$
and $G_{\rm in}$ have the same initial data at ${\cal J}^-$, and are both
evolved forward using the homogeneous wave equation with respect to
the appropriate metric ($g_{ab}$ for $G$, $\eta_{ab}$ for $G_{\rm
in}$).

Now, let $G_{\rm in,0}(x,y)$ denote the two-point function of the
ordinary vacuum state, $\omega_{\rm in,0}$, in Minkowski spacetime
$(M,\eta_{ab})$, and let $G_0(x,y)$ be two-point function of the
corresponding state, $\omega_0$, in $(M, g_{ab})$ which approaches
$G_{\rm in,0}(x,y)$ in the asymptotic past; in other words, let
$G_0(x,y)$ be the two-point function of the ``in vacuum state'' in
$(M, g_{ab})$. For any Hadamard state, $\omega$, in $(M, g_{ab})$ with
two-point function $G(x,y)$, we define
\begin{equation}
F(x,y) = G(x,y) - G_0 (x,y).
\label{newF}
\end{equation}
Then, clearly, $F$ contains the
same information as the bidistribution $G$. However, the Hadamard
condition on $G$ now corresponds simply to the statement that $F(x,y)$
is a smooth function on $M\times M$. Furthermore, Eq.~(\ref{Gform})
is equivalent to the statement that $F$ is symmetric and
real-valued. In addition,
$F(x,y)$ satisfies the wave equation (\ref{waveeqn}) in each variable
\begin{equation}
{\cal D}_x F(x,y) = {\cal D}_y F(x,y) =0,
\label{basic2}
\end{equation}
by virtue of the fact that the bi-distributions $G(x,y)$ and
$G_0 (x,y)$ each satisfy this equation. The only other restriction on
$F$ is
the positivity condition arising from the corresponding condition
(\ref{positivity}) on $G$.

Since the expected stress tensor $\langle T_{ab} \rangle_\omega$
has a linear dependence on the two-point
distribution, $G$, associated with $\omega$, we
have, for any Hadamard state, $\omega$,
\begin{equation}
\langle T_{ab} \rangle_\omega = \langle T_{ab} \rangle_{\omega_0}
+ \langle T_{ab} \rangle_F.
\label{newTab}
\end{equation}
Here as above $\omega_0$ denotes the ``in'' vacuum state, and
\begin{equation}
\langle T_{ab} \rangle_F \equiv
\lim_{y \to x} {\cal D}_{ab} F(x,y),
\label{TabF}
\end{equation}
with ${\cal D}_{ab}$ being the same differential operator as appeared in
Eq.~(\ref{Tabformal}).  The vacuum polarization term $\langle T_{ab}
\rangle_{\omega_0}$
is functional of the spacetime metric $g_{ab}$ alone. For massless
fields,
it has been evaluated by Horowitz \cite{Horowitz} to first order in the
perturbed metric about flat spacetime, and we will make use of
Horowitz' formula in our analysis below.

Thus, in our reformulation of semiclassical theory, the
independent variables consist of a smooth metric $g_{ab}(x)$ and
a smooth, symmetric, real-valued function $F(x,y)$ which satisfies the
positivity
condition arising from Eq.~(\ref{positivity}).  The dynamical evolution
equations are simply that $F(x,y)$ satisfy the wave equation
(\ref{waveeqn}) in each variable with respect to the metric $g_{ab}$,
and
that $g_{ab}$ and $F$ satisfy the semiclassical Einstein equation
(\ref{basic}) with stress-energy tensor given
by Eqs.~(\ref{newTab}) and (\ref{TabF}). We remark
that a purely classical theory, where gravity is coupled to a
statistical ensemble of scalar field configurations, would differ
from this version of semiclassical gravity only in the following two
respects. First, in the classical theory, the term
$\langle T_{ab} \rangle_{\omega_0}$ would be absent from
Eq.~(\ref{newTab}). One may
view this term, which is proportional to $\hbar$, as
describing the vacuum polarization effects occurring in the
quantum theory.  Second, the quantum
mechanical positivity condition on $F$ arising from
Eq.~(\ref{positivity}) (which can be viewed as
a restriction on the space of allowed initial data for $F$ on $\Sigma
\times \Sigma$ for any Cauchy surface $\Sigma$) is less restrictive
than the corresponding classical positivity condition.

Under our above assumptions about the asymptotic behavior of states,
it follows from Eq.~(\ref{newF}) that in the asymptotic past
$F(x,y)$ approaches the smooth function
$F_{\rm in}(x,y)$ on Minkowski spacetime defined by
\begin{equation}
F_{\rm in}(x,y) = G_{\rm in}(x,y) - G_{\rm in,0}(x,y).
\label{newFin}
\end{equation}
Note that
$F_{\rm in}$ is just the usual ``regularized'' two-point function of
the state $\omega_{\rm in}$ in Minkowski spacetime.  It satisfies
a positivity condition which ensures that
$G_{\rm in}$ is the two-point distribution of a state in Minkowski
spacetime. We may view
$F_{\rm in}(x,y)$ and a corresponding quantity describing the incoming
classical gravitational radiation [see Eq.~(\ref{Delh}) above] as the
freely specifiable ``initial data'' for semiclassical gravity. In the
next section, we will develop a systematic perturbation expansion for
semiclassical solutions in terms of this ``initial data''.

We conclude this section by reminding the reader
of our notation for the four different states
under consideration here: (i) the state of interest, $\omega$,
on the curved spacetime $(M,g_{ab})$
(ii) the ``incoming state'' $\omega_{\rm in}$ on Minkowski
spacetime $(M,\eta_{ab})$, such that the n-point functions of $\omega$
approach those of $\omega_{\rm in}$ in the asymptotic past (iii) the
vacuum state, $\omega_{\rm in,0}$ in Minkowski spacetime, and (iv) the
corresponding ``in-vacuum'' state $\omega_0$ in the curved spacetime
$(M,g_{ab})$, such that the n-point functions of $\omega_0$ approach
those of $\omega_{\rm in,0}$ in the asymptotic past.  The two-point
functions of these states are, respectively, $G$, $G_{\rm in}$,
$G_{\rm in,0}$, and $G_0$.

\section{Perturbation theory about flat spacetime}
\label{deriveperteqns}

\subsection{Derivation of the equations}

In this section we derive the explicit form of
the perturbation expansion of the
semiclassical equations off of the Minkowski spacetime/vacuum
solution, valid to second order in deviation from flatness.  Let
$\left( M, g_{ab}(\varepsilon), \omega(\varepsilon) \right)$ be
a smooth one parameter family of solutions  of
the semiclassical equations discussed in the last section, such that
$(M,g_{ab}(0)=\eta_{ab})$ is
Minkowski spacetime and $\omega(0)$ is the Minkowski vacuum
state, $\omega_{\rm in,0}$. We expand all relevant quantities about
$\varepsilon = 0$ as follows:
\begin{eqnarray}
g_{ab}(\varepsilon) &=& \eta_{ab} + \varepsilon h^{(1)}_{ab} + \varepsilon^2
h^{(2)}_{ab} + O(\varepsilon^3) \nonumber \\
\omega_{\rm in} &=& \omega_{\rm in,0} + \varepsilon \,\omega_{\rm
in}^{(1)}
+ \varepsilon^2 \omega_{\rm in}^{(2)} + O(\varepsilon^3) \nonumber \\
F &=& \varepsilon F^{(1)} + \varepsilon^2 F^{(2)} + O(\varepsilon^3) \nonumber
\\
F_{\rm in} &=& \varepsilon F_{\rm in}^{(1)} + \varepsilon^2 F_{\rm in}^{(2)} +
O(\varepsilon^3) \nonumber
\\
{\cal D} &=& {\cal D}^{(0)} + \varepsilon {\cal D}^{(1)} +
O(\varepsilon^2) \nonumber \\
{\cal D}_{ab} &=& {\cal D}_{ab}^{(0)} + \varepsilon
{\cal D}_{ab}^{(1)} + O(\varepsilon^2),
\label{expandall}
\end{eqnarray}
where $\omega_{\rm in}$, $F$, $F_{\rm in}$, ${\cal D}$, and
${\cal D}_{ab}$ were defined in the previous subsection.  For the
remainder of this paper the operators ${\cal D}_{ab}$, ${\cal
D}_{ab}^{(1)}$ and ${\cal D}_{ab}^{(2)}$ will be implicitly understood to
include the operation of taking the coincidence limit $y
\to x$; these operators thus act on functions on $M \times M$ and take
values in
the space of tensors on $M$.
Note that it would not make sense to introduce an expansion in
$\varepsilon$ of the family of states $\omega(\varepsilon)$, as
for different values of $\varepsilon$, these states
act on the different algebras of observables corresponding to
the different spacetimes. For this reason, we view the state
$\omega(\varepsilon)$ as a functional of the
``in" state $\omega_{\rm in}(\varepsilon)$ on Minkowski spacetime
and of the spacetime metric
$g_{ab}(\varepsilon)$, i.e., $\omega = \omega[\omega_{\rm in},g_{cd}]$,
and we then expand $\omega_{\rm in}(\varepsilon)$.
Note that for a
fixed metric $g_{ab}$, the state $\omega$ is a linear function of
$\omega_{\rm in}$.

We write
\begin{equation}
\langle T_{ab}[g_{cd}] , \omega_{\rm in} \rangle \equiv \langle
T_{ab}[g_{cd}] \rangle_{\omega[\omega_{\rm in},g_{cd}]}.
\label{notationdefine}
\end{equation}
This defines the tensor $T_{ab}$ on the left hand side as a linear map
on the space of states on Minkowski spacetime which takes values in
the space of conserved tensors on the spacetime $(M,g_{ab})$.
Using the expansion
(\ref{expandall}) for $g_{ab}(\varepsilon)$, we express this linear
map on Minkowski spacetime states in the form as
\begin{eqnarray}
&& T_{ab}[g_{cd}(\varepsilon)] = T_{ab}^{(0)} + \varepsilon
T_{ab}^{(1)}[h^{(1)}] \nonumber \\
\mbox{}& &+ \varepsilon^2 \left\{ T_{ab}^{(1)}[h^{(2)}] +
T_{ab}^{(2)}[h^{(1)},h^{(1)}]
\right\} + O(\varepsilon^3),
\label{STexpand}
\end{eqnarray}
where $T_{ab}^{(0)}$ is the usual stress tensor operator in Minkowski
spacetime.  Equation (\ref{STexpand}) defines the tensors
$T_{ab}^{(1)}$ and $T_{ab}^{(2)}$ which act on metric
perturbations and pairs of metric perturbations respectively.

The perturbative semiclassical Einstein equations are obtained by
inserting
the expansions (\ref{expandall}) and (\ref{STexpand})
into Eq.~(\ref{basic}). We obtain at first order,
\begin{equation}
\kappa G_{ab}^{(1)}[h^{(1)}] = \langle T_{ab}^{(0)}, \omega_{\rm in}^{(1)}
\rangle +
\langle T_{ab}^{(1)}[h^{(1)}], \omega_{\rm in,0} \rangle
\label{firstorder}
\end{equation}
and at second order, we get
\begin{eqnarray}
&&\kappa G_{ab}^{(1)}[h^{(2)}] + \kappa G_{ab}^{(2)}[h^{(1)},h^{(1)}] =
\langle T_{ab}^{(0)}, \omega_{\rm in}^{(2)} \rangle \nonumber \\
\mbox{} & & + \langle T_{ab}^{(1)}[h^{(1)}], \omega_{\rm in}^{(1)} \rangle +
\langle T_{ab}^{(1)}[h^{(2)}], \omega_{\rm in,0} \rangle \nonumber \\
\mbox{} & & + \langle T_{ab}^{(2)}[h^{(1)},h^{(1)}], \omega_{\rm in,0} \rangle.
\label{secondorder}
\end{eqnarray}
Here the tensors $G^{(1)}$ and $G^{(2)}$ are defined by the identity
\begin{equation}
G_{ab}[\eta + \alpha h] = \alpha G^{(1)}_{ab}[h] + \alpha^2 G^{(2)}_{ab}[h,h]
+ O(\alpha^3),
\label{G12def}
\end{equation}
which holds
for any tensor $h_{cd}$, i.e.,~they are the linear and quadratic parts
of the Einstein tensor.  Explicit expressions for $G^{(1)}$ are given in
for example Ref.~\cite{Wald}, and for $G^{(2)}$ in Ref.~\cite{Habison}.

Before discussing the explicit form of the terms appearing on the right
sides of Eqs.~(\ref{firstorder}) and (\ref{secondorder}), it may be
useful
to write the equations in the notation appropriate to the case where
the one parameter family
of incoming states $\omega_{\rm in}(\varepsilon)$ corresponds to a one
parameter family of density matrices ${\hat \rho}(\varepsilon)$ on
the usual Fock space ${\cal H}$ of states on Minkowski spacetime.
If this family of density matrices is expanded as
\begin{equation}
{\hat \rho}(\varepsilon) = {\left| 0 \right>} {\left< 0 \right|} + \varepsilon
{\hat \rho}^{(1)} +
\varepsilon^2 {\hat \rho}^{(2)} + O(\varepsilon^3),
\label{rhoexpand}
\end{equation}
then Eqs.~(\ref{firstorder}) and (\ref{secondorder}) can be rewritten
as
\begin{equation}
\kappa G_{ab}^{(1)}[h^{(1)}] = {\left< 0 \right|} T^{(1)}_{ab}[h^{(1)}] {\left|
0 \right>} + {\rm tr}\left[
{\hat \rho}^{(1)} T^{(0)}_{ab} \right].
\label{firstorderI}
\end{equation}
and
\begin{eqnarray}
&&\kappa  G_{ab}^{(1)}[h^{(2)}] + \kappa G_{ab}^{(2)}[h^{(1)},h^{(1)}]
\nonumber \\
\mbox{}&=& {\rm tr}\left[ {\hat \rho}^{(2)} {T}^{(0)}_{ab} \right]
+ {\rm tr}\left[ {\hat \rho}^{(1)} {T}^{(1)}_{ab}[h^{(1)}] \right] \nonumber
\\
\mbox{} &&  + {\left< 0 \right|}{T}^{(1)}_{ab}[h^{(2)}] {\left| 0 \right>} +
{\left< 0 \right|} {T}^{(2)}_{ab}[h^{(1)},h^{(1)}]
{\left| 0 \right>}.
\label{secondorderI}
\end{eqnarray}

In Eqs.~(\ref{firstorder}) and (\ref{secondorder}), the terms involving
$\omega_{\rm in,0}$ are the ``vacuum polarization" terms, corresponding
to the first term on the right hand side of Eq.~(\ref{newTab}). The term
$\langle T_{ab}^{(1)}[h^{(1)}], \omega_{\rm in,0} \rangle$ has been computed
by Horowitz \cite{Horowitz}
in the massless case, and we will review Horowitz' results
in subsection \ref{STlinear} below. The term
$\langle T_{ab}^{(2)}[h^{(1)},h^{(1)}], \omega_{\rm in,0} \rangle$ has not
been computed, and, to avoid having to do so, we will eventually
pass to an approximation in which this term can be neglected
(see subsection \ref{scaling1} below).

We now derive explicit expressions for the non-vacuum-polarization
terms that appear in Eqs.~(\ref{firstorder}) and (\ref{secondorder}).
{}From Eqs.~(\ref{TabF}) and (\ref{expandall}) we obtain
\FL
\begin{equation}
\langle T_{ab} \rangle_F = \varepsilon {\cal D}^{(0)}_{ab} F^{(1)} +
\varepsilon^2 \left[ {\cal D}_{ab}^{(0)} F^{(2)} + {\cal D}_{ab}^{(1)} F^{(1)}
\right]
+ O(\varepsilon^3).
\end{equation}
Furthermore, from Eqs.~(\ref{basic2}) and (\ref{expandall}), we obtain
\begin{eqnarray}
{\cal D}_x^{(0)} F^{(1)}(x,y) &=& 0 \nonumber \\
{\cal D}_x^{(0)} F^{(2)}(x,y) &=& - {\cal D}_x^{(1)} F^{(1)}(x,y),
\end{eqnarray}
together with similar equations involving $y$-derivatives.
Since $F \to F_{\rm in}$ in the asymptotic past, the solutions to
these equations are
\begin{eqnarray}
F^{(1)} &=& F_{\rm in}^{(1)} \nonumber \\
F^{(2)} &=& F_{\rm in}^{(2)} + {\cal E}\left[ - {\cal D}_x^{(1)} F_{\rm
in}^{(1)}, -
{\cal D}_y^{(1)} F_{\rm in}^{(1)} \right].
\label{F1F2solns}
\end{eqnarray}
Here the quantity ${\cal E}[s_x,s_y]$ is defined  for any
sources $s_x(x,y)$ and $s_y(x,y)$ satisfying
${\cal D}^{(0)}_y s_x = {\cal D}^{(0)}_x s_y$, by
\FL
\begin{eqnarray}
{\cal E}[s_x,s_y](x^\prime,y^\prime) &=& \int_M d^4x \,  G^{(0)}_{\rm
ret}(x^\prime,x) s_x(x,y^\prime) \nonumber \\
\mbox{} && + \int_M d^4y \,  G^{(0)}_{\rm ret}(y^\prime,y)
s_y(x^\prime,y)
\nonumber \\
\mbox{} && - \int_M d^4x \int_M d^4y \,  G^{(0)}_{\rm
ret}(x^\prime,x) \nonumber \\
\mbox{} && \times \, G^{(0)}_{\rm ret}(y^\prime,y) {\tilde s}(x,y),
\label{deltaF}
\end{eqnarray}
where ${\tilde s} = {\cal D}^{(0)}_y s_x = {\cal D}^{(0)}_x s_y$, and
$G^{(0)}_{\rm
ret}$ is
the retarded Greens function for the differential operator ${\cal
D}^{(0)}$.
Combining all of the above results together with Eqs.~(\ref{newTab}),
(\ref{firstorder}) and (\ref{secondorder}) yields the fairly obvious
relations
\begin{eqnarray}
\langle T_{ab}^{(0)}, \omega_{\rm in}^{(1)} \rangle &=& {\cal D}^{(0)}_{ab}
F_{\rm
in}^{(1)} \nonumber \\
\mbox{} \langle T_{ab}^{(0)}, \omega_{\rm in}^{(2)} \rangle &=& {\cal
D}^{(0)}_{ab}
F_{\rm
in}^{(2)},
\label{obvious}
\end{eqnarray}
together with
\begin{eqnarray}
\langle T_{ab}^{(1)}[h^{(1)}], \omega_{\rm in}^{(1)} \rangle &=& {\cal
D}^{(1)}_{ab}\,
F^{(1)}_{\rm in} \nonumber \\
\mbox{} &&+ {\cal D}^{(0)}_{ab} \, {\cal E}\left[ - {\cal D}_x^{(1)} F_{\rm
in}^{(1)}, -
{\cal D}_y^{(1)} F_{\rm in}^{(1)} \right].
\label{hardterm1}
\end{eqnarray}

Finally, we note that the perturbative semiclassical equations have
the following structure: we can specify arbitrarily the incoming state
perturbations $\omega_{\rm in}^{(1)}$ and $\omega_{\rm in}^{(2)}$ (or, more
precisely, just their two-point functions), as well as the incoming
metric perturbations $h_{ab}^{(1) \, ,{\rm in}}$ and $h_{ab}^{(2) \,
,{\rm in}}$ [see Eq.~(\ref{Delh}) above].  We then may solve
Eqs.~(\ref{firstorder}) and (\ref{secondorder}) to obtain the metric
perturbations $h_{ab}^{(1)}$ and $h_{ab}^{(2)}$ \cite{finiteenergy}.

\subsection{Pure states versus mixed states}
\label{pureversusmixed}

As discussed in the introduction, our results concerning ANEC depend
crucially on whether the incoming state $\omega_{\rm in}$ is pure or
mixed to first order in $\varepsilon$.  Consider first the
case where all the states, $\omega_{\rm in}(\varepsilon)$,
correspond to density matrices, ${\hat \rho}(\varepsilon)$, in the usual
Fock space. Then the perturbed
state will be pure to first order if and only if
\begin{equation}
{\hat \rho}(\varepsilon) = \left| \Psi(\varepsilon) \right> \left<
\Psi(\varepsilon) \right| + O(\varepsilon^2),
\label{pureexpand}
\end{equation}
where
\begin{equation}
\left|\Psi(\varepsilon)\right> \ = \ \left|0\right> + \varepsilon
\left|\psi\right>^{(1)}  + O(\varepsilon^2).
\label{purestateexpand}
\end{equation}
Thus the state is pure to first order if and only if
\begin{equation}
{\hat \rho}^{(1)} = {\left| 0 \right>} {\left< \psi \right|}^{(1)} + {}^{(1)}
{\left| \psi \right>} {\left< 0 \right|},
\label{rho1pure}
\end{equation}
for some ${\left| \psi \right>}^{(1)} \in {\cal H}$ with $\left< 0 | \psi
\right>^{(1)}=0$.
By contrast, the most
general first order density matrix perturbation is of the form
\FL
\begin{equation}
{\hat \rho}^{(1)} = {\left| 0 \right>} {\left< \psi \right|}^{(1)} + {}^{(1)}
{\left| \psi \right>} {\left< 0 \right|} + {\hat P} - ({\rm
tr} {\hat P}) {\left| 0 \right>} {\left< 0 \right|},
\label{rho1mixed0}
\end{equation}
where ${\hat P}$ is a positive, Hermitian trace class operator on
${\cal H}$ such that ${\hat P} {\left| 0 \right>} = 0$.  Thus, the perturbed
state is pure to first order if and only if ${\hat P}=0$.

By inspection, when Eq.~(\ref{rho1pure}) holds,
it can be seen that the two-point
function $F^{(1)}_{\rm in}$ will have the property that its
mixed-frequency part [i.e., the part that is positive frequency
with respect to one variable and negative frequency with respect to
the other; see Eq.~(\ref{F1explicit}) below] vanishes.  Moreover, the
converse is also true, since if
the mixed-frequency part vanishes we have
\begin{equation}
{\rm tr}\left[ {\hat \rho}^{(1)} \, {\hat \Phi}_-(u) {\hat \Phi}_+(u)
\right] =0
\label{vanish2}
\end{equation}
for any test function $u$, where we have used the decomposition
\begin{equation}
{\hat \Phi} = {\hat \Phi}_+ + {\hat \Phi}_-
\label{posneg}
\end{equation}
of the field operator into its positive and negative frequency parts.
[In a conventional mode expansion, ${\hat \Phi}_+$ would consist of the
annihilation operators and ${\hat \Phi}_-$ the creation operators.]
Now since the operator ${\hat P}$
is trace class, it is compact.  Thus there will exist an orthonormal
basis $\left|\psi_j\right>$ of the space of states orthogonal to
${\left| 0 \right>}$ such that
\begin{equation}
{\hat P} = \sum_{j=0}^\infty p_j \left|\psi_j\right> \left<
\psi_j\right|
\end{equation}
for some $p_j\ge0$, $j=0,1,2 \ldots$, where the convergence is in the
operator norm topology.  Equations (\ref{rho1mixed0}) and
(\ref{vanish2}) now imply that
\begin{equation}
\sum_j p_j \, || {\hat \Phi}_+(u) \left| \psi_j\right> \, ||^2 = 0.
\end{equation}
However, if ${\hat \Phi}_+(u) \left| \psi_j\right> = 0$ for all $u$,
then
$\left|\psi_j\right> = {\left| 0 \right>}$, which contradicts the fact that
$\left|\psi_j\right>$ is orthogonal to ${\left| 0 \right>}$.
Therefore $p_j=0$ for all $j$, and the incoming state is pure to first
order.

In addition, it follows immediately from Eq.~(\ref{rho1mixed0}) that
if ${\hat \rho}^{(1)}$ is a possible
first order state perturbation, then $- {\hat \rho}^{(1)}$ will be an
allowable first order state perturbation if and only if the state is
pure to first order.  This has the implication, which we discussed
in the introduction, that if the ANEC integral is to be
non-negative generally, it must vanish
at first order for pure states, but not necessarily for mixed states.

For general, algebraic states there is a more abstract notion of
purity, which defines a state $\omega$ to be pure if it cannot be
written in the form $c \omega_1 + (1-c) \omega_2$ where
$\omega_1$ and $\omega_2$ are distinct states and $0 < c < 1$
(see, e.g.,~Ref.~\cite{Wald94}). Thus, the pure states are extreme
boundary points of the convex linear space of all states. It should
follow
that for general algebraic states, if $\omega_{\rm in}^{(1)}$ is a possible
first order state perturbation off of a pure state,
then $- \omega_{\rm in}^{(1)}$ will be an
allowable first order state perturbation if and only if the
perturbed state is pure to first order.
In addition, the positivity
condition, Eq.~(\ref{positivity}), applied to both
$\omega_{\rm in,0} + \varepsilon \omega_{\rm in}^{(1)}$
and $\omega_{\rm in,0} - \varepsilon \omega_{\rm in}^{(1)}$
should then imply the
vanishing of the mixed-frequency part of the two-point function for
general first-order pure states.
Thus, the conclusions of the previous paragraph
should continue to hold for general, algebraic states, although we have
not attempted to give a rigorous proof of these results.

Our analysis of ANEC given in Sec.~\ref{generalformulae} below will
divide into two cases, depending upon whether or not the mixed
frequency part of the perturbed two point function vanishes.  In the
remainder of this paper, we shall use the terminology ``pure perturbed
two point function'' to mean a perturbed two point function for which
the mixed frequency part vanishes, and ``mixed perturbed two point
function'' to mean one for which the mixed frequency part does not
vanish. For states in the usual Fock space -- and, presumably, also
for general algebraic states -- a perturbed state will be pure to
first order if and only if its perturbed two-point function is pure.

\subsection{Gauge freedom and transformations}

In this subsection we analyze the gauge freedom in the perturbation
equations. The gauge freedom in the one parameter
family of exact solutions $\left( M, g_{ab}(\varepsilon),
\omega(\varepsilon) \right)$ consists simply of one
parameter families of diffeomorphisms
$\varphi_\varepsilon : M \to M$, where $\varphi_0$ is the identity
map \cite{caveat3b}. Here, these diffeomorphisms act simultaneously
on $g_{ab}$ and the two-point function $G$ of $\omega$.

It can be shown \cite{Geroch} that for an arbitrary one-parameter
family of diffeomorphisms $\varphi_\varepsilon$ with $\varphi_0$
being the identity map,
there exist unique vector fields $\xi^{a \, (1)}$ and $\xi^{a \,
(2)}$ on $M$, such that to order $\varepsilon^2$
\begin{equation}
\varphi_\varepsilon = {\cal D}_{\xi^{(2)}}(\varepsilon) \circ {\cal
D}_{\xi^{(1)}}(\varepsilon^2/2).
\label{gaugefreedom1}
\end{equation}
Here ${\cal
D}_\tau(\lambda) : \, M \to M$ denotes the
one parameter group of diffeomorphisms generated by $\tau^a$
The vector fields $\xi^{a \, (1)}$ and $\xi^{a \,
(2)}$ are given by the following formulae in terms of
their actions on test functions $f \in C_0^\infty(M)$:
\begin{equation}
\xi^{a \, (1)} \nabla_a f = {d \over d \varepsilon} (f \circ
\varphi_\varepsilon)_{|\varepsilon=0}
\end{equation}
and
\begin{equation}
\xi^{a \, (2)} \nabla_a f = {d^2 \over d \varepsilon^2} (f \circ
\varphi_\varepsilon)_{|\varepsilon=0} - {\cal L}_{\xi^{(1)}} \, {\cal
L}_{\xi^{(1)}} \,f,
\end{equation}
where ${\cal L}$ denotes the Lie derivative.
Thus, the gauge freedom in the second order perturbation equations can
be parameterized by pairs of vector fields on $M$.

Now let $T(\varepsilon)$ be any one parameter family of tensor fields
on $M$ (we suppress tensor indices), which has the expansion
\begin{equation}
T(\varepsilon) = T^{(0)} + \varepsilon T^{(1)} + \varepsilon^2 T^{(2)} +
O(\varepsilon^3).
\end{equation}
Then from Eq.~(\ref{gaugefreedom1}) we can calculate the
transformation properties of the expansion coefficients $T^{(0)}$, $T^{(1)}$
etc.  We find
\begin{equation}
\varphi^*_\varepsilon T(\varepsilon) = T^{(0)} + \varepsilon {\bar T}^{(1)} +
\varepsilon^2 {\bar T}^{(2)} + O(\varepsilon^3),
\end{equation}
where
\begin{equation}
{\bar T}^{(1)} = T^{(1)} + {\cal L}_{\xi^{(1)}} T^{(0)},
\end{equation}
and
\begin{eqnarray}
{\bar T}^{(2)} &=& T^{(2)} + {1\over2}{\cal L}_{\xi^{(2)}} T^{(0)}  + {1\over2}
{\cal
L}_{\xi^{(1)}} {\cal L}_{\xi^{(1)}} T^{(0)} \nonumber \\
\mbox{} && + {\cal L}_{\xi^{(1)}} T^{(1)}.
\end{eqnarray}
Let us denote gauge transformed quantities by overbars.  Then we find
\FL
\begin{eqnarray}
{\bar h}^{(1)} &=& h^{(1)} + {\cal L}_{\xi^{(1)}} \eta \nonumber \\
{\bar h}^{(2)} &=& h^{(2)} + {1\over2} {\cal L}_{\xi^{(2)}}\eta + {1\over2}
{\cal
L}_{\xi^{(1)}} {\cal L}_{\xi^{(1)}} \eta  + {\cal L}_{\xi^{(1)}} h^{(1)}
\nonumber \\
{\bar G}^{(1)} &=& G^{(1)} \nonumber \\
{\bar G}^{(2)} &=& G^{(2)} + {\cal L}_{\xi^{(1)}} G^{(1)} \nonumber \\
{\bar F}^{(1)}(x,y) &=& F^{(1)}(x,y) \nonumber \\
{\bar F}^{(2)}(x,y) &=& F^{(2)}(x,y) + {\cal L}^x_{\xi^{(1)}} F^{(1)}(x,y)
\nonumber \\
\mbox{} && + {\cal L}^y_{\xi^{(1)}} F^{(1)}(x,y).
\label{gaugeaction}
\end{eqnarray}
We use these formulae in Sec.~\ref{secondorderanalysis} below.

\subsection{The linearized stress tensor and the explicit form of the
first order perturbation equation}
\label{STlinear}

In this subsection we analyze the vacuum polarization term
appearing in the first order
perturbation equation (\ref{firstorder}) in the massless case, $m=0$.
This has been calculated by Horowitz \cite{Horowitz} using an
axiomatic approach, who obtained
\FL
\begin{eqnarray}
\left< {T}^{(1)}_{ab}[h^{(1)}](x), \omega_{\rm in,0} \right>  &=& \int_{M} d^4y
\bigg\{
H_\lambda(x-y) \nonumber \\
\mbox{} && \times \left[a A^{(1)}_{ab}(y) + b B^{(1)}_{ab}(y)\right] \bigg\}
\nonumber \\
\mbox{} && + \alpha A^{(1)}_{ab}(x) + \beta B^{(1)}_{ab}(x).
\label{linearvacpol}
\end{eqnarray}
Here $A^{(1)}_{ab}$ and $B^{(1)}_{ab}$ are the linearized versions of the
local curvature tensors (\ref{Adefgeneral}) and (\ref{Bdefgeneral}):
\begin{eqnarray}
A^{(1)}_{ab} &=& - 2 \Box G^{(1)}_{ab} + {2\over3} \nabla_a \nabla_b R^{(1)} -
{2\over3} \eta_{ab} \Box R^{(1)} \nonumber \\
B^{(1)}_{ab} &=&  -2 \eta_{ab} \Box R^{(1)} +2 \nabla_a \nabla_b R^{(1)},
\label{ABdef}
\end{eqnarray}
where $R^{(1)}$ is the linearized Ricci scalar and the derivative operators
$\nabla^a$ and $\Box$ are the zeroth order derivative operators
associated with the flat metric $\eta_{ab}$.
The coefficients $a$ and
$b$ in Eq.~(\ref{linearvacpol}) are constants which depend on the
curvature coupling $\xi$.
These coefficients were given by Horowitz for the cases $\xi =0$
and $\xi = 1/6$; the general formulae can be derived from point
splitting and are
\begin{eqnarray}
a &=& {1 \over 4 \pi (960 \pi^2)} \nonumber \\
\mbox{} b &=& {(1-6 \xi)^2 \over 4 \pi (576\pi^2)}.
\label{abdef}
\end{eqnarray}
Note that the coefficient $a$ is positive for all values of $\xi$, and
moreover the corresponding coefficient for other fields such as
Maxwell and neutrino fields is also positive \cite{Horowitz}.  This
fact will be be relevant in our analysis below.
The quantities $\alpha$, $\beta$ in Eq.~(\ref{linearvacpol})
are free parameters, c.f., the discussion after
Eq.~(\ref{STgeneral}).  The parameter $\lambda$
is a free parameter with dimensions of length
corresponding to the lengthscale $\lambda_0$ discussed in subsection
\ref{status}.  The
quantity $H_\lambda$ is a distribution with support on the past
light cone, rather like the retarded Greens function solution to the
massless wave equation.  An explicit formula for $H_\lambda$ is
\FL
\begin{equation}
H_\lambda(x)  = \lim_{\alpha \to 0^-} \left[\,
- \delta^\prime(\sigma - \alpha) \Theta_+(x) + 2 \pi \ln[-\alpha /
\lambda^2]\,
\delta^4(x)\, \right],
\label{Hlambda}
\end{equation}
where $\sigma = x_a x^a/2$, $\Theta_+(x)$ takes the value $1$ inside
the past light cone and vanishes elsewhere, and it is understood
that the limit $\alpha
\to 0^-$ is taken after integrating against a test function.  See
Horowitz \cite{Horowitz} or Jordan \cite{Jordan} for more details.

The expression (\ref{linearvacpol}) is essentially a special case of the
general formula (\ref{STgeneral}), but linearized and specialized to the
incoming vacuum state.  As such it contains a linearized version of the
two parameter local curvature ambiguity described by the parameters
$\alpha$ and $\beta$.  Also the logarithmic scaling described by
Eq.~(\ref{logscalegeneral}) has a counterpart in
Eq.~(\ref{linearvacpol}): the distribution (\ref{Hlambda}) has
the property that
\begin{equation}
H_{\lambda^\prime}(x) - H_\lambda(x) = 4 \pi \ln(\lambda /
\lambda^\prime) \delta^4(x).
\label{Hscaling}
\end{equation}
Therefore the free parameters $\alpha$, $\beta$ and $\lambda$ are not
independent.  In Sec.~\ref{status} we chose to make $\lambda = L_p$,
thus fixing the values of $\alpha$ and $\beta$.  In the linearized
analysis here and below, we follow Horowitz \cite{Horowitz} and choose
that value
of $\lambda$ which makes $\alpha=0$.  Thus, the two independent free
parameters in the linearized stress tensor are $\beta$ and $\lambda$.
Note that our assumption discussed in Sec.~\ref{status} that $\alpha$
and $\beta$ are ``of order unity at the Planck scale'' translates into
the assumption that
\begin{eqnarray}
\lambda &\sim& L_p \nonumber \\
\beta &\sim& 1.
\label{parameterassumption}
\end{eqnarray}

{}Formula (\ref{linearvacpol}) completes the explicit specification of all
the terms appearing in the linearized semiclassical Einstein equation
(\ref{firstorder}). Thus, using Eq.~(\ref{obvious}) and setting
$\alpha=0$, the complete, explicit form of this equation
is
\FL
\begin{eqnarray}
\kappa G_{ab}^{(1)}[h^{(1)}] &=& \hbar {\cal D}^{(0)}_{ab} F_{\rm in}^{(1)} +
\hbar
\beta
B^{(1)}_{ab}(x)
\nonumber \\
\mbox{} && + \hbar \int_M d^4y \bigg\{
H_\lambda(x-y) \nonumber \\
\mbox{} && \times
\left[a A^{(1)}_{ab}(y) + b B^{(1)}_{ab}(y)\right] \bigg\}.
\label{firstorderexplicit}
\end{eqnarray}
For later convenience, we have explicitly inserted the factors of
$\hbar$ appearing in this equation \cite{dimensions}.  It can be seen
that the right hand side of Eq.~(\ref{firstorderexplicit}) contains
terms
involving fourth derivatives of the perturbed metric, so the
linearized semiclassical Einstein equation has the nature of a fourth
order integro-differential equation rather than a second order
differential equation.  As previously remarked, it can be solved by
specifying the source term
\begin{equation}
s_{ab} = \langle T_{ab}^{(0)} , \omega_{\rm in}^{(1)} \rangle =
 {\cal D}_{ab}^{(0)} F_{\rm in}^{(1)}
\label{source0}
\end{equation}
and solving for the metric perturbation $h_{ab}^{(1)}$.  Its exact
solutions have been discussed in detail by Horowitz \cite{Horowitz},
in the special case of the homogeneous version of the equation,
without the source term (\ref{source0}).
In Appendix \ref{firstordersolutions} we
obtain all solutions to Eq.~(\ref{firstorderexplicit}) whose spatial
Fourier transforms exist, thereby generalizing the analysis of
Horowitz to allow for a non-vacuum incoming quantum
state.

As found in Appendix \ref{firstordersolutions} and in
Ref.~\cite{Horowitz}, the linearized semiclassical Einstein equation
has, in effect, more degrees of freedom than the corresponding
classical equation, and new ``run-away" solutions exist.  Thus,
presumably, not all of the exact solutions should be regarded as
physical. Section \ref{simonstuff} below will be devoted to the issue
of how to extract physical information from
Eq.~(\ref{firstorderexplicit}) and equations of a similar character.

\subsection{Explicit form of the second order perturbation equation in
the ``long-wavelength'' limit}
\label{scaling1}

As we have previously indicated, we will find that the ANEC integral
vanishes identically for pure states to first order in deviation from
flatness. This is a key necessary condition for the validity of ANEC,
but it does not, by itself, provide strong evidence that any version
of ANEC actually holds for pure states.  In order to investigate this
issue further, it is necessary to go (at least) to second order
perturbation theory. However, we are unable to do this for general
perturbations because we do not have an explicit expression for the
term $\langle T_{ab}^{(2)}[h^{(1)},h^{(1)}], \omega_{\rm in,0} \rangle$
appearing
in Eq.~(\ref{secondorder}). Nevertheless, as we now shall describe,
there are limiting circumstances under which this unknown term will be
negligible compared with the other terms appearing in
Eq.~(\ref{secondorder}).  These limiting circumstances correspond to
the case of ``long wavelengths" (compared with the Planck scale) of
the field and perturbed metric, together with the condition that the
first order perturbed metric not be dominated by incoming
gravitational radiation. These conditions should encompass a wide
range of physically interesting and potentially achievable
situations. In Section \ref{secondorderanalysis} below, we will perform
an
analysis of the validity of ANEC for pure states to second order in
perturbation theory in this limit.

We now give a precise description of the long wavelength limit in
terms of one-parameter families of solutions to the semiclassical
equations, in which the characteristic lengthscale ${\cal L}$ of a
solution satisfies ${\cal L} \to \infty$, while the length scales that
determine the semiclassical theory, $L_p$ and $\lambda$ [the
lengthscale appearing in Eq.~(\ref{firstorderexplicit})], are kept
fixed.  However, we remark that this limit
is equivalent to a limit in which $\hbar = L_p^2 \to 0$ with
${\cal L}$ and $\lambda / L_p$ fixed, i.e., the long-wavelength
limit is equivalent to a $\hbar \to 0$ limit, as long as the
lengthscale $\lambda$ is taken to scale proportionally to $L_p$.

As previously discussed in subsection \ref{status}, under
the scaling $g_{ab} \to \alpha^2 g_{ab}$ of the spacetime
metric, the natural scaling of the two-point function (corresponding to
keeping the particle content fixed but going to longer wavelengths
by the factor $\alpha$) is
\begin{equation}
G(x,y) \to \alpha^{-2} G(x,y).
\end{equation}
Thus, in the context of perturbation theory off of flat spacetime, the
transformation corresponding to keeping the incoming
particle content the same but increasing the wavelength of the particles
by the factor $\alpha$ is given by
$\eta_{ab} \to \alpha^2 \eta_{ab}$ and
$F_{\rm in}{^{(n)}} \to \alpha^{-2} F_{\rm in}{^{(n)}}$ for all $n$.
Equivalently, by applying the diffeomorphism
$x^\mu \to \alpha x^\mu$ to this transformation, we see that the
``long wavelength limit" should correspond to the large $\alpha$ limit
of solutions to the perturbative semiclassical equations
on a fixed background Minkowski spacetime, $(M, \eta_{ab})$,
with a one-parameter family of states whose $n$th order (in
$\varepsilon$) perturbed incoming initial data varies as
\begin{equation}
F_{\rm in}{^{(n)}}(x,y; \alpha) = \alpha^{-2}
{\bar F}_{\rm in}{^{(n)}}(x/\alpha,y/\alpha),
\label{Finalpha}
\end{equation}
where ${\bar F}_{\rm in}{^{(n)}}(x,y)$ is the perturbed incoming
initial data of some fixed (i.e., $\alpha$ independent) state
\begin{equation}
{\bar \omega}_{\rm in}(\varepsilon) = \omega_{\rm in,0} + \varepsilon
\,{\bar \omega}_{\rm in}^{(1)}
+ \varepsilon^2 {\bar \omega}_{\rm in}^{(2)} + O(\varepsilon^3).
\label{rhoexpand1}
\end{equation}
Without loss of generality, we may
choose ${\bar \omega}_{\rm in}$ to have a characteristic lengthscale,
${\cal L}_0$, of order the Planck length, ${\cal L}_0
\sim L_p$.  We may then interpret the parameter $\alpha$ as
measuring the
characteristic lengthscale ${\cal L}$ of the initial state in units of
the
Planck length.

The quantity
\begin{equation}
s_{ab} = \langle T_{ab}^{(0)} , \omega_{\rm in}^{(1)} \rangle
\label{source}
\end{equation}
acts as a source term for the first order metric perturbation,
$h_{ab}^{(1)} (x; \alpha)$, in the linearized
semiclassical Einstein equation (\ref{firstorder}). For the
family of incoming states given by Eq.~(\ref{Finalpha}),
$s_{ab}$ scales as
\begin{equation}
s_{ab}(x; \alpha) =
\alpha^{-4} \langle T_{ab}^{(0)} , {\bar \omega}_{\rm in}^{(1)}
\rangle(x/\alpha).
\label{salpha}
\end{equation}
In order to examine the
behavior of Eq.~(\ref{firstorder}) under
scaling, it is convenient to make the change of variables
${\hat h}^{(1)} (x; \alpha) \equiv h^{(1)} (\alpha x; \alpha)$.
If we substitute Eq.~(\ref{salpha}) into Eq.~(\ref{firstorder}) and use
Eqs.~(\ref{linearvacpol}) and (\ref{Hscaling}), we find that the
explicit
$\alpha$ dependence of the resulting equation is given by
\begin{eqnarray}
\kappa G_{ab}^{(1)} [{\hat h}^{(1)}]  &=&  {1\over \alpha^2}
\left[ \langle T^{(0)}_{ab} , {\bar \omega}^{(1)}_{\rm in} \rangle
+ \langle T^{(1)}_{ab}[{\hat h}^{(1)}], \omega_{\rm in,0} \rangle \right]
\nonumber \\
\mbox{} &&+ {\ln \alpha \over \alpha^2} Z_{ab}^{(1)} [{\hat h}^{(1)}].
\label{alphadep}
\end{eqnarray}
Here
\begin{equation}
Z_{ab}^{(1)} \equiv 4 \pi \left[ a A^{(1)}_{ab} + b B^{(1)}_{ab} \right]
\label{Zabdef}
\end{equation}
is the linearized anomalous scaling contribution to the vacuum
polarization discussed in subsections \ref{status} and \ref{STlinear}
above; see Visser \cite{Visser} for extensive further discussion.
Note that the $\alpha$ dependence in Eq.~(\ref{alphadep}) exactly
mirrors the $\hbar$ dependence in Eq.~(\ref{firstorderexplicit}), provided
that we assume that $\lambda \propto L_p$ [see Eq.~(\ref{Hscaling})
above].  This is as expected from our remark above concerning the
equivalence of the $\hbar \to 0$ and long-wavelength limits, since
\begin{equation}
{1\over \alpha^2} \sim {{L_p}^2 \over {\cal L}^2} = {\hbar \over
{\cal L}^2}.
\label{alphahbar}
\end{equation}

The solution to the first order perturbation equation with source
(\ref{salpha}) can be written as [c.f., Eq.~(\ref{Delh}) above]
\begin{equation}
h_{ab}^{(1)}(x,\alpha) = h_{ab}^{(1) \, ,{\rm in}}(x,\alpha) + \Delta
h_{ab}^{(1)}(x,\alpha),
\label{Delh1}
\end{equation}
where $h_{ab}^{(1)\, ,{\rm in}}$ is the homogeneous solution to
Eq.~(\ref{alphadep}) with the same initial data as $h_{ab}^{(1)}(x,\alpha)$
at ${\cal J}^-$ and $\Delta h_{ab}^{(1)}$ is the retarded
solution of that equation with source (\ref{salpha}). [Note, however,
that
on account of the higher derivative nature of Eq.~(\ref{alphadep}),
there
will be more initial data to specify for $h_{ab}^{(1)}(x,\alpha)$ than
occurs for
the classical linearized Einstein equation. This situation will be
rectified
when we replace Eq.~(\ref{alphadep}) with the reduced order equation
(\ref{firstordermixed3}) in subsection \ref{reductionoforder} below.]
To specify a one-parameter family of
solutions, we are free to chose any scaling with $\alpha$ we wish for
the initial data for $h_{ab}^{(1)\, ,{\rm in}}$  \cite{caveatqg}.
However, a natural choice of
scaling is suggested by the following considerations.  If we were
solving the classical linearized Einstein equation with source
(\ref{salpha}), the retarded solution would scale as $\Delta {h^{\rm
cl}}_{ab}(x; \alpha) = \alpha^{-2} \Delta{{\bar h}^{\rm cl}}_{ab} (x/
\alpha)$.  The retarded solution $\Delta h_{ab}^{(1)}$ to the linearized
semiclassical Einstein equation will not scale this simply on account
of the anomalous scaling appearing in Eq.~(\ref{alphadep}), but the
dominant scaling behavior will still be of this form [see
Eqs.~(\ref{ansatz}) and (\ref{chiexpand}) below]. Thus, if we
scale the initial data for the
incoming gravitational radiation so that near ${\cal J}^-$ we have
\begin{equation}
h_{ab}^{(1) \, ,{\rm in}}(x; \alpha) =
\alpha^{-2} {\bar h}_{ab}^{(1) \, ,{\rm in}} (x/ \alpha),
\label{gravradncondt}
\end{equation}
then from Eq.~(\ref{Delh1}) the first order semiclassical metric
perturbation will not be dominated by incoming gravitational radiation
in the long wavelength limit $\alpha \to \infty$. This is the
situation we wish to consider with regard to the second order
perturbation equations.

Before considering the second order equations, it is useful to make the
change of variables
\begin{equation}
\chi_{ab}^{(1)}(x,\alpha) \equiv \alpha^2 h_{ab}^{(1)}(\alpha x, \alpha).
\label{ansatz}
\end{equation}
{}From Eq.~(\ref{alphadep}), we obtain
\begin{eqnarray}
\kappa G_{ab}^{(1)}[\chi^{(1)}] &=&
\langle T^{(0)}_{ab} , {\bar \omega}^{(1)}_{\rm in} \rangle
+ {\ln \alpha \over \alpha^2} Z_{ab}^{(1)}[\chi^{(1)}]
\nonumber \\
\mbox{} &&+ {1\over \alpha^2}\, \langle T^{(1)}_{ab}[\chi^{(1)}] , \omega_{\rm
in,0} \rangle.
\label{firstordermixed1}
\end{eqnarray}
With the above ansatz for the form of the incoming
gravitational radiation, the $\alpha$ dependence of $\chi^{(1)}$ will be
given
by
\begin{equation}
\chi^{(1)} = \chi^{(1,0)} + {\ln \alpha \over \alpha^2} \chi^{(1,1)} +
{1\over \alpha^2} \chi^{(1,2)} + O[(\ln
\alpha)^2/\alpha^4],
\label{chiexpand}
\end{equation}
where the $\chi^{(1,j)}$ are independent of $\alpha$.

We now turn to the second order perturbation equation
(\ref{secondorder}).  Note that once $h_{ab}^{(1)}$ has been obtained,
this equation as an equation for $h_{ab}^{(2)}$ has exactly the same
structure as the first order equation, except that the source term now
includes additional pieces constructed from the first order quantities:
\begin{equation}
\kappa G_{ab}^{(1)}[h^{(2)}] = s_{ab}^{(2)} + \langle T_{ab}^{(1)}[h^{(2)}],
\omega_{\rm
in,0}
\rangle,
\label{secondorder2}
\end{equation}
where
\FL
\begin{eqnarray}
s_{ab}^{(2)} &=&
\langle T_{ab}^{(0)}, \omega_{\rm in}^{(2)} \rangle
+ \langle T_{ab}^{(1)}[h^{(1)}], \omega_{\rm in}^{(1)} \rangle \nonumber \\
\mbox{} && + \langle T_{ab}^{(2)}[h^{(1)},h^{(1)}], \omega_{\rm in,0} \rangle -
\kappa G_{ab}^{(2)}[h^{(1)},h^{(1)}].
\label{secondordersource}
\end{eqnarray}
Consider now the behavior of the source term (\ref{secondordersource})
under the scaling given by Eqs.~(\ref{Finalpha}), (\ref{rhoexpand1})
and (\ref{gravradncondt}).  If we make the change of variables
\begin{equation}
{\bar s}_{ab}^{(2)}(x,\alpha) \equiv \alpha^4 s_{ab}^{(2)}(\alpha x,\alpha)
\label{s2alphadep}
\end{equation}
(which is chosen to make ${\bar s}_{ab}^{(2)}$ independent of $\alpha$ in
the large $\alpha$ limit), then we find that
\FL
\begin{eqnarray}
{\bar s}_{ab}^{(2)} &=& \langle T_{ab}^{(0)}, {\bar \omega}_{\rm in}^{(2)}
\rangle
+ {1 \over \alpha^2} \langle T_{ab}^{(1)}[\chi^{(1)}], {\bar \omega}_{\rm
in}^{(1)} \rangle \nonumber \\
\mbox{} && - {1 \over \alpha^2} \kappa
G_{ab}^{(2)}[\chi^{(1)},\chi^{(1)}] + {\ln \alpha \over \alpha^4}
Z_{ab}^{(2)}[\chi^{(1)},\chi^{(1)}] \nonumber \\
\mbox{} &&+ {1 \over \alpha^4}
\langle T_{ab}^{(2)}[\chi^{(1)},\chi^{(1)}], \omega_{\rm in,0} \rangle.
\label{secondordersource1}
\end{eqnarray}
Here the quantity $Z_{ab}^{(2)}$ is just the second
order part of the anomalous scaling tensor
\begin{equation}
Z_{ab} \equiv 4 \pi \left[ a A_{ab} + b B_{ab} \right];
\label{Zabdef1}
\end{equation}
see Eqs.~(\ref{logscalegeneral}), (\ref{ABdef}) and (\ref{Zabdef}).
In our calculation of the second order perturbation to the ANEC
integral in Sec.~\ref{secondorderanalysis} below, we shall work to
$O(1/\alpha^2)$ beyond leading order, and consequently we can drop the
last two, vacuum polarization terms from the right hand side of
Eq.~(\ref{secondordersource1}).  In particular, the uncalculated term
$\langle T_{ab}^{(2)}[\chi^{(1)},\chi^{(1)}], \omega_{\rm in,0} \rangle$ may be
dropped. Moreover, in this equation we can
replace $\chi^{(1)}$ by its leading order approximation $\chi^{(1,0)}$ in
the
large
$\alpha$ limit, given by
\begin{equation}
\kappa G_{ab}^{(1)}[\chi^{(1,0)}] = \langle T_{ab}^{(0)},{\bar \omega}_{\rm
in}^{(1)}\rangle
\label{chi10eqn}
\end{equation}
(see Eqs.~(\ref{firstordermixed1}) and (\ref{chiexpand}) above).
Making the change of variables ${\hat
h}^{(2)} (x; \alpha) \equiv h^{(2)} (\alpha x; \alpha)$, the resulting
equation is
\FL
\begin{eqnarray}
\kappa G^{(1)}_{ab}[{\hat h}^{(2)}] &=& {1 \over \alpha^2} \langle
T^{(0)}_{ab},
{\bar \omega}^{(2)}_{\rm in} \rangle + {\ln \alpha \over \alpha^2}
Z^{(1)}_{ab}[{\hat h}^{(2)}] \nonumber \\
\mbox{} && + {1 \over \alpha^2} \langle T^{(1)}_{ab}[{\hat
h}^{(2)}],\omega_{\rm in,0} \rangle + {1 \over \alpha^4} \langle
T_{ab}^{(1)}[\chi^{(1,0)}],{\bar \omega}_{\rm in}^{(1)} \rangle \nonumber \\
\mbox{} && - {\kappa \over \alpha^4}
G_{ab}^{(2)}[\chi^{(1,0)},\chi^{(1,0)}] + O\left( {\ln \alpha \over
\alpha^6} \right).
\label{secondorderapprox}
\end{eqnarray}

\section{Dealing with the pathological solutions: Simon's
prescription(s).}
\label{simonstuff}

As seen in the previous section, a key feature of the linearized
semiclassical Einstein equation is that it fails to remain a second
order partial differential equation; rather it contains fourth order
time derivatives of the metric.  Consequently, the semiclassical
equations have more ``degrees of freedom'' than the classical
equations, i.e., additional free functions --- namely, the second and
third time derivatives of the metric --- can be specified as initial
data.  Directly related to the presence of these additional degrees of
freedom is the existence --- for all values of the free parameters
$\xi$, $\beta$ and $\lambda$ --- of new, ``pathological'' solutions of
the semiclassical equations, which grow exponentially in time on a
timescale of order the Planck time (see Appendix
\ref{firstordersolutions}
and Ref.~\cite{Horowitz}). In addition, for
all $\xi$, $\beta$, and $\lambda$ except for $\xi = 1/6$ and $\lambda >
\lambda_{\rm crit}$, there exist solutions with
spatial wavelength ${\cal L} \gg L_p$ which oscillate in time at a
frequency of order the Planck frequency.

What attitude should one take to these solutions? If one justifies the
semiclassical equations via the $1/N$ approximation (see
Sec.~\ref{status} above) and formally takes the limit $N \to \infty$,
one could view the semiclassical equations as holding exactly. In that
case, one should take all of its solutions seriously and conclude that
flat spacetime is unstable in semiclassical gravity.  However, as
already indicated in Sec.~\ref{status} above, we do not adopt this
view here.  Rather, although we view the semiclassical equations as
accurate -- for sufficiently large $N$ -- to arbitrarily high order in
a long wavelength expansion, we nevertheless view them as approximate
equations, with domain of validity is ${\cal L} \gg L_p$. Hence, in
our view, the solutions which grow exponentially on a timescale of
order the Planck time or oscillate in time (with significant
amplitude) at frequencies of order the Planck frequency lie outside
the domain of validity of the approximation.  From this point of view,
the additional degrees of freedom admitted by the semiclassical
equations are merely artifacts of the semiclassical approximation. In
particular, from this point of view the exponentially growing
solutions to the semiclassical equations are spurious, and are not
indicative of any physical instability of flat spacetime.  A similar
attitude toward the additional degrees of freedom admitted by the
semiclassical equations would, of course, result from viewing the
semiclassical equations as approximate equations, with unknown higher
order correction terms of relative magnitude $\sim L_p^2 / {\cal L}^2$
as discussed in Sec.~\ref{status} above and in Ref.~\cite{Simon1}.

However, two nontrivial issues result from this viewpoint towards the
``pathological'' solutions, with regard to the extraction of physical
predictions from the semiclassical equations.  First, if the domain of
validity of the semiclassical equations is ${\cal L} \gg L_p$, then
since the corrections to the {\it classical} equations which appear in
the semiclassical equations are of relative magnitude $L_p^2 /{\cal
L}^2$, it might seem that solutions to the semiclassical equations
cannot accurately describe any situation where the deviation from a
classical solution becomes large.  In other words, in any circumstance
where semiclassical theory makes a prediction significantly different
from that of the classical theory, it might be expected to be highly
inaccurate. (Note that this difficulty would appear to be even more
severe if one takes the view that the semiclassical equations have
unknown corrections of order $\sim L_p^4 / {\cal L}^4$.)
If that were
the case, there would be little point in studying the detailed
properties of solutions to the semiclassical equations.  However, we
shall argue in subsection \ref{proposal1} below that this is not the
case: Semiclassical theory should be able to accurately describe
phenomena where the deviations from classical behavior are locally
small, but where, nevertheless, long term cumulative effects result in
very large global deviations from classical solutions.  Furthermore,
this ability of semiclassical theory to accurately describe such
phenomena should the hold even in the case of finite $N$ when unknown
correction terms of order $\sim L_p^4 / {\cal L}^4$ appear in the
semiclassical Einstein equation. While this viewpoint on the domain of
validity of the semiclassical Einstein equation is fairly widespread
--- e.g., it is commonly assumed that semiclassical gravity will give
an accurate description of the black hole evaporation process until
the stage where the black hole mass approaches the Planck mass --- we
will devote subsection \ref{proposal1} to giving a clear justification
for it and to distinguishing this use of the semiclassical equations
from using them to obtain ``approximate perturbative solutions".

The second issue that arises is that if the semiclassical equations
admit spurious solutions, by precisely what criteria do we determine
whether a given solution is ``physical'' or not? This issue has
recently been addressed by Simon in a series of papers
\cite{Simon1,Simon2,Simon3}.  Simon actually makes a number of
independent suggestions with regard to the semiclassical equations.  In
this section we will discuss all these suggestions in detail. We will
argue that, in a general context, all of the suggestions have
shortcomings. However, we also will argue that
in the special case of perturbation theory about Minkowski
spacetime, Simon's ``reduction of order'' proposal \cite{Simon3}
yields a satisfactory prescription for extracting physical predictions
from the semiclassical equations.  In the remaining sections of this
paper, we then will investigate the validity of ANEC for solutions to
the ``reduced order'' semiclassical equations.

The above mathematical and physical issues which arise from the
``higher derivative'' nature of the
semiclassical equations are not unique to this context. Indeed,
in this regard,
the semiclassical Einstein equation is closely analogous to the
Abraham-Lorentz equation of motion for a classical charged point
particle including radiation reaction, which also has exponentially
growing, ``runaway'' solutions \cite{Simon3}.
For completeness and also to aid our analysis of the semiclassical
Einstein equation, we discuss this (much simpler and
well studied) example in
subsection \ref{ALeq}. In subsection \ref{proposal1}, we explain how
solutions to the semiclassical equations can be accurate even when they
predict large (global) deviations from classical theory. The final two
subsections critically analyze the ``perturbative solutions'' and the
``reduction of order'' proposals discussed by Simon
\cite{Simon,Simon1,Simon2,Simon3}.

\subsection{The analogy to radiation reaction of point particles}
\label{ALeq}

We begin this subsection with a discussion of the nature and range
of validity of the Abraham-Lorentz equation, analogous to our
discussion in Sec.~\ref{status} of the
status of the semiclassical Einstein equation.
It is well known that the idealization of classical, charged point
particles is inconsistent, and that this inconsistency is the source of
the well known difficulties with the radiation reaction equation.
More specifically, consider a finite distribution of charge with some
physical size $\sim {\cal L}$.  In the nonrelativistic limit the
radiation reaction force on the charge distribution will be given by
\cite{Jackson}
\begin{equation}
{{\bf F}_{\rm react}\over m}  =   {\tau} {\dot {\bf a}} + O\left[
\tau {\ddot {\bf a}} ({\cal L}/c)\right],
\label{radiationreactionforce}
\end{equation}
where ${\bf a}$ is the acceleration, $m$ is the mass, $q$ is the
charge, $c$ is the speed of light and
\begin{equation}
\tau \equiv {2\over3} (q^2 / m c^3).
\label{tau}
\end{equation}
One might think that the ``point particle limit'' can simply be
obtained by letting ${\cal L} \to 0$, yielding the usual radiation
reaction force ${\bf F}_{\rm react} = m {\tau} {\dot {\bf a}}$,
without any unknown correction terms.  The difficulty with this is of
course that all physical, finite distributions of charge satisfy
\begin{equation}
{\cal L} \agt c \tau,
\label{restriction}
\end{equation}
assuming only that the electromagnetic self energy $\sim q^2 /{\cal
L}$ cannot exceed the externally measured mass-energy $m c^2$ (i.e.,
that the ``bare mass'' for a finite distribution of charge is always
positive).  Therefore the limit ${\cal L} \to 0$ for fixed mass $m$
and charge $q$ is unphysical.  The limit ${\cal L} \to 0$, $q \to 0$
with $q^2/{\cal L}$ fixed presumably will exist and presumably will
yield the unique ratio ${\bf F}_{\rm react} / q^2$ given by
Eq.~(\ref{radiationreactionforce}) (without correction terms) in the
limit. However, for fixed mass $m$ and (nonzero) charge $q$, there
will always be an ambiguity in the radiation reaction force that
depends on the structure of the particle, which is of fractional
magnitude $\agt \tau / \tau^*$, where $\tau^*$ is the timescale over
which the acceleration is changing.

It should be noted that the restriction (\ref{restriction}) --- and,
correspondingly, the presence of unknown correction terms of order
$O(\tau^2)$ in the radiation reaction force
(\ref{radiationreactionforce}) --- should occur in any theory where
classical electromagnetic fields are coupled to some other degrees of
freedom with a continuous distribution of charge.  In particular, in
semiclassical QED, a one-particle, nonrelativistic electron state is
effectively a finite distribution of charge with width ${\cal L} \sim
c \tau / \alpha$, where $\alpha$ is the fine structure constant, since
the expected value of the current operator will not be concentrated at
a point but will typically have a width of order ${\cal L}$.  Hence,
any ``derivation'' of Eq.~(\ref{radiationreactionforce}) for electrons
from semiclassical QED should also give rise to correction terms of
the order indicated.

Thus, the status of the equation of motion for a non-relativistic,
charged particle in a given, fixed external electric field ${\bf E}$,
\begin{equation}
{\ddot {\bf x}} = {q\over m} \,  {\bf E}({\bf x},t) + {\tau}
{\stackrel{{\bf \ldots}}{{\bf x}}},
\label{radiationreaction}
\end{equation}
is closely analogous to that of the semiclassical Einstein equation.
First, the domain of validity of this equation
is limited to the regime $\tau^* \gg  \tau$, similar to the domain of
validity ${\cal L} \gg L_p$ of the semiclassical Einstein equation.
Second, as in the case of the semiclassical
Einstein equation, the small parameter appearing in
Eq.~(\ref{radiationreaction}) multiplies a term
containing higher-order time
derivatives than originally appeared in the equation of motion, thereby
effectively increasing the ``number of degrees of freedom'' of the
system.
This higher-order time derivative term $\tau {\stackrel{{\bf
\ldots}}{{\bf x}}}$ is
responsible for the existence of so-called runaway solutions,
which grow exponentially in time.

Consider now the space of solutions of the radiation reaction equation
(\ref{radiationreaction}). If the electric field ${\bf E}$ is of
compact support in time, the runaway solutions will have the property
that ${\ddot {\bf x}} \propto \exp(t/\tau)$ for large $t$.  Solutions
which do not manifest this run-away behavior form a 6-dimensional
submanifold of the 9-dimensional manifold of solutions to
Eq.~(\ref{radiationreaction}), since the non-runaway solutions have
vanishing acceleration at late times (for ${\bf E}$ of compact support
in time).  Indeed, when the electric field is bounded above, it can be
shown that given any initial position and velocity, there exists a
unique initial acceleration generating a non-runaway solution.  In
particular, in the special case where the electric field is
independent of position ${\bf x}$, the general solution of
Eq.~(\ref{radiationreaction}) for an arbitrary initial acceleration
${\bf a}_0$ at time $t=0$ is given by
\begin{eqnarray}
{\bf a}(t) = e^{t/\tau} \left[ {\bf a}_0 - {q \over m} \int_0^\infty
ds \, {\bf E}(\tau s) e^{-s} \right] \nonumber \\
+ {q\over m} \int_0^\infty ds \, {\bf E}(t + \tau s) e^{-s},
\label{specialsolution}
\end{eqnarray}
where ${\bf a} = {\ddot {\bf x}}$.
It is clear that the choice of initial acceleration, ${\bf a}_0$, which
makes the first term vanish is the unique choice which
generates a non-runaway solution.  Moreover, as is well
known, the non-runaway solutions can be characterized as those
solutions which satisfy the integro-differential equation given
in Jackson \cite{Jackson}
\begin{equation}
{\dot {\bf v}} = {q\over m} \int_0^\infty ds \, {\bf E}({\bf x}(t+\tau
s),t+\tau s) \, e^{-s}.
\label{integralradiationreaction}
\end{equation}

The runaway solutions quickly evolve into a regime where the unknown
corrections to the radiation reaction force will become as large as
the radiation reaction force itself, i.e., where $\tau^* \sim \tau$ in
the notation used above.  Therefore these solutions lie outside the
domain of validity of the Abraham-Lorentz equation and are normally
deemed to be ``unphysical''.  It is conventional to take the space of
``physical solutions'' to be the six-dimensional space of non-runaway
solutions satisfying Eq.~(\ref{integralradiationreaction}).  However,
these solutions have the unphysical property that ``pre-acceleration''
is required at early times (before the electric field is turned on) in
order to avoid run-away behavior at late times \cite{preacceleration}.

In summary, there
are close parallels between the radiation reaction equation and the
semiclassical Einstein equation.  In our discussions in the remainder
of this section, we will use the radiation reaction equation
as a simple example and model for the issues that arise.  We will
return to the issue of obtaining ``physical solutions'' to the
radiation reaction equation and of the semiclassical Einstein equation
in the last two subsections of this section.  However, we first
address an important issue concerning the accuracy of solutions to
such equations.

\subsection{The physical relevance of solutions to the semiclassical
equations}
\label{proposal1}

As we have discussed, the equations we are considering have only a
limited domain of validity.
More specifically, Eq.~(\ref{radiationreaction}) holds only when
$\tau^* \gg  \tau$, whereas Eq.~(\ref{basic}) [as well as its linearized
version Eq.~(\ref{firstorderexplicit})] has the domain of
validity ${\cal L} \gg L_p$.  Indeed, the situation with regard to
Eq.~(\ref{radiationreaction}) is even worse in that
there are unknown correction terms of order $\tau^2$ in that
equation. [As we have previously discussed, similar
unknown correction terms also would appear in Eq.~(\ref{basic})
if one justifies that equation via the one-loop approximation or takes
$N$ to be finite in the $1/N$ approximation; see Sec.~\ref{status}
above.]
Since Eq.~(\ref{radiationreaction}) differs from the corresponding
equation without radiation reaction only by a term of order
$\tau / \tau^*$, and Eq.~(\ref{basic}) differs from the
corresponding classical equation only by terms of order
$(L_p/ {\cal L})^2$, it might seem that Eqs.~(\ref{radiationreaction})
or (\ref{basic}) can never be valid in any circumstance
where
they predict large deviations from the corresponding classical behavior,
i.e., in any circumstance where they have the
potential to make dramatically new predictions.

Indeed, it might appear that, at best, the only useful and reliable
information that ever could be justifiably extracted from solutions to
equations like the ones we are considering would be the information
contained in ``approximate, perturbative solutions" to some finite
order.
To explain what is meant by this, let us focus attention on the
radiation
reaction equation (\ref{radiationreaction}).
Suppose that we attempt to perturbatively solve this equation
order by order in an expansion in the small
parameter $\tau$ appearing in that equation.  Thus,
we seek approximate solutions of the form
\begin{equation}
{\bf x}_J(t) = \sum_{j=0}^J {\bf x}^{(j)}(t) \tau^j,
\label{xN}
\end{equation}
where each ${\bf x}^{(j)}(t)$ will satisfy a differential equation that
is second order in time with sources constructed from the externally
applied fields and from the ${\bf x}^{(k)}$ with $k<j$.
Note that there is no problem
with runaway effects in constructing the approximate solutions ${\bf
x}_J(t)$, and, furthermore, for any given order $J$, an initial position
and velocity will determine a unique approximate solution ${\bf x}_J$
(unlike the situation with the exact solutions) \cite{Campanelli}.

The key question here is: How can we be
justified in keeping the higher order
corrections, ${\bf x}^{(j)}(t)\, \tau^j$ for $j \ge 2$, to the
solutions, when the equation of motion itself is ambiguous at
$O(\tau^2)$?  More specifically, the equations for
${\bf x}^{(j)}(t)$ for all $j > 1$ are completely ambiguous because of
the unknown correction terms appearing in Eq.~(\ref{radiationreaction}).
Thus, there is apparently no justification for going beyond the
lowest order approximate solution
\begin{equation}
{\bf x}_1(t) = {\bf x}^{(0)}(t) + \tau {\bf x}^{(1)}(t),
\label{rrsol1}
\end{equation}
since unknown terms of order unity appear in the equation for ${\bf
x}^{(2)}$.  If we stop with the approximate solution (\ref{rrsol1}),
the difficulties with the existence of additional degrees of freedom
and the presence of ``pathological'' solutions to the radiation
reaction equations would not arise, since the perturbative equations
are not problematical. However, solutions of the type
(\ref{rrsol1}) would not be
of much interest, since -- whatever precise form the exact radiation
reaction equations take -- one could not expect Eq.~(\ref{rrsol1}) to
be a good approximation whenever $\tau {\bf x}^{(1)}(t)$ becomes of
order ${\bf x}^{(0)}(t)$, since then the unknown higher order
corrections, ${\bf x}^{(j)}(t) \tau^j$ with $j > 1$, to the solution
should also be comparably large. In other words, Eq.~(\ref{rrsol1})
should be a poor approximation whenever radiation reaction has any
significant effect upon the motion of the particle.

Completely parallel remarks would apply to the case of the
semiclassical Einstein equation (\ref{basic})
if one justifies that equation via the one-loop approximation, so that
there are unknown corrections of order $\hbar^2$ \cite{caveat8}.
However,
even when we take our viewpoint of justifying this equation via
the $1/N$ approximation and thereby treat Eq.~(\ref{basic}) as being
valid to all orders in $\hbar$, the situation is not significantly
different.
Although we would be formally justified in going beyond the
first order (in $\hbar$) approximate perturbative solution, there
apparently would be little point in doing so, since
the equation is valid only when ${\cal L} \gg L_p$, and, in that regime,
the higher order perturbative solutions should
merely make tiny additional
corrections to the first order perturbative solution. Again, however,
first order approximate perturbative solutions would be of little
interest.

Simon \cite{Simon1,Simon2,Simon3} has argued
that one is not justified in going beyond approximate solutions of the
type (\ref{rrsol1}), and that this therefore solves the problem of the
pathological solutions \cite{noteS}.  As we now discuss, we disagree
with this conclusion. With regard to Eq.~(\ref{radiationreaction})
it is indeed true that in generic situations the higher order
corrections, ${\bf x}^{(j)}(t)$ for $j \ge 2$, will be sensitive to the
unknown higher order corrections to the equation of motion, and that
therefore only the approximate solution (\ref{rrsol1}) will be
physically meaningful.  However, as we now explain, there {\it are}
situations where the higher order corrections ${\bf x}^{(j)}$ for all
$j \ge 2$ are not sensitive to the higher order corrections to the
equation of motion, and where, correspondingly, physically meaningful
solutions to Eq.~(\ref{radiationreaction}) can be obtained which go
well beyond the approximation (\ref{rrsol1}) and predict large
radiation reaction effects. Similarly, there are situations where
Eq.~(\ref{basic}) may predict large deviations from classical behavior
even though ${\cal L} \gg L_p$ everywhere. We now explain these
comments in detail.

These situations where the higher order corrections to the solutions
to Eq.~(\ref{radiationreaction})
are not sensitive to the higher order corrections to the equation of
motion (\ref{radiationreaction})
itself arise when radiation reaction effects are ``locally small'' but
accumulate secularly, so that they become large at late times. This
occurs, in particular, when the evolution timescale is set by
radiation reaction.  A good example is the case of an
electromagnetically bound particle in a Coulomb field undergoing a
radiation-reaction-driven inspiral.  Clearly the approximate solution
(\ref{rrsol1}) will provide a poor description of the motion once the
radius of the orbit has shrunk by a factor of two.  However, an
accurate description of the motion will be provided by an appropriate
solution of the exact equation (\ref{radiationreaction}), provided only
that the timescale of the inspiral is much larger than the orbital
period, so that the radiation reaction effects are locally small.

To see this more concretely, suppose that the exact equation of
motion (including all higher order corrections) were of the form
\begin{equation}
{\bf a} = {\bf a}_{\rm ext} + \tau_1 {\dot {\bf a}} + \alpha_1 \tau_2^2
{\ddot {\bf a}}.
\end{equation}
Here ${\bf a}_{\rm ext}$ is the acceleration due to the externally
applied electric field, $\alpha_1$ is an unknown numerical coefficient
of order unity, and $\tau_1 = \tau_2 = \tau$.  (We have temporarily
distinguished the $\tau$'s that appear in the radiation reaction
acceleration, and in the next order correction to this acceleration,
to aid the following discussion.)  Then it is clear that effects that
are quadratic and higher order in $\tau_1$ will be important in
describing the inspiral.  However, contributions of order $\tau_2^2$
to the solution will give rise to small corrections to the inspiral of
relative magnitude $\alt \tau / \tau_p$, where $\tau_p \gg \tau$ is the
initial orbital period.  This can be seen by solving the exact
equation with $\alpha_1=0$, and checking a posteriori that the
$\alpha_1 \tau_2^2 {\ddot {\bf a}}$ term is always a small correction
to the equation of motion when evaluated on this solution
\cite{note2}.

Indeed, the above type of situation -- where
non-perturbative effects in a small
parameter are large but can be reliably calculated even
though the equation is known only to first order in that parameter --
actually occurs quite commonly in physics. A good example is provided by
Newtonian hydrodynamics.  Dissipative terms
in the hydrodynamic equations normally have an effect on the fluid
motion that is smaller than the effects of the non-dissipative terms
by a factor $\sim \epsilon$, where $\epsilon$ is the ratio of a
microscopic lengthscale to a macroscopic lengthscale.  Derivations of
the hydrodynamic equations from statistical
mechanics throw away small correction
terms of order $\epsilon^2$.  However, effects that are
non-perturbative in $\epsilon$ in solutions to the equations will be
meaningful when the macroscopic evolution timescale is determined by
dissipative effects.  In this case the dissipative terms and the
higher order corrections are, in effect,
boosted from being $O(\epsilon)$ and
$O(\epsilon^2)$ respectively, to being $O(1)$ and $O(\epsilon)$
respectively (relative to the non-dissipative terms).

To illustrate this claim with a simple, concrete example, consider the
one
dimensional heat equation
\begin{equation}
{\partial T \over \partial t} = \sigma {\partial^2 T \over \partial
x^2},
\label{heat}
\end{equation}
where $\sigma = l^2 / \tau$, and $l$ and $\tau$ are some microscopic
length and time scales. First, we note that it is clear that effects
which are
non-perturbative in the ``small parameter''
$\sigma$ are very significant for the solutions.
Indeed, if we choose initial data of compact support,
then at later times the temperature $T$ will be nonzero outside the
support of the initial data.  However, the approximate perturbative
solutions analogous to (\ref{xN})
(generated by expanding in $\sigma$) will be
nonzero only in the support of the initial data.  Therefore none of
the approximate perturbative solutions are
even qualitatively accurate;
non perturbative effects are vitally important.

Now suppose that there were a correction term to Eq.~(\ref{heat}) of the
form $- \alpha_1 (l^4 / \tau) \partial^4 T / \partial x^4$, where
$\alpha_1$ is an unknown numerical coefficient of order unity.  Suppose
that the initial data is of the form $T(x,t=0) = f(x/{\cal L})$, where
${\cal L}$ is the macroscopic lengthscale over which the initial data
varies.  We make the following rescaling of variables:  let $x =
{\cal L} \rho$, and let $t = T s$, where $T = ({\cal L} / l)^2 \tau$
is the macroscopic evolution timescale associated with the heat
conduction.  Then the
modified heat equation takes the form
\begin{equation}
{\partial T \over \partial s} = {\partial^2 T \over \partial \rho^2} -
\alpha_1 \epsilon {\partial^4 T \over \partial \rho^4},
\end{equation}
where $\epsilon = (l / {\cal L})^2 \ll 1$, and where the initial
conditions are $T(\rho,0) = f(\rho)$.  From the form of this equation
it is clear that the solutions will have
important non-perturbative contributions from the first
term on the right hand side, but that the second term
(for $\alpha_1 > 0$) will be a
small correction of order $\epsilon$.

The situation with regard to the semiclassical Einstein equation is
closely analogous.  Approximate perturbative solutions similar to
Eq.~(\ref{xN}) to any finite order in $\hbar$
will not be adequate to describe such
phenomena as the evaporation of a black hole over timescales
long enough for the black hole to lose a significant fraction of its
initial
mass.  To describe this process, it will
be necessary to consider effects that are non perturbative in
$\hbar$.  It should be possible to calculate these effects reliably
from the semiclassical equations provided only that ${\cal R} \gg L_p$
everywhere in the region of interest, where ${\cal R}$ is the local
radius of curvature. Even if there were unknown corrections to the
semiclassical equations that are higher order in $\hbar$, these
corrections
should be qualitatively unimportant in the black hole evaporation
process -- except near the singularity and near the final moments of
evaporation, where these unknown corrections become locally large.

To summarize, given an equation of motion whose range of validity
restricts a ``correction term" in that equation
to be locally very small, there, nevertheless, can be a
wide range of circumstances where this equation can reliably predict
phenomena where this correction term is responsible for
producing large deviations from the uncorrected motion.
The approximate, perturbative solutions (\ref{xN})
to any finite order are
completely inadequate for describing such phenomena.
Even if the equation being considered is itself valid only to first
order in some small parameter, it can occur that solutions which are
non-perturbative in this parameter are physically meaningful.
Therefore, it is of critical importance to have a means of determining
which solutions to these equations should be viewed as ``physically
relevant'' and which solutions should be deemed to be ``spurious''.
The next two subsections examine two proposals for extracting the
physically relevant solutions.

\subsection{Extraction of a preferred subclass of ``physical
solutions''}
\label{proposal2}

A possible method for dealing with the additional degrees of
freedom of the modified equations
is to identify a preferred subclass of the space of exact
solutions. This can be done for the radiation reaction equation
by simply discarding the run-away solutions, although the remaining
solutions have the unphysical feature of ``pre-acceleration''. However,
it is less clear what should be done when some of the additional
degrees of freedom are associated with oscillatory solutions instead
of exponential solutions, as
occurs in the linearized semiclassical equations. It is even less
clear what should be done in the case of nonlinear equations
(such as the full semiclassical equations),
where the solutions might not cleanly separate into subclasses
of the ``correct size'' on the basis of their late and/or early time
behavior.

One proposal for identifying a preferred subclass of ``physical
solutions'' is to admit only those solutions which are
``perturbatively expandable'' \cite{Simon} in the appropriate small
parameter $\epsilon$ (where $\epsilon$ would be $\tau$ in the
radiation reaction case, and ${\hbar}$ or, equivalently, $1/\alpha^2$,
in the semiclassical case). By ``perturbatively expandable'', it is
meant that the solution can be expressed as a convergent power series
of the form (\ref{xN}) (with $J = \infty$), with each ${\bf x}^{(j)}$
satisfying the appropriate $j$th order perturbation equation.
Equivalently, the requirement is that the solution should be connected
to a solution with $\epsilon=0$ by a one parameter family of exact
solutions with parameter $\epsilon$ which is analytic in $\epsilon$.
Note that it is essential that analyticity in $\epsilon$ be imposed,
since it should be possible to connect {\em every} solution at finite
$\epsilon$ to a solution with $\epsilon = 0$ with a one-parameter
family which is merely {\em smooth} in $\epsilon$ \cite{note2b}.

This proposal would appear to be of the correct
character, since the perturbative equations have the correct number
of degrees of freedom, and the ``pathological solutions'' do not have
analytic behavior in the small parameter $\epsilon$ at $\epsilon = 0$.
This subsection is devoted to a critical examination of this proposal.

To begin, consider the radiation reaction equation, and
suppose that the electric field ${\bf E}({\bf
x},t)$ is analytic and independent of position.  In this case it can
be seen from Eq.~(\ref{radiationreaction})
that the series generated by solving order by order in $\tau$ is
\begin{equation}
{\bf a}(t) = \sum_{n=0}^\infty \, {q\over m} \tau^n {d^n \over dt^n}
{\bf E}(t).
\label{pp}
\end{equation}
This is precisely the expansion in $\tau$ of the non-runaway solution
[the second term in Eq.~(\ref{specialsolution})]. Thus, when the
series converges, the two coincide.  It seems plausible that
the ``perturbative expandability'' criterion also will select the
non-runaway solutions in the
more general case of an analytic ${\bf E}$ which is position dependent.

However, the criterion of ``perturbative expandability''
fails in the case of smooth but non-analytic ${\bf
E}$. In particular, under the circumstances where
Eq.~(\ref{specialsolution}) is applicable, it
can be seen that no are no solutions which are analytic in $\tau$.
In this case the series (\ref{pp}) will not
converge.  Indeed, when the electric field is
smooth and of compact support in
time (and hence non-analytic), each term in the perturbative expansion
will consist of straight line motion both before and after ${\bf E}$ is
``turned on''.  Therefore the summed series, if it converges, must also
have this property.  However, the summed series must also satisfy
Eq.~(\ref{radiationreaction}), and all exact solutions of this
equation which have vanishing acceleration at late times will exhibit
pre-acceleration at early times. Consequently, the series which
attempts to define the ``perturbatively expandable'' solutions cannot
converge.

The criterion of ``perturbative expandability'' appears to fail much
more
dramatically when (at least some of)
the additional degrees of freedom are oscillatory
in character, as is the case for the semiclassical equations. As a simple model
of
this phenomenon, consider the differential equation
\begin{equation}
\left(\epsilon^2 {d^2  \over d t^2} + 1\right) g(t) = \rho(t),
\label{model}
\end{equation}
in the limit $\epsilon \to 0$.  Note that the linearized semiclassical
Einstein
equation for individual spatial Fourier modes of the metric perturbation
is closely analogous to this equation with $g$ of the form
$g(t) = f^{''}(t) + \omega_0^2 f(t)$.
The general solution of Eq.~(\ref{model}) is given by
\begin{eqnarray}
g(t) &=& {1\over2} \int ds \,\sin(|s|) \,\rho(t + s \epsilon) \nonumber
\\
\mbox{} && + A(\epsilon)\sin(t/\epsilon) + B(\epsilon)
\cos(t/\epsilon).
\label{gensolnmodel}
\end{eqnarray}
The inhomogeneous, first term in this solution can also be written as
\begin{equation}
g_{\rm inhom}(t) = {\rm P.V.}\ \int {d \omega \over 2 \pi} {e^{i \omega
t} \over 1 - \epsilon^2 \omega^2} {\tilde \rho}(\omega),
\label{pv}
\end{equation}
where P.V.~means ``the principal value of''.  This
term is just the average of the
advanced and retarded solutions.

If ${\tilde \rho}(\omega)$ is of compact support in
$\omega$ --- which is a
{\em much} stronger requirement than $\rho(t)$ being an
analytic function of $t$ --- then it is easy to show
from Eq.~(\ref{pv}) that $g_{\rm inhom}(t;\epsilon)$ is analytic in
$\epsilon$. It follows that when ${\tilde \rho}$ is of compact support,
Eq.~(\ref{pv}) yields the unique ``perturbatively expandable'' solution
to Eq.~(\ref{model}). However, when ${\tilde \rho}$ fails to
be of compact support in $\omega$, it appears that there do not exist
{\em any} perturbatively expandable solutions
to Eq.~(\ref{model}). In essence, the fact that one has a pole on the
real axis of the $\omega$-plane in the integrand of Eq.~(\ref{pv})
-- which is associated with the existence of new
oscillatory modes -- makes
the analyticity behavior in $\epsilon$ at $\epsilon = 0$ much worse than
in the case where the new degrees of freedom correspond to
exponentially growing or decaying modes (i.e., when the poles
occur away from the real $\omega$ axis).  Although we have not obtained
a complete proof that no perturbatively expandable solutions exist
when ${\tilde \rho}(\omega)$ fails to be of compact support, we have
verified
that the series
\begin{equation}
g(t) = \sum_{n=0}^\infty \, (-1)^n \epsilon^{2 n} {d^{2n} \rho \over d
t^{2 n}}(t)
\end{equation}
which defines the perturbatively expandable solutions
fails to converge for some simple, analytic, very well behaved choices
of $\rho$, including Gaussian behavior in $t$.

Thus, it appears that the criterion of perturbative expandability is of
very limited applicability. Even in cases where the additional degrees
of freedom have an exponentially growing and/or decaying character, the
criterion may fail. However, when the additional degrees of freedom have
an oscillatory character, it appears that
perturbatively expandable solutions will exist
only in very exceptional cases.

One might seek some other criterion which would single out a
preferred subclass of ``physical solutions''. In the case where the
additional degrees of freedom have an exponentially growing and/or
decaying character, the non-runaway solutions are, of course a natural
candidate for this preferred subclass \cite{notecc},
although even in this case, these ``physical solutions'' have
unphysical features like ``pre-acceleration'' \cite{preacceleration}.
However, when the additional degrees of freedom
have an oscillatory character, there seems little hope of singling out
a preferred subclass of solutions on any physical grounds.  The
difficulties encountered in doing this can be seen in our above model
(\ref{model}): The issue of picking out a preferred ``physical
solution'' is essentially equivalent to picking out a preferred Greens
function for the differential operator appearing in Eq.~(\ref{model}).
Although it is possible to mathematically identify preferred Green's
functions (e.g., the retarded Green's function, the advanced Green's
function, or their average) there does not appear to be any grounds
for arguing that any one of these is ``better behaved'' or ``more
physical'' than the others.

Note that the different solutions obtained by choosing different
Greens functions will all be tangent to the same approximate
perturbative solution (in the sense of having the same derivatives with
respect to $\epsilon$ at $\epsilon=0$).  They will differ by a
function which is smooth in $\epsilon$, but which is also
non-perturbative in $\epsilon$ in the sense that all of its
derivatives with respect to $\epsilon$ vanish at $\epsilon=0$.  For
example, in our simple model (\ref{model}), the difference between the
advanced and retarded solutions is just
\begin{equation}
g_{\rm adv} - g_{\rm ret} = - {1 \over \epsilon} \, {\rm Im} \, \left[
e^{- i t / \epsilon} {\tilde \rho}(1/\epsilon) \right],
\end{equation}
which is smooth in $\epsilon$ as $\epsilon \to 0$ if $\rho(t)$ is
smooth.

We now turn our attention to a quite different idea: the modification of
the equations themselves so that all of their solutions will be
``physical''.

\subsection{Reduction of order --- modifying the original equation}
\label{reductionoforder}

In this section, we analyze the method of ``self-consistent reduction
of order'' \cite{Simon3} as a means for obtaining physical predictions
from the radiation reaction or semiclassical equations. Instead of
seeking to identify a subset of ``physical solutions'' to the given
equation, this approach generates a modified, second order equation,
which is ``as accurate" or ``nearly as accurate" as the original
equation, but whose solutions are all well behaved and can be
interpreted as being ``physical''.  The idea of reduction of order is
quite old --- it has been advocated in the context of the radiation
reaction equation by Landau and Lifshitz \cite{Landau}, Teitelboim
\cite{Teitelboim}, and Ford {\it et.~al.~}\cite{Ford-OConnell}.  It is
also a standard procedure that is used in the derivation of
post-Newtonian equations of motion in classical relativity, see, e.g.,
Ref.~\cite{Damour}.  It has been used in the context of classical,
higher derivative theories of gravity by Bel {\it et. al.} \cite{Bel}, and
more recently it has been discussed in detail in a wide variety of
contexts, and in particular advocated in the context of semiclassical
gravity by Simon \cite{Simon3}.

The justification for this method can be understood as follows. We are
given an equation of motion which is believed to accurately describe
phenomena with sufficiently large length and/or time scales, e.g.,
$\tau^* \gg \tau$ for Eq.~(\ref{radiationreaction}) or ${\cal L} \gg
L_p$ for Eqs.~(\ref{basic}) and (\ref{firstorderexplicit}). However,
the given equation (presumably) does not predict even qualitatively
correct behavior outside of its range of validity.
Now, generically, any solution to the given
equation will have some non-vanishing Fourier components which lie
outside the equation's domain of validity.  For some solutions, these
Fourier
components behave in such a pathological manner that the entire
solution is dominated by the qualitatively incorrect behavior, as
occurs for the ``run-away" solutions.
However, a good remedy for this difficulty would be to
modify the given equation so that it is equivalent --- to the desired
accuracy --- to the given equation at large length and/or time scales
but does not predict any pathological behavior at
small scales (and, thus, presumably,
is at least qualitatively correct in this regime). In the situations
where
it is applicable, the reduction of order method achieves this goal.

The reduction of order algorithm for an ordinary or partial
differential equation may be stated as follows. We start with an
equation
(or system of equations) for the unknown variable $x$ of the general
form
\begin{equation}
{d^{n} x \over dt^{n}} = P + \tau Q
\label{redord}
\end{equation}
where $\tau$ is a ``small parameter''. Here
we assume that $P$ contains terms
involving no more that $(n-1)$ time derivatives of $x$, but that $Q$
contains terms involving $m \ge n$ time derivatives of $x$, so that the
``small correction'', $\tau Q$, introduces time derivatives of the
same or higher differential order as appeared in the original equation.
To apply the reduction of order procedure, we
differentiate Eq.~(\ref{redord}) $(m-n)$ times with respect to $t$,
and substitute the resulting formula for $d^{m} x / dt^{m}$ into the
expression for $Q$. We then discard the resulting terms in $Q$ which
are quadratic and higher order in $\tau$. We thereby obtain a new
equation which is formally equivalent to Eq.~(\ref{redord}) to order
$\tau$ and which has the same general form as Eq.~(\ref{redord}), but
for which the term which plays the role of $Q$ now contains at most
$(m-1)$ time derivatives of $x$. We then continue to iterate this
procedure until the maximum number of time derivatives of $x$
appearing in $Q$ is reduced to $(n-1)$, at which point no further
reduction of differential order of the time derivatives can be
achieved. The resulting equation is then of the same differential
order in time as the original equation Eq.~(\ref{redord}) with $\tau =
0$. Thus, we end up with an equation which, formally, is ``as accurate
as'' Eq.~(\ref{redord}) to order $\tau$,
but which does not contain any new ``degrees
of freedom''. Note that this final,
reduced order equation is {\it uniquely determined} by the
requirements: (i) that it should contain only terms that are zeroth
order or first order in $\tau$, (ii) that it should be formally
equivalent to
Eq.~(\ref{redord}) to $O(\tau^2)$, and
(iii) that it should be of the form (\ref{redord}) where the
right hand side does not contain any derivatives with respect to time
of order higher than $n-1$.

However, although the reduction of order algorithm is uniquely defined
for any equation of the form (\ref{redord}) for a given choice of
variables, it should be noted that
some ambiguities in the algorithm can be introduced by making a
$\tau$-dependent change of variables: If one introduces a new variable
$y = y(x; \tau)$, re-writes Eq.~(\ref{redord}) as an equation for $y$,
and then neglects the terms of order $\tau^2$ and higher, the resulting
reduced order equation for $y$ need not be precisely equivalent to the
reduced order equation for $x$ \cite{notcommute}. However, this inequivalence of the
equations can occur only at order $\tau^2$ and higher, and, thus, should
not have an important effect on the behavior of solutions in regimes
where reduction of order can be justified [c.f., the discussion in
Sec.~\ref{proposal1} above].

The radiation reaction equation (\ref{radiationreaction}) provides a
good illustration of how this procedure works and of its justification.
By following the steps described above, one
obtains the equation \cite{Landau,Teitelboim}
\begin{eqnarray}
{\ddot {\bf x}} = {q\over m} \,\left[  {\bf E}({\bf x},t) + {\tau}
{\partial {\bf E} \over \partial t}({\bf x},t) + \tau \left({\dot {\bf
x}}
\cdot {\bf \nabla}\right) {\bf E}({\bf x},t) \right].
\label{radiationreactionmodified}
\end{eqnarray}
This modified equation of motion is formally equivalent to
Eq.~(\ref{radiationreaction}) up to order $O(\tau)$, and differs from
it at order $O(\tau^2)$.  Since the unmodified equation
(\ref{radiationreaction}) has unknown corrections at order
$O(\tau^2)$, the modified radiation reaction equation
(\ref{radiationreactionmodified}) gives a description of the motion
whose expected accuracy in the regime $\tau^* \gg \tau$ is just as
high as that of the original equation (\ref{radiationreaction}).
[Indeed, Eq.~(\ref{radiationreactionmodified}) differs from
Eq.~(\ref{radiationreaction}) at order $O(\tau^2)$ merely by the term
$\tau^2 {\ddot {\bf a}}$, which is of the same order of magnitude as
the expected corrections to Eq.~(\ref{radiationreaction}) discussed
in Sec.~\ref{ALeq} above due to the finite size effects \cite{caveat5}.]
However, Eq.~(\ref{radiationreactionmodified})
suffers from none of the problems of the original equation. The
modified equation of motion is second order in time, so there are no
``new degrees of freedom'' present.  All of its solutions
are well behaved, i.e, there are no runaway solutions nor any
pre-acceleration effects. We conclude that, in this case,
the problem of pathological solutions can therefore be overcome
by adopting Eq.~(\ref{radiationreactionmodified}) as the equation of
motion.

A refined version of the reduction of order method can be applied
when the original equation is known to higher than first order in
the small parameter $\tau$, as occurs in semiclassical gravity
when the equations are justified via the ``$1/N$''
approximation, as discussed in Sec.~\ref{status} above. To illustrate
this refined version, consider, again, Eq.~(\ref{redord}), but in the
case where this equation is known to $O(\tau^2)$, so that there is
possibly an additional explicit correction term present of the form $\tau^2
Q'$.  We wish the reduced
order equation also to be valid to $O(\tau^2)$. To achieve this, we
eliminate
the higher order derivatives from the term $\tau Q$ exactly as before,
except that we now discard only the new terms which are cubic
or higher order in $\tau$. The resulting equation will then be of the
desired form at order $\tau$, but there will remain a term of the form
$\tau^2 {\tilde Q}$, where ${\tilde Q}$ contains higher derivatives.
However, these higher derivative contributions
to ${\tilde Q}$ can then be eliminated by
applying the same procedure to ${\tilde Q}$ as was previously
applied to $Q$. Clearly, this procedure can be generalized to any finite
order in $\tau$.

Consider now the application of the reduction of order procedure to
the linearized semiclassical Einstein equation
(\ref{firstorderexplicit}), where the small parameter is $\hbar$.
Equation (\ref{firstorderexplicit}) is an integro-differential equation
for the metric perturbation
rather than a local partial differential equation. In general, the
reduction of order procedure could be ambiguous for such equations,
since it may be possible to alter the apparent differential order of
terms in an integrand via integration by parts. However, since the
right hand side of Eq.~(\ref{firstorderexplicit}) involves only
derivatives of
the linearized Einstein tensor, there is an obvious procedure in this
case for
obtaining a reduced order equation valid to
order $\hbar$ [or, equivalently, to order $1/\alpha^2 = (L_p/ {\cal
L})^2$, c.f., Eqs.~(\ref{firstorderexplicit}) and (\ref{alphadep})
above]:
We merely substitute $G_{ab}^{(1)} = 0$ on the right hand side of
Eq.~(\ref{firstorderexplicit}). However, the resulting equation is too
trivial in that it does not incorporate any of the effects of the curved
spacetime. In order to see these effects -- and, thus, the dominant
curvature-related contributions to ANEC at small curvatures
and long wavelengths -- we must go to second order in $\hbar$. To do
so, we apply to Eq.~(\ref{firstorderexplicit}) the above ``refined
version"
of the reduction of order algorithm to order $\hbar^2$. This corresponds
to substituting
$\hbar s_{ab}/\kappa$ for $G_{ab}^{(1)}$ everywhere on the right hand side
of
Eq.~(\ref{firstorderexplicit}), where
$s_{ab} \equiv {\cal D}^{(0)}_{ab} F_{\rm in}^{(1)}$. Using
Eq.~(\ref{Hscaling}), the resulting reduced
order equation is
\begin{eqnarray}
\kappa G_{ab}^{(1)}(x) &=&  \hbar s_{ab}
+  2 \pi {\hbar^2 \ln \hbar } \left[ a {\cal
A}_{ab}(x) +
b {\cal B}_{ab}(x) \right] \nonumber \\
\mbox{} &+& \hbar^2 \bigg\{ \beta {\cal B}_{ab}(x) +
 \int_{M} d^4 x^\prime H_{\lambda/L_p}(x - x^\prime) \nonumber \\
\mbox{} && \times  \left[ a {\cal A}_{ab}(x^\prime) + b
{\cal B}_{ab}(x^\prime)\right] \bigg\} \nonumber \\
\mbox{} &+& O[\hbar^3 (\ln \hbar)^2],
\label{firstordermixed3}
\end{eqnarray}
where ${\cal A}_{ab}$ and ${\cal B}_{ab}$ are given by
Eqs.~(\ref{ABdef}) with $G_{ab}^{(1)}$ replaced by $s_{ab}/\kappa$.
As discussed in Sec.~\ref{scaling1} above, the quantity $\lambda/L_p$
in Eq.~(\ref{firstordermixed3}) is a $\hbar$-independent constant.
This modified equation is second order in time and simply has the form
of the classical linearized Einstein equation with a given source.
Thus, it has no new ``degrees of freedom'' nor does it admit any
solutions with pathological behavior. Furthermore,
the exact solutions to this reduced order equation
(\ref{firstordermixed3}) will fail to satisfy the
unmodified linearized semiclassical equation (\ref{firstorderexplicit})
only by terms of order $O(\hbar^3)$.
Thus, we shall adopt Eq.~(\ref{firstordermixed3}) as the equation of
motion for linearized semiclassical gravity in our subsequent analysis.

Several facts should be noted concerning the above reduction of order
of the first order semiclassical equation.  First, solutions to the
reduced order equation (\ref{firstordermixed3}) actually correspond
precisely to the second order ``approximate perturbative solutions" of
subsection \ref{proposal1} [i.e., the approximate perturbative
solutions obtained by retaining terms of $O(1)$, $O(\hbar \ln \hbar)$,
and $O(\hbar)$].  This very special situation arises because all terms
of order $O(\hbar)$ which involve $h^{(1)}$ in the unmodified equation
(\ref{firstorderexplicit}) are proportional to $G_{ab}^{(1)}$, and, thus,
vanish in the classical limit. In more general situations, even for
linear equations, solutions to reduced order equations will differ
significantly from approximate perturbative solutions.  As discussed
above, in situations where the solutions do differ and where the
reduction of order procedure can be justified, solutions of the
reduced order equations should give a much better description of
physical phenomenon than approximate perturbative solutions.

Second, although we have
formally treated $\hbar$ as the small parameter, we could equivalently
have started from Eq.~(\ref{alphadep}) and used $1/\alpha^2$ instead;
c.f., Eq.~(\ref{alphahbar}) above and associated discussion.  A
closely analogous equivalence applies to the radiation reaction
equation:
If we consider a one parameter family of electric fields given by
${\bf E}({\bf x},t; \alpha) = \alpha^{-2} {\bf E}_0({\bf x},t/\alpha)$
and define ${\bf X}(t;\alpha) = {\bf x}(\alpha t;\alpha)$, then ${\bf
X}$ satisfies the differential equation
\begin{equation}
{\ddot {\bf X}} = {q\over m} \,  {\bf E}_0({\bf X},t) + {{\tau} \over
\alpha} {\stackrel{{\bf \ldots}}{{\bf X}}}.
\label{radiationreactionrescaled}
\end{equation}
It is clear that reducing order treating $1/\alpha$ as the small
parameter is equivalent to treating $\tau$ as the small parameter.

Third, we can only justify going to order $\hbar^2$ (or higher) in
the reduction of order procedure in the context of
the $1/N$ limit.  This is because, in the case of finite $N$, there
will be unknown corrections to Eq.~(\ref{firstorderexplicit}) at the
same order [$O(\hbar^2)$] as terms that we have retained.  The $1/N$
limit is still necessary even if we specialize to the situation,
discussed in Sec.~\ref{scaling1} above, that the incoming
gravitational radiation does not dominate the first order metric
perturbation.  To see this, let us write the unknown corrections to
the right hand side of Eq.~(\ref{firstorderexplicit}), for finite $N$,
as
\begin{equation}
{\hbar^2 \over N} L_{ab}[h^{(1)}] + {\hbar^2 \over N}
K_{ab}[\omega_{\rm in}^{(1)}] + O(\hbar^3,1/N^2),
\label{unknown}
\end{equation}
where $L_{ab}$ and $K_{ab}$ are linear but otherwise unknown
functionals of $h^{(1)}$ and $\omega_{\rm in}^{(1)}$ respectively.  Now if we
assume that the incoming gravitational radiation satisfies the
condition (\ref{gravradncondt}), then for all solutions of the
equation, $h_{ab}^{(1)} \propto \hbar$ [c.f., Eq.~(\ref{chiexpand}) above],
and the first term in Eq.~(\ref{unknown}) can be neglected.  However,
the second term will still be present unless we let $N \to \infty$.

Our final remark is that, in the large $N$ limit, the original
equation (\ref{firstorderexplicit}) is formally known to all orders in
$\hbar$, whereas the reduced order equation is valid only to order
$\hbar^2$. Therefore, the reduced order equation
(\ref{firstordermixed3}) is slightly less accurate than the original
equation.  However, this slight loss of accuracy is unimportant
since -- for the reasons
explained in Sec.~\ref{proposal1} above -- the effect of $O(\hbar^3)$
corrections should be negligible in the long wavelength limit.
Note that if more accurate equations were needed, it would be
straightforward to iterate the
reduction of order procedure to obtain an equation
accurate to any desired order in $\hbar$.  In Appendix \ref{exactanec}
below we shall effectively carry out reduction of order to arbitrarily high
order in $\hbar$.

We now consider the second order semiclassical Einstein equation
(\ref{secondorder}).  As we have previously noted, the explicit form
of this equation is not known, since the term $\langle
T_{ab}^{(2)}[h^{(1)},h^{(1)}], \omega_{\rm in,0} \rangle$ has not been
evaluated. However, in subsection \ref{scaling1} above, we derived the
explicit approximate form (\ref{secondorderapprox}) of this equation,
which is valid for long wavelengths and when the incoming
gravitational radiation does not dominate the first order metric
perturbation.  The approximate equation (\ref{secondorderapprox}) also
has the character of possessing higher derivative terms multiplied by
a small parameter (namely, $1/\alpha^2$), and has unknown correction
terms
of order $O(\ln \alpha / \alpha^6)$.  Thus, we may apply the reduction
of order algorithm directly to this equation to obtain an equation
which should be as accurate as Eq.~(\ref{secondorderapprox}) at long
wavelengths but which has none of the pathological behavior at short
wavelengths.  We obtain
\FL
\begin{eqnarray}
\kappa G^{(1)}_{ab}[{\hat h}^{(2)}] &=& {1 \over \alpha^2} \langle
T^{(0)}_{ab},
{\bar \omega}^{(2)}_{\rm in} \rangle + {\ln \alpha \over \alpha^4}
Z^{(1)}_{ab}[\chi^{(2,0)}] \nonumber \\
\mbox{} && + {1 \over \alpha^4} \langle
T^{(1)}_{ab}[\chi^{(2,0)}],\omega_{\rm in,0} \rangle + {1 \over \alpha^4}
\langle
T_{ab}^{(1)}[\chi^{(1,0)}],{\bar \omega}_{\rm in}^{(1)} \rangle \nonumber \\
\mbox{} && - {\kappa \over \alpha^4}
G_{ab}^{(2)}[\chi^{(1,0)},\chi^{(1,0)}] + O\left( {(\ln \alpha)^2 \over
\alpha^6} \right).
\label{secondordermixed3}
\end{eqnarray}
Here $\chi^{(2,0)}$ denotes the retarded solution to the equation
\begin{equation}
\kappa G^{(1)}_{ab}[\chi^{(2,0)}] =  \langle T^{(0)}_{ab},
{\bar \omega}^{(2)}_{\rm in} \rangle.
\label{chi20eqn}
\end{equation}
We shall use Eq.~(\ref{secondordermixed3}) in our analysis of the
validity of ANEC to second order in $\varepsilon$ in
Sec.~\ref{secondorderanalysis} below.  Note that solutions to
Eq.~(\ref{secondordermixed3}) also coincide with second order
approximate perturbative solutions.

As we have just argued, the reduction of order procedure is applicable
to our perturbation analysis and, in a completely satisfactory manner,
it solves the problem of the existence of extra degrees of freedom and
pathological solutions possessed by the unmodified equations. However,
in general, the method of reduction of order has some important limits
to its applicability, and we now briefly mention two of these.

First, the method is directly applicable only to local, ordinary or
partial, differential equations, although we were able to extend it in
a natural manner to the integro-differential equation
(\ref{firstorderexplicit}).  However, the full, nonlinear semiclassical
Einstein equation is a highly nonlocal equation, which is not known to
be even of an integro-differential type. Thus, it is not obvious if
and/or how the reduction of order procedure could be applied to the
full, nonlinear semiclassical Einstein equation.

Second, although the reduction of order algorithm can be applied
to any system of local differential
equations of the form (\ref{redord}), in the case
of partial differential equations, the procedure is guaranteed only
to reduce the differential order of the time derivatives, not of the
spatial derivatives.
Consider for example the equation in Minkowski spacetime
\begin{equation}
\Box \Phi = \rho + \epsilon H^{abc} \nabla_a \nabla_b \nabla_c \Phi
\label{eghard}
\end{equation}
for a scalar field $\Phi$, where $H^{abc}$ is a fixed tensor and
$\epsilon$ is a small parameter.  The reduction of order procedure can
be used to eliminate the third order time derivative of $\Phi$ from
the equation, but it does not eliminate the third order spatial
derivatives or all of the third order mixed spatial and time
derivatives.  In particular, the resulting reduced order equation is
not hyperbolic, and presumably would not have a well posed initial value
formulation. Furthermore, in circumstances where this happens, the
reduction procedure will, in general, necessitate breaking Lorentz
covariance, i.e., one will obtain inequivalent reduced order equations
by carrying out the procedure with respect to different choices of time
coordinate. Thus, it is only in the happy circumstance --- such as in
the case of Eq.~(\ref{firstorderexplicit}) --- where the reduction of
order procedure simultaneously eliminates all of the higher order time
{\em and} space derivatives that this procedure is likely to yield an
equation with satisfactory mathematical properties.

Fortunately, the above difficulties need not concern us here. As we
have discussed above, Eqs.~(\ref{firstordermixed3})
and (\ref{secondordermixed3}) appear to provide a
completely satisfactory solution to the problem of extracting physical
predictions from perturbative semiclassical gravity. The remainder of
this paper will be devoted to investigating whether ANEC holds for
solutions to these reduced order equations.

\section{The generalized ANEC integral}

We now turn to the second of the two principal purposes of this paper,
which is to analyze the positivity of the ANEC integral in solutions
of the perturbative, reduced order semiclassical equations.
Specifically, given the metric perturbations $h_{ab}^{(1)}(x)$ and
$h_{ab}^{(2)}(x)$, we would like to show that the ANEC integral along any
complete, achronal, null geodesic in the spacetime $(M, \eta_{ab} +
\varepsilon h^{(1)}_{ab} + \varepsilon^2 h^{(2)}_{ab})$ is non-negative to
order $\varepsilon^2$.  However, this desired positivity property
actually fails, as explained in the introduction (see also below).
Nevertheless, we do obtain a positivity result involving a
transversely smeared version of the ANEC integral, in which the
null-null component of the stress tensor is averaged transversely to
the geodesic as well as along the geodesic.  This transversely smeared
ANEC integral plays a key role in our main results.  In this section
we define a third integral which we call a generalized ANEC integral,
which is an integral over all of spacetime, and which reduces to the
transversely smeared ANEC integral in a certain limit
[c.f., Eq.~(\ref{generalaneclimit1}) below].  This generalized ANEC
integral
is useful as a technical tool in our proofs below.  In this section we
define the transversely smeared and generalized ANEC integrals in
general spacetimes.  We also
derive the perturbative expansion in $\varepsilon$ of the usual and
generalized ANEC integrals.

\subsection{Definition of the generalized ANEC integral}
\label{generalizedanec}

Let $\gamma$ be any inextendible null geodesic
in an arbitrary spacetime, $(M,g_{ab})$.  To begin, fix a
smooth, positive function $S({\bf x})$ on ${\bf R}^2$, with $\int
S({\bf x}) \, d^2{\bf x}=1$, which depends only on the magnitude
$|{\bf x}|$ of ${\bf x}$.  This
{\it smearing function} will control the transverse smearing.
Let $\lambda$ be an affine parameter for $\gamma$, and
denote by $\lambda^a$ the null tangent vector $\partial / \partial
\lambda$.  Let ${\cal P}$ be a fixed point on the geodesic, and
introduce an orthonormal basis $\lambda^a$, $\zeta^a$, $e_A^a$ at
${\cal P}$, $A=1,2$, where $\zeta^a \zeta_a =0$, $\lambda^a
\zeta_a=-1$, and $e_A^a e_B^b g_{ab} = \delta_{AB}$.  Extend this basis
by
parallel transport to all of $\gamma$.  Introduce Fermi-Walker type
coordinates $x = (\lambda, \zeta, x_T^1, x_T^2)$ in a neighborhood
${\cal N}$ of $\gamma$, such that the exponential map takes the vector
$\zeta
\zeta^a + x_T^A e_A^a$ at the point $x(\lambda)$ on $\gamma$ to the
point with coordinates $(\lambda,\zeta,x_T^A)$.  Then the vector field
$\lambda^a
\equiv (\partial / \partial \lambda)^a$ is a vector field on ${\cal
N}$ which is an extension of the tangent to the geodesic.  Let $\chi$ be
some
smooth function which is unity in a
neighborhood of $\gamma$ and which vanishes outside ${\cal N}$.  Fix
lengths
$\Lambda$, $\Lambda_L$ and $\Lambda_T$ and define the function
$\Theta_\gamma \in C^\infty(M)$ by
\FL
\begin{eqnarray}
\Theta_\gamma(\lambda, \zeta, {\bf x}_T) &=& {\chi(x) \over \sqrt{2 \pi}
\Lambda_L }
\exp\left\{ - {1 \over 2} \left[ \lambda^2 /
\Lambda^2 + \zeta^2 / \Lambda_L^2 \right] \right\} \nonumber \\
\mbox{} && \times {1 \over \Lambda_T^2} S({\bf x}_T/ \Lambda_T),
\label{geodesicfunction}
\end{eqnarray}
where ${\bf x}_T = (x_T^1,x_T^2)$.
By virtue of the truncating function $\chi$, the function
$\Theta_\gamma$ is well
defined even where the Fermi-Walker coordinates do not exist.
We define the generalized ANEC integral to be
\FL
\begin{equation}
I_s(\Lambda, \Lambda_L, \Lambda_T) = \int_M d^4x \sqrt{-g(x)} \, \,
\Theta_\gamma(x) G_{ab}(x) \lambda^a \lambda^b.
\label{generalanec}
\end{equation}

Clearly this quantity depends on our arbitrary choices of ${\cal P}$,
$\chi$, $\zeta^a$, $S$ etc.  However, there are two separate cases in
which this arbitrariness becomes unimportant.  First, the limit
\begin{equation}
\lim_{\Lambda \to \infty} \ \lim_{\Lambda_L, \Lambda_T \to 0} \,
I_s(\Lambda, \Lambda_L, \Lambda_T),
\label{generalaneclimit}
\end{equation}
when it exists, should be independent of these arbitrary choices, and
for
sufficiently well behaved stress tensors should reduce to the usual ANEC
integral.  Some of our results below will apply to the quantity
(\ref{generalaneclimit}).

The second case in which we can obtain something which does not depend
on our arbitrary choices of ${\cal P}$ and $\chi$ is when
we specialize the definition to perturbation theory about flat spacetime.
Here we consider the transversely smeared ANEC integral
\begin{equation}
{\bar I}_s(\varepsilon;\Lambda_T) = \lim_{\Lambda \to \infty} \
\lim_{\Lambda_L \to 0} \,
I_s(\Lambda, \Lambda_L, \Lambda_T),
\label{generalaneclimit1}
\end{equation}
where $\Lambda_T$ is fixed at a value of the order of the Planck
length.
In perturbation theory about flat space
we can choose ${\cal N}$ to be the entire spacetime and $\chi$ to be
unity, and obtain at each order in $\varepsilon$ quantities which depend
only on
(i) the length $\Lambda_T$, (ii) the smearing function $S$, and (iii)
the choice of parallel
propagated null vector $\zeta^a$ along $\gamma$ with $\zeta^a
\lambda_a = -1$.  In subsection \ref{anecexpansion1} below we  derive
explicit formulae for these quantities.

\subsection{Expansion of the ANEC integral}
\label{anecexpansion}

We now derive the expansion in $\varepsilon$ of the usual ANEC integral;
the expansion of the generalized ANEC integral will be considered in
the following subsection.  Suppose that $\gamma(\varepsilon)$ is a one
parameter family of curves on $M$ such that $\gamma(\varepsilon)$ is
a null geodesic with respect to the metric $g_{ab}(\varepsilon)$.
This one parameter family can be represented by a map $\Gamma: {\bf
R} \times (-\varepsilon_0,\varepsilon_0) \to M: (\lambda,\varepsilon)
\to
x^a(\lambda,\varepsilon)$, where for each $\varepsilon$, the parameter
$\lambda$ is an affine parameter for the corresponding geodesic.  To
order $\varepsilon$, this one parameter family of geodesics is
characterized by the zeroth order geodesic $\gamma = \gamma(0)$ (a
null geodesic in Minkowski spacetime), and by the vector field $v^c =
(\partial / \partial \varepsilon)^c$ on $\gamma$.  This vector field
cannot be
completely freely specified on $\gamma$ but must satisfy the equation
\begin{equation}
\lambda^b \lambda^c \nabla_b \nabla_c v^a + C^{(1)\, a}_{\ \ \ \ bc}
\lambda^b \lambda^c =0,
\label{jacobi}
\end{equation}
in order that $\gamma(\varepsilon)$ be a geodesic to order
$\varepsilon$.  Here the tensor $C^{(1)}$ is given by
\begin{equation}
C^{(1)\, a}_{\ \ \
\ bc} = - \nabla^a h^{(1)}_{bc} + 2 \nabla_{(b} h^{(1) \,
a}_{\,\,\,\,\,c)}.
\label{C1abc}
\end{equation}
Under the gauge transformation
(\ref{gaugefreedom1}), we have $\Gamma \to {\bar \Gamma} =
\varphi_{\varepsilon}^{-1} \circ \Gamma$, and correspondingly
\begin{equation}
v^a \to {\bar v}^a = v^a - \xi^{(1) a}.
\label{vgauge}
\end{equation}
The gauge transformation properties (\ref{gaugeaction}) and
(\ref{vgauge}) are consistent with differential equation
(\ref{jacobi}).

Now let
\begin{equation}
I(\varepsilon) = \int_{\gamma(\varepsilon)} d \lambda \
G_{ab}(\varepsilon)
\lambda^a(\varepsilon) \lambda^b(\varepsilon),
\label{anecbasic}
\end{equation}
where $\lambda^a(\varepsilon)$ is the tangent to
$\gamma(\varepsilon)$.  Let $\lambda^a = \lambda^a(0)$ be the tangent
to $\gamma$.  Then it is easy to show that
\begin{equation}
I(\varepsilon) = \varepsilon I^{(1)} + \varepsilon^2 I^{(2)} +
O(\varepsilon^3),
\end{equation}
where
\begin{equation}
I^{(1)} = \int_\gamma G_{ab}^{(1)} \lambda^a \lambda^b,
\label{firstorderanec}
\end{equation}
and
\begin{equation}
I^{(2)} = \int_\gamma \left[ G_{ab}^{(2)}  + {\cal L}_v G_{ab}^{(1)} \right]
\lambda^a \lambda^b.
\label{secondorderanec}
\end{equation}
Here ${\cal L}_v$ denotes the Lie derivative, and $G^{(2)}_{ab}$ is the
complete second order perturbation in the Einstein tensor, given by
\begin{equation}
G^{(2)}_{ab} = G^{(1)}_{ab}[h^{(2)}] + G^{(2)}_{ab}[h^{(1)},h^{(1)}].
\label{Gsdef}
\end{equation}
The second term in
Eq.~(\ref{secondorderanec}) can be thought of as reflecting the fact
that the metric perturbations cause a change in the geodesic.  Note
that the quantities $I^{(1)}$ and $I^{(2)}$ are gauge invariant (in the
``active'' sense in which we are considering the gauge freedom here),
but that $I^{(2)}$ does depend on the gauge covariant vector field $v^a$
on $\gamma$.

\subsection{Expansion of the generalized ANEC integral}
\label{anecexpansion1}

Consider now a corresponding analysis of the perturbative expansion of
the generalized ANEC integral (\ref{generalanec}).  For each finite
$\varepsilon$, the construction of the coordinate system described in
Sec.~\ref{generalizedanec} yields scalar fields
$\lambda(\varepsilon)$, $\zeta(\varepsilon)$ and
$\sigma_T(\varepsilon) \equiv \delta^{AB} x^T_A x^T_B$ on the
spacetime $(M,g_{ab}(\varepsilon))$, and also the vector field
$\lambda^a(\varepsilon) \equiv (\partial / \partial
\lambda)^a(\varepsilon)$.  Note that although these objects are
defined in terms of an algorithm to obtain a coordinate system, they
are themselves coordinate-independent scalar and tensor fields.  Their
domain of definition is however restricted to some neighborhood of the
geodesic $\gamma$.  They can be expanded as
\begin{eqnarray}
\lambda(\varepsilon) &=& \lambda^{(0)} + \varepsilon \lambda^{(1)} +
O(\varepsilon^2)\nonumber
\\
\zeta(\varepsilon) &=& \zeta^{(0)} + \varepsilon \zeta^{(1)} +
O(\varepsilon^2) \nonumber \\
\sigma_T(\varepsilon) &=& \sigma_T^{(0)} + \varepsilon
\sigma_T^{(1)} + O(\varepsilon^2) \nonumber  \\
\lambda^a(\varepsilon) &=& \lambda^{(0)\,a} + \varepsilon
\lambda^{(1)\,a} + O(\varepsilon^2),
\end{eqnarray}
where the expansion coefficients on the right hand side are
defined on all of $M$.
Similarly the volume 4-form can be expanded as
$\epsilon_{abcd}(\varepsilon) = \epsilon^{(0)}_{abcd} + \varepsilon
\epsilon^{(1)}_{abcd} + O(\varepsilon^2)$, where $\epsilon^{(1)} =
h^{(1)\,e}_{\,\,\,\,\,\,e} \epsilon^{(0)}/2$.  Inserting these expansions
into
Eq.~(\ref{generalanec}) yields
\begin{equation}
I_s(\varepsilon) = \varepsilon I_s^{(1)} + \varepsilon^2
I_s^{(2)} + O(\varepsilon^3),
\end{equation}
where
\FL
\begin{equation}
I_s^{(1)}(\Lambda, \Lambda_L, \Lambda_T) = \int_{M} \epsilon^{(0)}_{cdef}  \
\Theta^{(0)}_\gamma(x) G^{(1)}_{ab}(x) \lambda^a \lambda^b.
\label{firstorderanec1}
\end{equation}
Here $\Theta^{(0)}_\gamma(x)$ is given by the expression
(\ref{geodesicfunction}) with $\chi$ replaced by $1$, and where the
arguments of the expression (\ref{geodesicfunction}) are
$\lambda = \lambda^{(0)}, \zeta=\zeta^{(0)}$, etc., the inertial coordinates on
Minkowski spacetime.

The expression for $I^{(2)}_s$ is
\FL
\begin{eqnarray}
I_s^{(2)} &=& \int_{M} \epsilon^{(0)}_{cdef}  \
\Theta^{(0)}_\gamma(x) G^{(2)}_{ab} \lambda^a \lambda^b, \nonumber \\
\mbox{} && + \int_{M}
\epsilon^{(0)}_{cdef} G^{(1)}_{ab} \big[
 2 \lambda^{(0)\,a} \lambda^{(1)\,b} \Theta_\gamma^{(0)}
+ \lambda^{(0)\,a} \lambda^{(0) \,b} \Theta_\gamma^{(1)}
\nonumber\\ \mbox{} &&
+ {1\over2}
h^{(1)\,e}_{\,\,\,\,\,\,e} \lambda^{(0)\,a} \lambda^{(0) \,b}
\Theta_\gamma^{(0)} \big].
\label{Igamma2}
\end{eqnarray}
Since the smearing function $S$ in Eq.~(\ref{generalanec})
depends only on the magnitude $\sigma_T \equiv {\bf x}_T^2$ of ${\bf
x}_T$, the function $\Theta_\gamma^{(1)}$ that appears in
Eq.~(\ref{Igamma2}) is given by
\FL
\begin{equation}
\Theta_\gamma^{(1)}(x) = {\partial \Theta_\gamma \over \partial \lambda}
\lambda^{(1)}(x) + {\partial \Theta_\gamma \over \partial \zeta}
\zeta^{(1)}(x) + {\partial \Theta_\gamma \over \partial \sigma_T}
\sigma_T^{(1)}(x).
\end{equation}

As in the previous subsection, the quantities $I_s^{(1)}$ and $I_s^{(2)}$ are
gauge invariant.  We now specialize to a particular choice of gauge,
which is just that
associated with the coordinate system described in
Sec.~\ref{generalizedanec}.  Specifically, suppose that we are give a
1-parameter family of metrics $g_{ab}(\varepsilon)$ and a fixed choice
of gauge for each $\varepsilon$.  Now apply the diffeomorphism
$\varphi_\varepsilon$ given by identifying the coordinates
$\lambda,\zeta,x_T^A$ in the spacetimes $(M,g_{ab}(0))$ and
$(M,g_{ab}(\varepsilon))$ \cite{note8}.  This yields a choice of
gauge in which we have
\begin{equation}
\lambda^{(1)\,a} = \lambda^{(1)} = \zeta^{(1)} = \sigma_T^{(1)} = 0.
\end{equation}
The expression for $I_s^{(2)}$ in this gauge reduces to
\begin{eqnarray}
I_s^{(2)} &=& \int_{M} \epsilon^{(0)}_{cdef}  \
\Theta^{(0)}_\gamma(x) \lambda^a \lambda^b \nonumber \\
\mbox{} &&\times \left[ G^{(2)}_{ab}(x) + {1\over2}
h^{(1)\,e}_{\,\,\,\,\,\,e}(x) G^{(1)}_{ab}(x)\right].
\label{secondorderanec2}
\end{eqnarray}
In this gauge the vector field $v^a$ described in
Sec.~\ref{anecexpansion} vanishes.  
Moreover it is straightforward to show that $h_{ab}^{(1)}$ vanishes
along the geodesic $\gamma$, and that $h_{ab}^{(1)} \lambda^a
\lambda^b =0$ throughout $M$.  These consequences of our gauge choice
will be used in Sec.~\ref{vanishing} below.  Finally, we note that in
the limit 
$\Lambda \to \infty$, $\Lambda_L, \Lambda_T \to 0$ of no transverse
smearing, the expression (\ref{secondorderanec2}) reduces to the
previously obtained expression (\ref{secondorderanec}), since
$h^{(1)\,e}_{\,\,\,\,\,\,e}$ vanishes on $\gamma$ in our chosen gauge.

\section{The ANEC integral in first order perturbation theory}
\label{firstorderexact}

In this section we establish the results concerning the first order
contribution to the ANEC integral which were discussed in the
introduction.  We start in subsection \ref{states} by characterizing
the precise class of incoming states we are considering.  In subsection
\ref{solns1} we obtain the solutions of the reduced order
semiclassical equation (\ref{firstordermixed3}).  We derive a general
formula for the first order perturbation (\ref{firstorderanec1}) to
the transversely smeared ANEC integral for these solutions in
subsection \ref{generalformulae}.  Finally, we show that this vanishes
for pure incoming states in subsection \ref{puresec}, and that for
mixed states it is positive in the long-wavelength limit in subsection
\ref{mixedsec}.

In Appendix B below we consider a specific subclass of solutions of
the original, unmodified semiclassical equation (\ref{firstorder}),
given by the use of half advanced plus half retarded Greens functions.
We show that for these solutions, in the region $\lambda >
\lambda_{\rm crit}$ of parameter space, a transversely smeared ANEC
integral vanishes for pure states and is always strictly positive for
mixed states, even outside of the long wavelength limit.  Although, as
discussed in Sec.~\ref{proposal2} above, we do not attribute any
preferred physical status to these solutions, expanding this result in
$1/\alpha^2$ and $\ln \alpha / \alpha^2$ provides an alternative
demonstration of the results of this section for solutions of the
reduced order equation in the long wavelength limit, at least for
$\lambda > \lambda_{\rm crit}$.

\subsection{Characterization of incoming states}
\label{states}

In the remainder of the paper we do not deal with fully general
states.  Rather, we restrict attention to a subclass of states whose
two point functions have suitable differentiability properties and
fall-off properties at infinity when
restricted to spatial slices. More specifically, let $\Sigma$ denote
the hypersurface $t=0$ in Minkowski spacetime, and let ${\cal V}$ be
the class of smooth functions on $\Sigma \times \Sigma$ all of whose
derivatives are $L^1$ on $\Sigma \times \Sigma$.
We consider states whose two point function perturbations $F_{\rm
in}^{(1)}({\bf
x},t;{\bf x}',t')$ and $F_{\rm in}^{(2)}({\bf x},t;{\bf x}',t')$ and
associated time derivatives $\partial F_{\rm in}^{(j)} / \partial t$,
$\partial F_{\rm in}^{(j)} / \partial t'$, and $\partial^2 F_{\rm
in}^{(j)} / (\partial t \partial t')$, for $j = 1,2$ all lie in ${\cal
V}$ when restricted to $\Sigma \times \Sigma$.
Our choice of class of states is dictated mostly by convenience and is
not the most general class of states for which our results are valid;
however it is a sufficiently large class of states to be interesting.

We can express $F_{\rm in} = \varepsilon F_{\rm in}^{(1)} + \varepsilon^2
F_{\rm in}^{(2)}$ as
\begin{eqnarray}
F_{\rm in}(x,x^\prime) &=& \int d^3 {\bf k} \, \int d^3 {\bf
k}^\prime \,
f({\bf k},{\bf k}^\prime) \, e^{i k \cdot x} \, e^{i k^\prime \cdot
x^\prime} \nonumber \\
\mbox{} && + \int d^3 {\bf k} \, \int d^3 {\bf k}^\prime \,
g({\bf k},{\bf k}^\prime) \, e^{i k \cdot x} \, e^{- i k^\prime \cdot
x^\prime} \nonumber \\
\mbox{} && + \, {\rm c.c.}
\label{F1explicit}
\end{eqnarray}
Here as in Appendix \ref{FTstress}, bold face vectors are spatial,
three dimensional vectors, while $k = k^a$ denotes a 4-vector.
Also it is understood that $k^a = ({\bf k},\omega_{\bf k})$ where
$\omega_{\bf k} = |{\bf k}|$.
Equation (\ref{F1explicit}) essentially defines the functions $f$ and
$g$ as
suitable complex linear combinations of the spatial Fourier transforms
of the four functions $F_{\rm in}$, $\partial F_{\rm
in} / \partial t$, $\partial F_{\rm in} /
\partial t^\prime$ and $\partial^2 F_{\rm in} / (\partial t
\partial t^\prime)$ restricted to $\Sigma \times \Sigma$, where
$\Sigma$ is the Cauchy surface $t=0$; see Eq.~(\ref{explicitfgdef})
below.
Note that the functions $f$ and $g$ are formally related to the
conventional creation and annihilation operators ${\hat a}_{\bf
k}^\dagger$ and ${\hat a}_{\bf k}$ by
\begin{eqnarray}
f({\bf k},{\bf k}^\prime) &\propto& \langle {\hat a}_{\bf k} {\hat
a}_{{\bf
k}'} \rangle \nonumber \\
g({\bf k},{\bf k}^\prime) &\propto& \langle {\hat a}_{{\bf k}'}^\dagger
{\hat a}_{\bf
k} \rangle.
\end{eqnarray}
Note also that the part of the two point function that is purely
positive frequency or purely negative frequency is given by the
function $f$, while the ``mixed-frequency''part (the part that is
positive frequency with respect to one variable and negative frequency
with respect to the other) is given by the function $g$.

We expand $f$ and $g$ as
\begin{eqnarray}
f &=& \varepsilon f^{(1)} + \varepsilon^2 f^{(2)} + O(\varepsilon^3) \nonumber
\\
g &=& \varepsilon g^{(1)} + \varepsilon^2 g^{(2)} + O(\varepsilon^3).
\label{fgexpand}
\end{eqnarray}
In the terminology introduced in Sec.~\ref{pureversusmixed} above, we
refer to the first order perturbed two point function $F_{\rm in}^{(1)}$
as ``pure" if and only if $g^{(1)}=0$; otherwise
$F_{\rm in}^{(1)}$ is said to be ``mixed".

The functions $f$ and $g$ cannot be chosen arbitrarily but must
satisfy the positivity condition (\ref{positivity}), which for Fock
space states is just the statement that
\begin{equation}
\langle {\hat \Phi}(u) {\hat \Phi}(u)^\dagger
\rangle \ge 0
\end{equation}
for all complex smearing functions $u$.  For general, algebraic states
$\omega_{\rm in}$
the positivity condition is equivalent to the positivity of the operator
\begin{equation}
M_{ij} = \left[\begin{array}{cc}
 g & f \\
 f^* & g^* + J \end{array} \right],
\label{Mijdef}
\end{equation}
where $J({\bf k},{\bf k}') = \hbar \delta^3({\bf k} - {\bf k}') /
(16 \pi^3 \omega_{\bf k})$, in the sense that for all $u_j \in C_0^\infty(M)$,
$1\le j \le 2$,
\begin{equation}
\int d^3 {\bf k} \int d^3 {\bf k}' \, \, {\tilde
u}_i({\bf k},\omega_{\bf k})^* M_{ij}({\bf k},{\bf k}') {\tilde
u}_j({\bf k}',\omega_{{\bf k}'}) \ge 0.
\end{equation}
Thus, $M$ is a positive operator on
$L^2({\bf R}^3)\,\oplus \,L^2({\bf R}^3)$, which implies, in
particular, that $g$ is a positive operator on $L^2({\bf R}^3)$:
\begin{equation}
\int d^3 {\bf k} \int d^3 {\bf k}' \, g({\bf k},{\bf k}') \,{\tilde
u}(k)
{\tilde u}(k')^* \ge 0.
\label{gpos0}
\end{equation}
For Fock space states, Eq.~(\ref{gpos0}) is just the statement that $
\langle {\hat \Phi}_+(u)^\dagger {\hat \Phi}_+(u) \rangle \ge 0$ for
any test function $u$.  Note also that the corresponding classical
positivity condition, which requires that the two point function be the
expected value of $\Phi(x) \Phi(y)$ with respect to some positive
measure on the space of field configurations, is the stronger
condition that
\begin{equation}
\left[\begin{array}{cc}
 g & f\\
 f^* & g^*\\
\end{array} \right] \ge 0.
\end{equation}

We now insert the expansions (\ref{fgexpand}) into the positivity
condition (\ref{Mijdef}) and expand order by order in $\varepsilon$ to
determine the restrictions on the incoming state perturbations.
We obtain at first order that
\begin{equation}
g^{(1)} \ge 0
\label{gpos}
\end{equation}
in the sense of Eq.~(\ref{gpos0}), while $f^{(1)}$ can be chosen
arbitrarily.  At second order we obtain the restriction on $g^{(2)}$ that
\begin{equation}
\left[\begin{array}{cc}
 P & 0\\
 0 & 1\\
\end{array} \right]
\,\,
\left[\begin{array}{cc}
 g^{(2)} & f^{(1)}\\
 f^{(1)\,*} & J\\
\end{array} \right]
\,\,
\left[\begin{array}{cc}
 P & 0\\
 0 & 1\\
\end{array} \right] \ge 0,
\label{poscondt2}
\end{equation}
where $P$ is the projection operator onto the kernel of $g^{(1)}$.  In
particular, when $g^{(1)}=0$ this reduces to
\begin{equation}
\left[\begin{array}{cc}
 g^{(2)} & f^{(1)}\\
 f^{(1)\,*} & J\\
\end{array} \right]
\ge 0.
\label{poscondt2b}
\end{equation}

Next, we introduce an alternative, convenient set of coordinates
on the light cone in momentum space.
Recall that we have defined
inertial coordinates $\lambda,\zeta,x_T^1, x_T^2$ on Minkowski space
$M$, and also a null orthonormal basis $\lambda^a$, $\zeta^a$,
$e_1^a$, $e_2^a$, where $\lambda^a$ is the tangent to the zeroth order
geodesic $\gamma$.  Thus, $x^a = \lambda \lambda^a + \zeta \zeta^a +
{\bf x}_T$, where ${\bf x}_T = x_T^A {\bf e}_A$, $A=1,2$.  We
introduce corresponding coordinates $\gamma, \beta, {\bf k}_T$ on
momentum space such that
\begin{equation}
k^a = \gamma \lambda^a + \beta \zeta^a +{\bf k}_T.
\label{kspace}
\end{equation}
In these coordinates the positive light cone volume
element can be written as
\begin{eqnarray}
\Theta(k^0) \delta(k^2) d^4 k &=& {d^3 {\bf k} \over 2
\omega_{\bf k}} \delta( \omega - \omega_{\bf k}) d \omega \nonumber \\
&=& \Theta(\beta) {d \beta \over 2 \beta} d^2 {\bf k}_T
\delta\left(\gamma -
{{\bf k}_T^2 \over (2 \beta)}\right) d \gamma.
\label{massshellvol}
\end{eqnarray}
Therefore solutions of the wave equation can be represented by
functions of $\beta$ and of the two dimensional vector ${\bf k}_T$.
These null coordinates on momentum space, which
are specially adapted to the given null geodesic $\gamma$, will be
useful throughout our computations below.

The first order two point function $F_{\rm in}^{(1)}$ has an expansion
analogous to Eq.~(\ref{F1explicit}) but with $f$ and $g$ replaced by
$f^{(1)}$ and $g^{(1)}$.  We can rewrite this expansion
in terms of the coordinates introduced above as
\begin{eqnarray}
F_{\rm in}^{(1)}(x,x^\prime) &=& \int_0^\infty {d\beta \over \beta}\int d^2
{\bf
k}_T  \,
\int_0^\infty {d\beta^\prime \over \beta^\prime}\int d^2 {\bf k}_T^\prime
\nonumber \\
\mbox{} && \times \bigg[ {\hat f}(\beta,{\bf k}_T; \beta^\prime,{\bf
k}_T^\prime) \, e^{i k \cdot x} \, e^{i k^\prime \cdot
x^\prime} \nonumber \\
\mbox{} && + {\hat g}(\beta,{\bf k}_T; \beta^\prime,{\bf k}_T^\prime)
\, e^{i k \cdot x} \, e^{-i k^\prime \cdot
x^\prime} + {\rm c.c.}\,\bigg].
\label{F1explicit1}
\end{eqnarray}
Here ${\hat f} = \omega_{\bf k} \omega_{{\bf k}^\prime} f^{(1)}$, ${\hat g}
= \omega_{\bf k} \omega_{{\bf k}^\prime} g^{(1)}$, and
\begin{equation}
\omega_{\bf k} = {1 \over \sqrt{2}} \left( \beta + {{\bf k}_T^2 \over
2 \beta}\right).
\label{omegak}
\end{equation}
It it understood in these equations that $k = k^a$ is given by
Eq.~(\ref{kspace}) with $\gamma={\bf k}_T^2/(2 \beta)$.  Finally, as
briefly discussed in Appendix \ref{FTstress}, our assumed regularity
properties on the incoming state imply that ${\hat f}$ and ${\hat g}$
are continuous
as functions of $(\beta,{\bf k}_T,\beta^\prime,{\bf k}_T^\prime)$ and
satisfy, for any integer $N$,
\begin{equation}
{\rm max}\left\{|{\hat f}|,|{\hat g}|\right\}  \le {C_N
\over (1 + \omega_{\bf k}^2  +\omega_{{\bf k}^\prime}^2)^{N-1} }
\label{regular1}
\end{equation}
for some constant $C_N$, where $\omega_{\bf k}$ is given by
Eq.~(\ref{omegak}).

\subsection{Solutions of the reduced order equation}
\label{solns1}

In Appendix \ref{FTstress} we show that for states in the above class
the stress tensor (\ref{source}) which acts as a source in the
semiclassical equations is an $L^2$ tensor field on Minkowski
spacetime.  Consequently its spacetime Fourier transform exists as an
$L^2$ function.  In Appendix \ref{FTstress} we show that the Fourier
transform ${\tilde s}_{ab}$ is actually continuous everywhere away
from the light cone, and is bounded everywhere except for an
(integrable) divergence at the origin in momentum space.  Thus, we may
use Fourier transform methods to solve the (reduced order)
semiclassical equations.

The reduced order semiclassical equation (\ref{firstordermixed3})
expresses the linearized Einstein tensor
$G_{ab}^{(1)}[h^{(1)}]$ in terms of the source tensor (\ref{source}).  In
our analysis below we shall not need to solve (\ref{firstordermixed3})
for the metric perturbation $h^{(1)}_{ab}$; it will be sufficient to
work directly with the linearized Einstein tensor.  We now take the
Fourier transform of Eq.~(\ref{firstordermixed3}).
We use the following formula given by
Horowitz \cite{Horowitz} for the Fourier transform of the distribution
$H_\lambda$ 
\begin{equation}
{\tilde H}_\lambda(k) = - 2 \pi \left[ \ln \lambda^2 |k^2| + 2 \gamma
-1 - i \pi \, {\rm sgn}(k^0) \, \Theta(-k^2)\right],
\label{tildeHlambda}
\end{equation}
where $\gamma$ is Euler's constant and $\Theta$ is the step function.
Here and below tildes denote Fourier transforms.
Using Eqs.~(\ref{ABdef}) and (\ref{Hscaling}) we obtain
\FL
\begin{equation}
\kappa {\tilde G}_{ab}^{(1)}(k) = S_1(k) {\tilde s}_{ab}
 + S_2(k) (k_a k_b - \eta_{ab} k^2) {\tilde s}_c^{\ c},
\label{sol1a}
\end{equation}
where
\begin{equation}
S_1(k) = 1 + 16 \pi a L_p^2 k^2 {\tilde H}_{\lambda}(k),
\label{S1def}
\end{equation}
and
\begin{equation}
S_2(k) = (8 \pi) \left[({2\over3}a + 2 b) L_p^2 {\tilde H}_{\lambda}(k)
+
2 \beta L_p^2 \right],
\label{S2def}
\end{equation}
We have explicitly included factors of $G \hbar \equiv L_p^2$
in these formulae.  The tensor ${\tilde s}_{ab}$ is the spacetime
Fourier transform of the source tensor (\ref{source}) discussed above.
Note that Eqs.~(\ref{sol1a}) -- (\ref{S2def}) could
also be obtained by expanding the exact solutions (\ref{sol1}) --
(\ref{F3def}) given in Appendix \ref{firstordersolutions} in powers of
$\hbar$ and $\hbar \ln(\hbar)$ (assuming $\lambda \propto
\sqrt{\hbar}$ as discussed in Sec.~\ref{scaling1} above).

Next we rewrite the source tensor ${\tilde s}_{ab}$ in terms of the
regularized two-point function of the incoming state.
Using Eqs.~(\ref{FTdef}), (\ref{stclassical}) and (\ref{F1explicit}),
we find that
\begin{eqnarray}
{{\tilde s}_{ab}(l) \over (2 \pi)^4} &=& \int d^3 {\bf k} \int d^3 {\bf
k}^\prime \, \bigg[f^{(1)}({\bf k}, {\bf
k}^\prime) \sigma_{ab}(k,k^\prime) \delta^4(k+k^\prime-l) \nonumber \\
\mbox{} && + g^{(1)}({\bf k},{\bf k}^\prime) \sigma_{ab}(k,-k^\prime)
\delta^4(k-k^\prime -l) + {\rm c.c.}\bigg],
\label{source1}
\end{eqnarray}
where
\begin{eqnarray}
\sigma_{ab}(k,k^\prime) &=& ( 2 \xi-1) k_{(a} k^\prime_{b)} +
({1\over2} - 2 \xi)
 \eta_{ab} k_c k^{\prime \, c} \nonumber \\
\mbox{} && + \xi (k_a k_b + k^\prime_a k^\prime_b).
\label{sigmadef}
\end{eqnarray}

Inserting Eq.~(\ref{source1}) into Eq.~(\ref{sol1a}) yields
the null-null component of the linearized Einstein tensor
\begin{eqnarray}
\kappa {\tilde G}^{(1)}(l)_{ab} \lambda^a \lambda^b & =& \int d^3 {\bf k}
\int d^3 {\bf
k}^\prime f^{(1)}({\bf k}, {\bf
k}^\prime) J(k,k^\prime) \delta^4(k+k^\prime-l) \nonumber \\
\mbox{} && + g^{(1)}({\bf k},{\bf k}^\prime) J(k,-k^\prime)
\delta^4(k-k^\prime -l) \nonumber \\
\mbox{} && + {\rm c.c}.
\label{G1ans}
\end{eqnarray}
Here the function $J$ is given by
\FL
\begin{eqnarray}
{J(k,k^\prime) \over (2 \pi)^4} &=&
S_1(k+k^\prime) \bigg\{ (2 \xi-1) (\lambda\cdot k) (\lambda\cdot
k^\prime) \nonumber \\
\mbox{} && + \xi \left[ (\lambda\cdot k)^2 + (\lambda \cdot
k^\prime)^2\right] \bigg\}\nonumber \\
\mbox{} && + (1-6 \xi) (\lambda\cdot k + \lambda\cdot k^\prime)^2
(k\cdot k^\prime) \, S_2(k+k^\prime).
\label{Jfn}
\end{eqnarray}
By using the techniques of Appendix \ref{FTstress}, it is possible to
show that the linearized Einstein tensor (\ref{G1ans}) has the same
regularity properties as were proved in Appendix \ref{FTstress} for
the source tensor (\ref{source}).

\subsection{The first order ANEC integral: general formula}
\label{generalformulae}

We now calculate the first order perturbation to the generalized ANEC
integral (\ref{firstorderanec1}) and express it in terms of the
functions $f^{(1)}$ and $g^{(1)}$ characterizing the incoming state
$\omega_{\rm in}^{(1)}$.
Combining Eqs.~(\ref{firstorderanec1}) and (\ref{G1ans}) yields that
\begin{eqnarray}
I_s^{(1)} &=& {1\over \kappa} \int d^3 {\bf k} \int d^3 {\bf k}^\prime
\bigg[
f^{(1)}({\bf k}, {\bf k}^\prime) J(k,k^\prime) {\tilde
\Theta}^{(0)}_\gamma(k+k^\prime) \nonumber \\
\mbox{} && + g^{(1)}({\bf k},{\bf k}^\prime) J(k,-k^\prime) {\tilde
\Theta}^{(0)}_\gamma(k-k^\prime) \bigg] + {\rm c.c.}
\label{I1ansgeneral}
\end{eqnarray}
Here ${\tilde \Theta}^{(0)}_\gamma$ is the Fourier transform of the
smearing function (\ref{geodesicfunction}):
\begin{eqnarray}
{\tilde \Theta}^{(0)}_\gamma(\gamma,\beta,{\bf k}_T) &=& \sqrt{2 \pi}
{\Lambda }
\, \exp\left\{- {1 \over 2}\left[ \beta^2
\Lambda^2 + \gamma^2 \Lambda_L^2 \right]\right\}
\nonumber \\ \mbox{} &\times& {\tilde S}({\bf k}_T \Lambda_T).
\label{fsmear}
\end{eqnarray}
The existence of the integral (\ref{I1ansgeneral}) follows from the
regularity properties of the functions $f^{(1)}$ and $g^{(1)}$ discussed in
Appendix \ref{FTstress}: they are continuous away from the light cone,
and have an integrable divergence $\propto 1 / (\omega_{\bf k}
\omega_{{\bf k}'})$ at the origin.

Now by combining the alternative representation (\ref{F1explicit1}) of
the
two point function with Eqs.~(\ref{I1ansgeneral})
and (\ref{fsmear}), we obtain
\FL
\begin{eqnarray}
&&\lim_{\Lambda_L \to 0} I_s^{(1)}(\Lambda,\Lambda_L,\Lambda_T) =
\nonumber \\
&& {1\over  \kappa}\int_0^\infty {d\beta \over \beta} \int d^2 {\bf k}_T  \,
\int_0^\infty {d\beta^\prime \over \beta^\prime} \int d^2 {\bf k}_T^\prime \,
{\hat f}(k,k^\prime) J(k , k^\prime)
\nonumber \\ \mbox{} && \times
\sqrt{2 \pi} \Lambda  \exp\left[ - (\beta +
\beta^\prime)^2 \Lambda^2/2 \right] \, {\tilde
S}\left[ ({\bf k}_T + {\bf k}_T^\prime) \Lambda_T \right] \nonumber \\
&& + {1\over \kappa}\int_0^\infty {d\beta \over \beta} \int d^2 {\bf k}_T  \,
\int_0^\infty {d\beta^\prime \over \beta^\prime} \int d^2 {\bf k}_T^\prime \,
{\hat g}(k,k^\prime) J(k , -k^\prime)
\nonumber \\ \mbox{} && \times
\sqrt{2 \pi} \Lambda \exp\left[ - (\beta -
\beta^\prime)^2 \Lambda^2 /2 \right] \, {\tilde
S}\left[ ({\bf k}_T - {\bf k}_T^\prime) \Lambda_T \right] \nonumber \\
\mbox{} && + \, {\rm c.c.}
\label{I1full}
\end{eqnarray}
Note that Eqs.~(\ref{omegak}) and (\ref{regular1}) imply that for
${\bf k}_T \ne 0$, the integrands vanish more rapidly than any power
of $\beta$ as $\beta \to 0$, thereby assuring convergence of the
integrals in Eq.~(\ref{I1full}) despite the factors of $1/\beta$
coming from the light cone volume element.

\subsection{Pure incoming states}
\label{puresec}

We now show that in the limit $\Lambda \to \infty$, the pure frequency
contribution to the ANEC integral [the first term in
Eq.~(\ref{I1full})] vanishes.  This can be seen from the fact that in
the limit $\Lambda \to \infty$, the exponential factor in this term
becomes $\delta(\beta + \beta^\prime)$ and that therefore the entire
expression vanishes. Therefore, when $g^{(1)}=0$ we have
\begin{equation}
\lim_{\Lambda \to \infty} \ \lim_{\Lambda_L, \Lambda_T \to 0} \,
I^{(1)}_s(\Lambda, \Lambda_L, \Lambda_T) =0,
\label{key1}
\end{equation}
and thus whenever the usual ANEC integral exists, it must vanish.

There is a simple, intuitive way to understand this result.  The ANEC
integral is the integral along a line in position space, and thus
becomes the integral over a hyperplane (the hyperplane $\lambda^a
k_a=0$) of the Fourier transformed Einstein tensor ${\tilde G}_{ab}^{(1)}$
in momentum space.  Now, for states whose first order perturbed
two-point function is pure (i.e., for which $g^{(1)} = 0$), the linearized
Einstein tensor has support inside and on the light cone in momentum
space.  This can be seen from Eq.~(\ref{G1ans}) and the fact that if
$k$ and $k^\prime$ are future pointing null vectors, then $k+k^\prime$
is a future pointing null or timelike vector. Therefore, the only
possible contribution to the ANEC integral must be concentrated on the
null line $k^a \propto \lambda^a$ in momentum space.  Although this
line is of measure zero, {\it a priori} there still could be a
non-vanishing result since the (Fourier transformed) Einstein tensor
could have a distributional component on the line.  However, the
argument in Appendix \ref{FTstress} shows that there is no such
distributional component of the linearized Einstein tensor, so the
ANEC integral must vanish, as our calculation above shows explicitly.

As explained in the introduction, the result (\ref{key1}) that the
ANEC integral vanishes to first order for incoming states that are
pure to first order is one of the key results of this paper.  It
eliminates the counterexample to ANEC given in
Ref.~\cite{WaldYurtsever}.  Moreover, the vanishing of the ANEC
integral to first order is a necessary condition for the ANEC integral
to be positive generally; any nonzero first order contribution for
pure states could be arranged to be negative by choosing the sign of
the first order state perturbation appropriately.  Moreover, in this
situation any transverse smearing could not help.

The above result applies for solutions to the reduced order equation
which are accurate to $O(\hbar^2)$.  However, it is straightforward to
extend this result to all orders in $\hbar$, that is, to solutions of
reduced order equations which are accurate to higher order in $\hbar$.
The only difference is that the functions $S_1$ and $S_2$ appearing in
Eq.~(\ref{Jfn}) are slightly altered, which does not affect the
argument.  More precisely, these functions are replaced by expansions
to the appropriate order in $\hbar$ and $\hbar \, \ln \hbar $ of the
functions $1/F_1$ and $F_2 / (F_1 F_3)$, as can be seen from Appendix
\ref{exactanec} below.

Finally we remark that a limiting case of the above result in which
the backreaction is dialed to zero is just the fact that the ANEC
integral in Minkowski spacetime of a matrix element of the form ${\left< 0
\right|}
{\hat T}^{(0)}_{ab} {\left| \psi \right>}^{(1)}$ must vanish; see
Eq.~(\ref{rho1pure}) above.
This fact can also be deduced from the result previously established
by Klinkhammar \cite{Klinkhammar91} and by Wald and Yurtsever
\cite{WaldYurtsever} that the ANEC integral of the expected value of
${\hat T}^{(0)}_{ab}$ in Minkowski spacetime is nonnegative for a large class
of states.

\subsection{Mixed incoming states}
\label{mixedsec}

We now turn to the situation where we allow an arbitrary, mixed incoming
state.  Using Eqs.~(\ref{tildeHlambda}), (\ref{S1def}), (\ref{Jfn})
and (\ref{I1full}) we find that
\FL
\begin{eqnarray}
{\bar I}_s^{(1)}(\Lambda_T) &\equiv &\lim_{\Lambda \to \infty} \
\lim_{\Lambda_L \to 0} \,
I_s^{(1)}(\Lambda, \Lambda_L, \Lambda_T) \nonumber \\
\mbox{} &=&  \int {d^2 {\bf \Delta k}_T \over (2 \pi)^2} \ {\tilde
{\cal F}}({\bf
\Delta k}_T)\,
{\tilde  K}({\bf \Delta k}_T) \,
{\tilde S}(\Lambda_T {\bf \Delta k}_T),
\label{anec5}
\end{eqnarray}
where
\begin{equation}
{\tilde {\cal F}}({\bf \Delta k}_T) = {16 \pi^3 \over \kappa}
\int_0^\infty d
\beta \int d^2 {\bf k}_T \, {\hat g}(\beta,{\bf k}_T; \beta,{\bf k}_T
+ {\bf \Delta k}_T),
\label{calFdef}
\end{equation}
and
\begin{equation}
{\tilde  K}({\bf \Delta k}_T) = 1 - {{\bf \Delta k}_T^2 \over
\omega_c^2} \ln \left( {\hat \lambda^2} {\bf \Delta k}_T^2
\right).
\label{tildeKdef}
\end{equation}
Here $\omega_c^2 = 1/ (32 \pi^2 a L_p^2)$, ${\hat \lambda }=\lambda
\exp(\gamma - 1/2)$, and $\gamma$ is Euler's constant.
Note that the continuity of ${\hat g}$ and the fall-off property
(\ref{regular1}) is sufficient to guarantee the
existence of the integrals (\ref{anec5}) and (\ref{calFdef}).
Also, note that there is no longer any dependence on the curvature
coupling in Eq.~(\ref{anec5}), due to cancelations in Eq.~(\ref{Jfn})
when $k' = - k$.

The function ${\tilde K}$ in Eq.~(\ref{anec5}) is essentially the
factor $S_1$ that appears in Eq.~(\ref{sol1a}) (and is also related by
an expansion in $\hbar$ to the Greens function $1/F_1$ that appears in
Appendix \ref{firstordersolutions}).  The second term involving $S_2$
in Eq.~(\ref{sol1a}) does not contribute to the ANEC integral
(\ref{anec5}), because it has a tensorial structure in momentum space
proportional to
\begin{equation}
k_a k_b - \eta_{ab} k^2.
\end{equation}
In the original Minkowski space coordinates this corresponds to the
differential operator $d^2 / d \lambda^2$, which gives rise to a
total derivative in the ANEC integral and gives a vanishing
contribution.  More precisely, the contribution to the first order
perturbation $I^{(1)}$ vanishes identically, and the the contribution to
the first order perturbation of the generalized ANEC integral
$I_s^{(1)}$ vanishes once the limit $\Lambda \to \infty$ is taken.

The formula (\ref{anec5}) has a simple physical interpretation in
terms of an ANEC integral computed in flat spacetime without
backreaction, as we now describe.  Let $\gamma^\prime$
denote the null geodesic in Minkowski spacetime obtained by displacing
the geodesic $\gamma$ transversely by an amount $-{\bf x}_T$.  Define
\begin{equation}
I_F({\bf x}_T) = \int_{\gamma^\prime}
\langle T_{ab}^{(0)} , \omega_{\rm in}^{(1)} \rangle \lambda^a \lambda^b.
\label{IFdef}
\end{equation}
This is just the ANEC integral obtained from the incoming state {\it
without} including the effects of backreaction, i.e., in the test field
approximation, and with no transverse smearing.  It is independent of
our choice of $\zeta$, and hence it is a function on the two dimensional
space of vectors perpendicular to $\lambda^a$, where two vectors are
identified if they differ by a multiple of $\lambda^a$.  Using
Eq.~(\ref{anec5}) in the limit where ${\tilde  K} \to 1$ and
${\tilde S} \to 1$, and applied to a transversely displaced state,
we obtain
\begin{equation}
I_F({\bf x}_T) = \int {d^2 {\bf \Delta k}_T \over (2 \pi)^2} \
e^{i {\bf \Delta k}_T \cdot {\bf x}_T} \ {\tilde {\cal F}}({\bf \Delta
k}_T),
\label{IFformula}
\end{equation}
i.e., $I_F$ is just the Fourier transform of the function ${\tilde {\cal
F}}$.  Therefore we can rewrite the formula (\ref{anec5}) as
\FL
\begin{eqnarray}
{\bar I}_s^{(1)}(\Lambda_T) &=&
\int {d^2 {\bf \Delta k}_T \over (2 \pi)^2} \ {\tilde
I_F}({\bf
\Delta k}_T) \,
{\tilde K}({\bf \Delta k}_T) \, \nonumber
\\
\mbox{} && \times {\tilde S}(\Lambda_T {\bf \Delta k}_T).
\label{anec6}
\end{eqnarray}
In other words, the ANEC integral with backreaction is just the test
field ANEC integral $I_F({\bf x}_T)$ in Minkowski spacetime convolved
against the smearing
function $S({\bf x}_T/\Lambda_T)$, and against the distribution $K({\bf
x}_T)$ obtained from ${\tilde K}({\bf k}_T)$ by an inverse Fourier
transform.

Note that it follows from the analysis in Appendix \ref{technical}
that the distribution $K({\bf x}_T)$ is given by a
smooth function away from the origin ${\bf x}_T=0$, but not at the
origin.  However, the convolution $(K \circ S)({\bf x}_T)$ is a smooth
function for all ${\bf x}_T$ for our choice of smearing function given
by Eq.~(\ref{smearingfn}) below.
We also remark that Eqs.~(\ref{omegak}), (\ref{regular1}) and
(\ref{calFdef}) imply that the function $|{\bf k}_T|^N {\tilde {\cal
F}}({\bf k}_T)$ is $L^1$ for any $N$, which by Eq. (6.24) shows that
the ``test field ANEC integral'' function $I_F({\bf x}_T)$ is smooth.
This fact will be important in our analysis in Sec.~\ref{long1} below.

\subsubsection{Nonnegativity of the ANEC integral in Minkowski
spacetime}
\label{ANECtestpos}

{}From previous analyses by Klinkhammar \cite{Klinkhammar91} and by Wald
and Yurtsever \cite{WaldYurtsever}, it is known that the test field
ANEC integral $I_F({\bf x}_T)$ is always nonnegative for a large class
of states.  This result forms a key element
in our proof below that the smeared ANEC integral (\ref{anec6}) is
always
nonnegative for suitable choices of the smearing function $S$.  We
give a short proof of the result here, in order to lay the foundations
for later analyses.

First we give a motivational non-rigorous argument, which applies only
to states in the usual Fock space.  Use the
decomposition (\ref{posneg})
of the field operator into positive frequency and negative frequency
parts.  Then we obtain from Eq.~(\ref{stclassical}) that, up to total
derivatives  with respect to $\lambda$,
\begin{equation}
:{\hat T}^{(0)}_{ab}: \lambda^a \lambda^b = ({\hat \Phi}_-^\prime)^2 +
({\hat
\Phi}^\prime_+)^2 + 2 {\hat \Phi}_-^\prime {\hat \Phi}_+^\prime,
\label{stresstensorformal}
\end{equation}
where primes denote derivatives with respect to $\lambda$.  The colons
on the left hand side denote normal ordering.  The first two terms on
the right hand side integrate to zero when we integrate along the
geodesic because this picks out the zero frequency part, and the last
term is a manifestly nonnegative operator.  Hence the ANEC
``operator'' is nonnegative.

We now give a rigorous proof of the positivity of $I_F({\bf x}_T)$; it
follows directly from our general formula (\ref{IFformula}) for $I_F$
and the positivity condition on the two point function expressed in
our coordinates $(\beta,{\bf k}_T)$.  From Eq.~(\ref{gpos}), choosing
$u$ to be a suitable Gaussian and taking an appropriate limit allows
us to deduce that
\begin{equation}
 \int d^2 {\bf k}_T \int d^2 {\bf k}_T^\prime\, {\hat
g}(\beta_0,{\bf k}_T; \beta_0,{\bf k}_T^\prime) \ \ \ge 0,
\label{gpos1}
\end{equation}
for all $\beta_0$.  Therefore, from Eqs.~(\ref{calFdef}) and
(\ref{IFformula}) the quantity $I_F(0)$ is nonnegative.  It is clear
that the same is true for $I_F({\bf x}_T)$.

\subsubsection{The long wavelength limit}
\label{long1}

We now specialize to the long wavelength limit discussed in
Sec.~\ref{scaling1} above.  We also assume that $g^{(1)} \ne 0$, i.e., that
the first order perturbed two-point function is mixed.  First we show that the
unsmeared ANEC integral may be negative, and then show that for
suitable choices of the smearing function $S$, the transversely
smeared ANEC integral is always positive (not merely nonnegative) in
the limit ${\cal L}/L_p \to \infty$.  Note that it would be
inconsistent to analyze the solutions to the reduced order
semiclassical equation (\ref{firstordermixed3}) outside of this limit,
as correction terms of $O(L_p^4 / {\cal L}^4)$ were thrown away in the
derivation of these equations.

If we assume an incoming state of the form (\ref{Finalpha}), then the
test field ANEC integral varies as
\begin{equation}
I_F({\bf x}_T;\alpha) = \alpha^{-3} {\bar I}_F({\bf x}_T/\alpha),
\label{IFalpha}
\end{equation}
where ${\bar I}_F$ is the test field ANEC integral of the state ${\bar
\omega}_{\rm in}^{(1)}$.  Consequently the Fourier transform scales as
${\tilde  I}_F({\bf \Delta k}_T;\alpha) = \alpha^{-1} {\tilde {\bar
I}}_F(\alpha {\bf \Delta k}_T)$.  Substituting this into
Eq.~(\ref{anec6}) and making a change of variable in the integral
yields
\FL
\begin{eqnarray}
{\bar I}_s^{(1)}(\Lambda_T) &=& \alpha^{-3}
\int {d^2 {\bf \Delta k}_T \over (2 \pi)^2} \ {\tilde
{\bar I}_F}({\bf
\Delta k}_T)\,
{\tilde K}\left[ {{\bf \Delta k}_T \over \alpha}\right] \nonumber \\
\mbox{} && \times \,
{\tilde S}\left[{\Lambda_T {\bf \Delta k}_T \over \alpha }\right].
\label{anec7}
\end{eqnarray}
We now choose the smearing function to be
\begin{equation}
S({\bf x}_T) = {1 \over 1 + |{\bf x}_T|^4}.
\label{smearingfn}
\end{equation}
Its Fourier transform has
the expansion for small ${\bf k}_T$
\begin{equation}
{\tilde S}({\bf k}_T) = 1 + \nu_0 {\bf k}_T^2 \ln(|{\bf k}_T|) + \nu_1
{\bf k}_T^2 + O(|{\bf k}_T|^3),
\label{sfnexpand}
\end{equation}
where $\nu_0>0$ \cite{ftnote}.  Substituting Eqs.~(\ref{tildeKdef})
and (\ref{sfnexpand}) into Eq.~(\ref{anec7}) and
expanding in $1/\alpha^2$ yields that
\FL
\begin{equation}
\kappa {\bar I}_s^{(1)}(\Lambda_T) =
  {A \over \alpha^3} + B {\ln \alpha \over
\alpha^5} + {C \over \alpha^5} + O\left[{(\ln \alpha)^2 \over
\alpha^7}\right],
\label{anecexpand0}
\end{equation}
where
\begin{equation}
A = \int_\gamma  \langle T_{ab}^{(0)}, {\bar \omega}_{\rm in}^{(1)} \rangle \,
\lambda^a \lambda^b \ = {\bar I}_F(0),
\label{coeffA}
\end{equation}
\begin{equation}
B = \nu_0 (\Lambda_T^2 - \Lambda_{T,{\rm crit}}^2) ( {\bf \nabla}_T^2
{\bar I}_F)(0),
\label{Bans}
\end{equation}
and
\begin{equation}
\Lambda_{T,{\rm crit}} = 8 \pi \sqrt{a \over \nu_0} L_p.
\label{Lambdacrit}
\end{equation}

Note that the sign of the coefficient $a$ in the formula
(\ref{logscalegeneral}) for the anomalous scaling of the stress energy
determines the sign of the coefficient $B$ without smearing.  As
previously mentioned, $a$ is positive for the scalar field we are
considering, for all values of the curvature coupling
[c.f.,Eq.~(\ref{abdef}) above], and is also
positive for neutrino and Maxwell fields \cite{Horowitz}.  If $a$ had
been negative, the coefficient $B$ would have been positive without
any transverse smearing.

As we have
just discussed, the leading coefficient (\ref{coeffA}) is always
nonnegative.  However, as we now show, there do exist states for which
${\bar I}_F(0)$ vanishes, and this opens up possibilities for violations
of ANEC.  Let $h(\beta,{\bf k}_T)$ be a function on the
positive light cone satisfying
\begin{equation}
\int_0^\infty {d\beta \over \beta} \int d^2 {\bf k}_T \,
|h(\beta,{\bf k}_T)|^2 \ =1,
\label{hnormalization}
\end{equation}
and let $|\sigma\rangle$ be the one particle state
\begin{equation}
|\sigma\rangle =
\int_0^\infty {d\beta \over \beta} \int d^2 {\bf k}_T \,
\sqrt{\omega_{\bf k}} h(\beta,{\bf k}_T) \, {\hat a}_{\bf k}^\dagger {\left| 0
\right>}.
\end{equation}
Choose the the
incoming state perturbation ${\bar \omega}_{\rm in}^{(1)}$ to be that
given by the density matrix perturbation
\begin{equation}
{\bar {\hat \rho}}^{(1)} = -| 0 \rangle \langle 0 | + |\sigma\rangle \langle
\sigma |.
\label{vanisheg}
\end{equation}
Then the function ${\hat g}$ is given by
\begin{equation}
{\hat g}(\beta,{\bf k}_T; \beta^\prime,{\bf k}_T^\prime) \,\propto\,
h(\beta,{\bf k}_T) \, h(\beta^\prime,{\bf k}_T^\prime)^*.
\label{ghatexample}
\end{equation}
If we now choose any smooth function $v({\bf x}_T)$ of compact
support, and choose $h(\beta,{\bf k}_T) = h_1(\beta) {\tilde v}({\bf
k}_T)$ for some suitable $h_1$, then from Eqs.~(\ref{calFdef}),
(\ref{IFformula}) and (\ref{ghatexample}) we find that
\begin{equation}
{\bar I}_F({\bf x}_T) \propto \, | v({\bf x}_T) |^2.
\label{IFexample}
\end{equation}
Hence in particular we can choose a state which achieves ${\bar
I}_F(0)=0$.

Now when $A = {\bar I}_F(0) = 0$, then the quantity $\nabla_T^2 {\bar
I}_F(0)$ is always positive or zero.  This is because ${\bar I}_F({\bf
x}_T) \ge 0$ always, ${\bar I}_F(0) = 0$, and ${\bar I}_F$ is smooth
as discussed above.
It is clear that we can find states for which ${\bar I}_F(0) =0$ but
$\nabla_T^2 {\bar I}_F(0) \ne 0$.
Therefore, we find
that (i) in the limit of no smearing [i.e., $\Lambda_T \to 0$, $S({\bf
x}_T)
\to \delta^2({\bf x}_T)$], $B$ can be negative and therefore for
sufficiently large $\alpha$, the ANEC integral can be negative, and
(ii) From Eq.~(\ref{Bans}), when the transverse smearing length is
larger than the critical length (\ref{Lambdacrit}), $B$ is always
nonnegative.

So far we have shown that the smeared, first order ANEC integral is
always non-negative to order $1/\alpha^2$ beyond leading order.
However,
from Eq.~(\ref{IFexample}) it is clear that there exist incoming
states for which $B=0$; for instance one can choose the function
$v({\bf x}_T)$ to vanish in a neighborhood of the origin.  When
$A=B=0$, the expression for the coefficient $C$ is
\FL
\begin{equation}
C = {1\over 2} \nu_0 (\Lambda_T^2 - \Lambda_{T,{\rm crit}}^2)
\int {d^2 {\bf k} \over (2 \pi)^2} \,\ln[ \Lambda_T^2 {\bf k}_T^2]
\,{\bf k}_T^2 \, {\tilde {\bar I}}_F({\bf k}_T).
\label{Cans}
\end{equation}
We now shall show that when $A=B=0$, for $\Lambda_T > \Lambda_{T,{\rm
crit}}$ we have $C \geq 0$, with equality holding
if and only if $g^{(1)} = 0$.

To prove this result, we note first that if $A=B=0$, then we have $I_F(0) =
\nabla_T^2 I_F(0) =0$.  However,
$I_F$ is a smooth, nonnegative function, and hence it follows that all
the derivatives of $I_F$ at the origin up to and including third order
vanish.  In Appendix \ref{technical}, we show that this implies that
the coefficient $C$ is always strictly positive unless the function
$I_F({\bf x}_T)$ vanishes identically.  We now show that ${\bar
I}_F({\bf x}_T)$ cannot vanish identically unless $g^{(1)} = 0$, i.e.,
unless the perturbed two-point function is pure to first order.  Thus,
in the mixed case, in order to establish positivity of the smeared
ANEC integral in the long wavelength limit for nearly flat spacetimes,
it is not necessary to continue the expansion (\ref{anecexpand0}) to
higher powers in $1/\alpha$, nor is it necessary to go to second order
in $\varepsilon$.

To prove that $I_F({\bf x}_T)$ vanishes identically if and only if
$g^{(1)}=0$, we start by writing
\begin{equation}
I_F({\bf x}_T) = \int_0^\infty d\beta \, I_F({\bf x}_T,\beta)
\end{equation}
where $I_F({\bf x}_T,\beta)$ is defined by Eqs.~(\ref{calFdef}) and
(\ref{IFformula}) with the integral over $\beta$ omitted.  Moreover
Eq.~(\ref{gpos1}) implies that $I_F({\bf x}_T,\beta) \ge 0$ always,
and since $I_F({\bf x}_T) = {\bar I}_F({\bf x}_T) =0$ we deduce that
$I_F({\bf x}_T,\beta)=0$ for all ${\bf x}_T$ and
$\beta$.  Hence
\begin{equation}
\int d^2 {\bf k}_T \, {\hat g}( \beta, {\bf k}_T; \beta, {\bf k}_T +
{\bf {\Delta k}}_T) = 0
\label{gvanishing}
\end{equation}
for all ${\bf \Delta k}_T$.

Next we obtain a canonical representation for the function
${\hat g}(k,k^\prime)$.  Let ${\cal K}$ be the measure space
$({\bf R}^3,d \mu)$, with measure
$d \mu = d \beta d^2 {\bf k}_T$.  Then ${\hat g}$ is a
continuous complex function on ${\cal K} \times {\cal K}$ which
satisfies ${\hat g}(k,k^\prime) = {\hat g}(k^\prime,k)^*$.  Moreover
Eq.~(\ref{regular1}) implies that ${\hat g}$ is both $L^1$ and $L^2$
on ${\cal K} \times {\cal K}$. The $L^2$ property of ${\hat g}$ together
with its positivity property (\ref{gpos1}) implies that it defines
a positive, compact, self adjoint operator ${\hat G}$ on
$L^2({\cal K})$
\cite{ReedSimon}.  Hence by the Hilbert-Schmidt theorem, there is a
complete orthonormal basis $\varphi_n$ of $L^2({\cal K})$ such that
\begin{equation}
{\hat g}(k,k^\prime) = \sum_{n=0}^\infty \, \lambda_n
\varphi_n(k) \varphi_n(k^\prime)^*,
\label{gdecompos}
\end{equation}
for some $\lambda_n \ge 0$, where the convergence is in the operator
norm on the space of bounded operators on $L^2({\cal K})$.
Now inserting the decomposition (\ref{gdecompos}) into Eq.~(\ref{gvanishing})
and
specializing to ${\bf \Delta k}_T=0$ we find that
\begin{equation}
\sum_{n=0}^\infty \, \lambda_n \int d^2 {\bf k}_T \left|
\varphi_n(\beta,{\bf k}_T) \right|^2 =0.
\end{equation}
Since $\lambda_n \ge 0$ for all $n$, we obtain that ${\hat g}=0$.

By combining all the results in this subsection we find that for
states whose perturbed two-point function is mixed,
the transversely smeared ANEC integral for $\Lambda_T >
\Lambda_{\rm T,crit}$ will always be strictly positive for
sufficiently large $\alpha$, for solutions of the reduced order, first
order semiclassical equation (\ref{firstordermixed3}).

\section{Pure incoming states and the second order perturbation
equations}
\label{secondorderanalysis}

The above results establish that for mixed incoming states, the
leading order contribution to the smeared ANEC integral is always
positive.  However, for pure incoming states the ANEC integral
vanishes to $O(\varepsilon)$, and therefore we need to investigate the
second order perturbation equation (\ref{secondordermixed3}).  In this
section we calculate the $O(\varepsilon^2)$ contribution
(\ref{secondorderanec2}) to the transversely smeared ANEC integral for
solutions to Eq.~(\ref{secondordermixed3}), and show that it is
positive in the long wavelength limit.

To calculate the second order contribution to the ANEC integral, we
shall need the second order contribution (\ref{Gsdef})
to the Einstein tensor.  Defining ${\bar G}^{(2)}_{ab}(x) = \alpha^4
G^{(2)}_{ab}(x/\alpha)$ we find from Eq.~(\ref{firstordermixed3}) that
\FL
\begin{eqnarray}
{\bar G}_{ab}^{(2)}&=&
\langle T^{(0)}_{ab} , {\bar \omega}_{\rm in}^{(1)} \rangle
+ {\ln \alpha \over \alpha^2}  Z_{ab}^{(1)}[\chi^{(2,0)}]
 \nonumber \\
\mbox{} && + {1\over \alpha^2} \left\{  \langle
T^{(1)}_{ab}[\chi^{(2,0)}],\omega_{\rm in,0} \rangle + \langle
T_{ab}^{(1)}[\chi^{(1,0)}],{\bar \omega}_{\rm in}^{(1)} \rangle \right\}
\nonumber \\
\mbox{} &&+  O[(\ln \alpha)^2/\alpha^4].
\label{secondorderIII}
\end{eqnarray}
Here we have used the definitions
\begin{equation}
h^{(j)}(x;\alpha) = {\hat h}^{(j)}(x/\alpha;\alpha) = \alpha^{-2}
\chi^{(j)}(x/\alpha;\alpha)
\label{hhatdefs}
\end{equation}
for $j = 1,2$, and have used the fact that $\chi^{(1)}$ can be replaced
with its leading order approximation $\chi^{(1,0)}$ to adequate
accuracy, c.f., Eq.~(\ref{chiexpand}) above.  The leading order metric
perturbations $\chi^{(1,0)}$ and $\chi^{(2,0)}$ are given by
Eqs.~(\ref{chi10eqn}) and (\ref{chi20eqn}).
Finally, we insert the expansion (\ref{secondorderIII}) into
Eq.~(\ref{secondorderanec2}) to obtain
\FL
\begin{eqnarray}
I_s^{(2)} &=& {1 \over \kappa \alpha^3} \int_{M} \epsilon^{(0)}_{cdef} \,
\Theta_\gamma^{(0)}( {\Lambda\over\alpha},
{\Lambda_T\over\alpha}, {\Lambda_L\over\alpha}) \lambda^a \lambda^b
\nonumber \\
\mbox{} && \times \bigg\{
\langle T_{ab}^{(0)}, {\bar \omega}_{\rm in}^{(2)} \rangle
+ {\ln \alpha \over \alpha^2}  Z_{ab}^{(1)}[\chi^{(2,0)}]
 \nonumber \\
\mbox{} && + {1\over \alpha^2} \langle
T_{ab}^{(1)}[\chi^{(2,0)}],\omega_{\rm in,0} \rangle
+ {1\over \alpha^2}
\langle T_{ab}^{(1)}[\chi^{(1,0)}],{\bar \omega}_{\rm in}^{(1)} \rangle
\nonumber \\
\mbox{} &&+ {1 \over 2 \alpha^2} \chi^{(1,0)\,c}_{\,\,\,\,\,\,c}
\langle T_{ab}^{(0)}, {\bar \omega}_{\rm in}^{(1)} \rangle
+ O[(\ln \alpha)^2/\alpha^4] \bigg\}.
\label{Igamma2b}
\end{eqnarray}

We now exploit the close similarity between the first and second order
perturbation equations.  For the rescaled incoming state
(\ref{Finalpha}), the first order, reduced order equation
(\ref{firstordermixed3}) can be alternatively written in a form more
closely parallel with Eq.~(\ref{secondordermixed3}):
\FL
\begin{eqnarray}
\kappa G^{(1)}_{ab}[{\hat h}^{(1)}] &=& {1 \over \alpha^2} \langle
T^{(0)}_{ab},
{\bar \omega}^{(1)}_{\rm in} \rangle + {\ln \alpha \over \alpha^4}
Z^{(1)}_{ab}[\chi^{(1,0)}] \nonumber \\
\mbox{} && + {1 \over \alpha^4} \langle
T^{(1)}_{ab}[\chi^{(1,0)}],\omega_{\rm in,0} \rangle
+ O\left( {(\ln \alpha)^2 \over \alpha^6} \right).
\label{firstordermixed4}
\end{eqnarray}
This equation when inserted into Eq.~(\ref{firstorderanec}) will
produce the expansion (\ref{anecexpand0}).  Therefore it can be seen
that the first three terms in Eq.~(\ref{Igamma2b}) are exactly
analogous to those obtained from the first order analysis, and that
consequently
Eq.~(\ref{Igamma2b}) can be rewritten as
\FL
\begin{eqnarray}
\kappa {\bar I}_s^{(2)}(\Lambda_T) &\equiv&
\lim_{\Lambda \to \infty}  \ \lim_{\Lambda_L \to 0} \,
\kappa I_s^{(2)}(\Lambda, \Lambda_L, \Lambda_T) \nonumber \\
&=&  {{\hat A} \over \alpha^3} + {\hat B} {\ln \alpha \over
\alpha^5} + {{\hat C} + \Delta C \over \alpha^5} + O\left[{(\ln
\alpha)^2 \over
\alpha^7}\right],
\label{anecexpand1}
\end{eqnarray}
in analogy with Eq.~(\ref{anecexpand0}).  Here the coefficients ${\hat
A}$, ${\hat B}$ and ${\hat C}$ are functions only of the state
perturbation ${\bar \omega}_{\rm in}^{(2)}$, and have the exact same
functional dependence on ${\bar \omega}_{\rm in}^{(2)}$ as the
coefficients $A$, $B$, and $C$ in Eq.~(\ref{anecexpand0}) have on
${\bar \omega}_{\rm in}^{(1)}$.  Furthermore, the relevant positivity
conditions on
${\bar \omega}_{\rm in}^{(2)}$ also are the same as corresponding
conditions on ${\bar \omega}_{\rm in}^{(1)}$; see subsection
\ref{secondorderpos} below.  The coefficient $\Delta C$ in
Eq.~(\ref{anecexpand1}) is defined to the contribution from the last
two terms in Eq.~(\ref{Igamma2b}), and depends both on ${\bar
\omega}_{\rm in}^{(1)}$ and ${\bar \omega}_{\rm in}^{(2)}$ as well as the
freely specifiable incoming piece of the metric perturbation
$\chi_{ab}^{(1,0)}$.

Our strategy for proving the positivity of the second order ANEC
integral is the following.  First, in subsection \ref{vanishing} we
show that $\Delta C =0$ whenever ${\hat A} = {\hat B} =0$.  In
subsection \ref{secondorderpos}, we show that the space of allowed
second order state perturbations is effectively the same as that of
the first order perturbations, and then to appeal to the first order
analysis.

\subsection{Vanishing of the additional terms}
\label{vanishing}

We now show that whenever ${\hat A} = {\hat B} = 0$, the last two
terms in Eq.~(\ref{Igamma2b}) vanish, and that consequently $\Delta
C=0$.  The arguments in this subsection will be mostly formal; we
believe that these formal arguments could be translated into rigorous
arguments along the lines of the analysis given in
Sec.~\ref{ANECtestpos} above, and using the positivity condition.
However, we have not attempted to do so.

It is not difficult to see that the last term in
Eq.~(\ref{Igamma2b}) vanishes.  Since to the
appropriate order in $1/\alpha^2$ the transverse smearing is
unimportant for this term, after the limit $\Lambda \to \infty$ is
taken it can be written as
\begin{equation}
{1 \over 2 \alpha^2} \int_\gamma d\lambda \,
\chi^{(1,0)\,c}_{\,\,\,\,\,\,c}
\langle T_{ab}^{(0)} , {\bar \omega}_{\rm in}^{(1)} \rangle \lambda^a
\lambda^b.
\end{equation}
However, our choice of gauge guarantees that $\chi^{(1)}_{ab}=0$ along
$\gamma$.  In particular this will be true to each order in the
$1/\alpha^2$ expansion of Eq.~(\ref{chiexpand}), so that
$\chi_{ab}^{(1,0)}$ also vanishes on $\gamma$, and therefore this term
vanishes.

The more interesting term in Eq.~(\ref{Igamma2b}) is the second to
last term, which can
be shown to be proportional (at the relevant order in $1/\alpha^2$) to
\FL
\begin{eqnarray}
&& \int_\gamma d\lambda \,
\langle T_{ab}^{(1)}[\chi^{(1,0)}],{\bar \omega}_{\rm in}^{(1)} \rangle
\lambda^a \lambda^b, =
\int_\gamma d\lambda \,
{\cal D}_{ab}^{(1)}[\chi^{(1,0)}] F_{\rm in}^{(1)} \lambda^a \lambda^b
\nonumber
\\
\mbox{} && + \int_\gamma d\lambda \,
{\cal D}^{(0)}_{ab} \, {\cal E}\left[ - {\cal D}_x^{(1)}[\chi^{(1,0)}] F_{\rm
in}^{(1)}, - {\cal D}_y^{(1)}[\chi^{(1,0)}] F_{\rm in}^{(1)}\right] \lambda^a
\lambda^b,
\label{hardterm}
\end{eqnarray}
where we have used Eq.~(\ref{hardterm1}).  The meaning of the notation
in the first term is that the function $F_{\rm in}^{(1)}$ is acted on by
the first order change in the operator ${\cal D}_{ab}$ induced by the
metric perturbation $\chi^{(1,0)}$, and similarly for the second term.
Now since $\chi^{(1,0)}_{ab}$ satisfies Eq.~(\ref{chi10eqn}), it has a
contribution both from the perturbation to
the incoming state and from incoming classical gravitational
radiation.  Thus, Eq.~(\ref{hardterm}) contains both a contribution
quadratic in the incoming state perturbation ${\bar \omega}_{\rm
in}^{(1)}$ and a cross term between ${\bar \omega}_{\rm in}^{(1)}$
and the incoming classical gravitational radiation.

Consider first the second term in Eq.~(\ref{hardterm}).  As mentioned
above we are assuming that the coefficients ${\hat A}$ and ${\hat B}$
vanish, since from Eq.~(\ref{anecexpand1}) this is the only case in
which the terms (\ref{hardterm}) are relevant.  Therefore
\begin{equation}
{\hat A}  = \int_\gamma
\langle T_{ab}^{(0)} , {\bar \omega}_{\rm in}^{(2)} \rangle
\lambda^a \lambda^b \ \ =0.
\label{vanish3}
\end{equation}
The key idea that we now use is that this condition imposes constraints
on
${\bar \omega}_{\rm in}^{(2)}$, which in turn imposes constraints on
${\bar \omega}_{\rm in}^{(1)}$ sufficient to ensure that the second term
in Eq.~(\ref{hardterm}) vanishes.

We now restrict attention to the case of Fock space states.  Since the
state is pure to first order, the first order perturbation to the
density matrix is given by
\begin{equation}
{\bar {\hat \rho}}^{(1)} = {\left| 0 \right>} {\left< \psi \right|}^{(1)} +
{}^{(1)} {\left| \psi \right>} {\left< 0 \right|},
\label{rho1pure2}
\end{equation}
for some ${\left| \psi \right>}^{(1)} \in {\cal H}$ with $\left< 0 | \psi
\right>^{(1)}=0$.
The most general second order density matrix perturbation is then
\FL
\begin{eqnarray}
{\bar {\hat \rho}}^{(2)} &=& \left| \psi \right>^{(1)} {}^{(1)}\left< \psi
\right|
+ {\left| 0 \right>} {\left< \psi \right|}^{(2)} + {}^{(2)} {\left| \psi
\right>} {\left< 0 \right|} \nonumber \\
\mbox{} && + {\hat Q} - (1 + {\rm
tr} {\hat Q} ) {\left| 0 \right>} {\left< 0 \right|},
\label{rho2mixed2}
\end{eqnarray}
for some ${\left| \psi \right>}^{(2)} \in {\cal H}$ with $\left< 0 | \psi
\right>^{(2)}=0$,
where ${\hat Q}$ is a positive, Hermitian trace class operator such
that ${\hat Q} {\left| 0 \right>} = 0$.  Since ${\hat Q}$ is trace class there
will
exist an orthonormal basis $\left|\psi_j\right>$ of the space of
states orthogonal to ${\left| 0 \right>}$ such that
\begin{equation}
{\hat Q} = \sum_{j=0}^\infty q_j \left|\psi_j\right> \left<
\psi_j\right|
\end{equation}
for some $q_j\ge0$, $j=0,1,2 \ldots$.  The overall density matrix
will be positive to $O(\varepsilon^2)$ whenever the perpendicular
projections of $\left| \psi \right>^{(1)}$ and $\left| \psi
\right>^{(2)}$ into the kernel of ${\hat Q}$ are orthogonal to each
other.

We now give a non rigorous argument for the vanishing of
the second term in Eq.~(\ref{hardterm}).  If we insert
Eq.~(\ref{rho2mixed2}) into
Eq.~(\ref{vanish3}), then the contribution from the two terms in
Eq.~(\ref{rho2mixed2})
involving $\left|\psi\right>^{(2)}$ will
vanish for the reason explained in Sec.~\ref{puresec}.
Therefore,
using the formula (\ref{stresstensorformal}) and following the
argument given in
Sec.~\ref{ANECtestpos}, we find that
\begin{eqnarray}
&& \sum_j q_j
\int d\lambda \,\, \left< \psi_j \bigg| \left( {d {\hat \Phi}_+
\over d \lambda}\right)^\dagger \left( {d {\hat \Phi}_+
\over d \lambda}\right) \bigg| \psi_j \right> \nonumber \\
\mbox{} && +
\int d\lambda \,\, \left< {}^{(1)}\psi \bigg| \left( {d {\hat \Phi}_+
\over d \lambda}\right)^\dagger \left( {d {\hat \Phi}_+
\over d \lambda}\right) \bigg| \psi^{(1)} \right> =0.
\end{eqnarray}
{}From the arguments given in Sec.~\ref{ANECtestpos}, it is clear that
all the terms on the left hand side are individually nonnegative, and
hence they all vanish.  It follows that
\begin{equation}
{\hat \Phi}_{+}^\prime(\lambda,0,0) \left|\psi\right>^{(1)}=0,
\label{vanishes3}
\end{equation}
where primes denote derivatives
with respect to $\lambda$ and the notation means that the operator is
evaluated on the geodesic $\gamma$.
Next, using Eq.~(\ref{rho1pure2}), we can rewrite the second term in
Eq.~(\ref{hardterm}) in an alternative notation as
\begin{equation}
2 {\rm Re}\, \int_\gamma d\lambda \,\,  \lambda^a \lambda^b
\left< 0 \big|  \nabla_{(a} {\hat \Phi}^{(0)} \nabla_{b)} {\hat \Phi}^{(1)}
\big| \psi\right>^{(1)},
\label{hardterm3}
\end{equation}
Here ${\hat \Phi}^{(1)}$ is the first order change to the field operator
induced by the metric perturbation $\chi_{ab}^{(1,0)}$. Next,
split up the zeroth order operator ${\hat \Phi}^{(0)}$ into its positive
and negative frequency parts.  The negative frequency parts will
annihilate the vacuum on the left, and the positive frequency part can
be commuted through the ${\hat \Phi}^{(1)}$ term (since the states
${\left| 0 \right>}$
and ${\left| \psi \right>}^{(1)}$ are orthogonal), giving a result which
vanishes by the
condition (\ref{vanishes3}).

Consider now the first term in Eq.~(\ref{hardterm}).  To calculate
this we need the explicit form of the operator ${\cal
D}_{ab}$ that enters into the point splitting prescription
(\ref{Tabformal}) for calculating the stress tensor.  The general
expression for ${\cal D}_{ab}$ is given by [c.f.,
Eq.~(\ref{stclassical}) above]
\FL
\begin{eqnarray}
{\cal D}_{ab} F_{\rm in}^{(1)}(x,x') &=& S_{ab}
-{1\over2} g_{ab} g^{cd} S_{cd} + \xi G_{ab} S
\nonumber \\
\mbox{} && - \xi ( \nabla_a \nabla_b S - g_{ab} \Box S),
\label{calDdef}
\end{eqnarray}
where
\begin{equation}
S(x) = \lim_{x'\to x} F_{\rm in}^{(1)}(x,x'),
\label{Sdef}
\end{equation}
\FL
\begin{eqnarray}
S_{ab}(x) &=& \lim_{x'\to x} \nabla_{a'} \nabla_b F_{\rm in}^{(1)}(x,x')
\nonumber \\
&=& -
\lim_{x'\to x} \nabla_{a} \nabla_b F_{\rm in}^{(1)}(x,x') + \nabla_a
\nabla_b S/2,
\label{Sabdef}
\end{eqnarray}
and $G_{ab}$ is the Einstein tensor.  [Note that we are using a
nonstandard notation in which the coincidence limit is implicitly
understood in the symbol ${\cal D}_{ab}$.]  Now the operator ${\cal
D}_{ab}^{(1)}[\chi^{(1,0)}]$ in the expansion (\ref{expandall}) will
contain pieces linear in $\chi^{(1,0)}_{ab}$ and pieces that are
linear in the derivative of $\chi^{(1,0)}_{ab}$.  The pieces that are
linear in $\chi^{(1,0)}_{ab}$ will give a vanishing contribution to
the first term in Eq.~(\ref{hardterm}) since $\chi^{(1,0)}_{ab}$
vanishes along $\gamma$ in our choice of gauge.  The remaining piece
of ${\cal D}_{ab}^{(1)}$ yields
\FL
\begin{eqnarray}
\int_\gamma d\lambda \,  &&
{\cal D}_{ab}^{(1)}[\chi^{(1,0)}] F_{\rm in}^{(1)} \lambda^a \lambda^b = - \xi
\int_\gamma d\lambda \, (\nabla_c S) C^{(1) \, c}_{ab} \lambda^a
\lambda^b \nonumber \\
\mbox{} && +
\xi \int_\gamma d\lambda \, G_{ab}^{(1)}[\chi^{(1,0)}] \, S \, \lambda^a
\lambda^b,
\label{hardterm2}
\end{eqnarray}
where $C^{(1)\,c}_{ab}$ is given by Eq.~(\ref{C1abc}) with $h_{ab}^{(1)}$
replaced by $\chi^{(1,0)}_{ab}$.
The first term in Eq.~(\ref{hardterm2}) can be seen to vanish using
Eq.~(\ref{C1abc}) and the
fact that $h^{(1)}_{ab}=0$ on $\gamma$ and that $ h^{(1)}_{ab}
\lambda^a \lambda^b$ vanishes identically in our choice of gauge.
Using Eqs.~(\ref{chi10eqn}), (\ref{stresstensorformal}) and
(\ref{rho1pure2}), we find that the second term in
Eq.~(\ref{hardterm2}) contains the factor
\FL
\begin{equation}
G_{ab}^{(1)}[\chi^{(1,0)}] \lambda^a \lambda^b =
(2/\kappa)\, {\rm Re}  \,\, \left< 0 \big| {\hat
\Phi}_{+}^\prime(\lambda,0,0)^2
\big| \psi\right>^{(1)},
\end{equation}
which vanishes by Eq.~(\ref{vanishes3}).  Therefore
the expression (\ref{hardterm}) should vanish.

\subsection{Smeared positivity result}
\label{secondorderpos}

We now explain how to adapt the perturbative smeared positivity result
of Sec.~\ref{firstorderexact} to the present situation.  The
coefficients $A$, $B$ and $C$ in Sec.~\ref{firstorderexact} were
expressed in terms of the function $I_F({\bf x}_T)$ (the flat
spacetime ANEC integral along transversely displaced geodesics in the
state ${\bar \omega}_{\rm in}^{(1)}$), and
the only properties of this function necessary to prove the result was
that it was a pointwise positive smooth function.  Correspondingly the
coefficients ${\hat A}$, ${\hat B}$ and ${\hat C}$ can be expressed in
terms of the analogous function
\begin{equation}
I_F^{(2)}({\bf x}_T) \equiv \int_{\gamma^\prime}
\langle T_{ab}^{(0)} , {\bar \omega}_{\rm in}^{(2)} \rangle
\lambda^a \lambda^b.
\label{IF2def}
\end{equation}
This quantity can be expressed in terms of the mixed frequency part
$g^{(2)}$ of the second order two point function by equations analogous to
Eqs.~(\ref{calFdef}) and (\ref{IFformula}).  However, from
Eq.~(\ref{poscondt2b}) we find that $g^{(2)}$ obeys a stronger positivity
condition than $g^{(1)}$ in the case we are considering
when $g^{(1)}=0$.  In particular, $g^{(2)}$ satisfies an analog of
Eq.~(\ref{gpos}) and it  follows that the function (\ref{IF2def}) must
be 
nonnegative.

Hence the coefficients ${\hat A}$, ${\hat B}$ and ${\hat C}$ obey the
same positivity conditions as the coefficients $A$, $B$ and $C$:
${\hat A} \ge 0$, ${\hat B} \ge 0$ whenever ${\hat A} = 0$, and ${\hat
C} \ge 0$ whenever ${\hat A} = {\hat B} = 0$.  Moreover, the case
${\hat A} = {\hat B} = {\hat C} =0$ can be excluded in the following
way.  If all these coefficients vanish, an argument similar to that
given in Sec.~\ref{long1} shows that the operator ${\hat Q}$ in
Eq.~(\ref{rho2mixed2}) must vanish, and that consequently ${\left|
\psi \right>} =
{\bar {\hat \rho}}^{(1)} =0$.  Therefore, by defining $\varepsilon^\prime
= \varepsilon^2$ we see that we are really dealing with a first order
perturbation instead of a second order perturbation.

Since we showed above that $\Delta C=0$ whenever ${\hat A} = {\hat B}
=0$, we conclude from Eq.~(\ref{anecexpand1}) that for sufficiently
large $\alpha$, for pure states the transversely smeared ANEC integral
is positive.

\subsection{Second order vacuum polarization}

As we have explained, the second order vacuum polarization term
$\langle T_{ab}^{(2)}[h^{(1)},h^{(1)}], \omega_{\rm in,0} \rangle$ is not
explicitly known, but
arises at sufficiently high order in our long-wavelength expansion
that it can be neglected, when we assume that incoming gravitational
radiation does not dominate the first order metric perturbation.
However, if we drop the assumption on incoming gravitational
radiation, and assume in addition an incoming vacuum state, then we
can derive a condition that this unknown vacuum term must satisfy in
order for ANEC not to be violated.  This condition provides an
additional test of the ANEC hypothesis which is independent of our
analysis above.  Moreover, if the condition is satisfied, then it can
be shown that without any assumptions restricting the incoming
classical gravitational radiation, that the transversely smeared ANEC
integral is always nonnegative for solutions of the reduced order
equations, for general, nonvacuum incoming states, in the long
wavelength limit.

The condition we find, by carrying out a reduction of order to the
appropriate order of the perturbative semiclassical equations, is the
following.  Let $h_{ab}^{(1)}$ be any solution of $G_{ab}^{(1)}[h^{(1)}]=0$,
and
let $h_{ab}^{(2)}$ be any solution of $G^{(1)}_{ab}[h^{(2)}] +
G^{(2)}_{ab}[h^{(1)},h^{(1)}]=0$.  Thus, the spacetime $(M,\eta_{ab} +
\varepsilon
h_{ab}^{(1)} + \varepsilon^2 h_{ab}^{(2)})$ satisfies the vacuum Einstein
equation to second order and consists of classical gravitational
waves.  Then the quantity
\begin{equation}
\langle T_{ab}^{(2)}[h^{(1)},h^{(1)}],\omega_{\rm in,0}\rangle + \langle
T_{ab}^{(1)}[h^{(2)}],\omega_{\rm in,0} \rangle
\label{conj}
\end{equation}
describes the expected in-vacuum stress-energy tensor of the quantum field to
second order on this spacetime.  Moreover, the stress tensor
(\ref{conj}) does not depend on which solution $h^{(2)}$ of
$G_{ab}^{(1)}[h^{(2)}]
+ G_{ab}^{(2)}[h^{(1)},h^{(1)}]=0$ is chosen, and thus is a function only of
$h^{(1)}$.
The condition is that the ANEC integral of the quantity (\ref{conj})
should always be nonnegative.  We conjecture that this is the case.

\section{Conclusions}

In this section we recap briefly our main assumptions and assess the
significance of our results.  We have examined the positivity of
(transversely smeared) ANEC integrals for solutions of the reduced
order semiclassical Einstein equation.  Three small parameters have
appeared in our analysis: $\varepsilon$, measuring the deviation of
the metric from the flat metric and of the quantum state from the
incoming vacuum; $\hbar$ or equivalently $L_p^2 / {\cal L}^2$, our
``long wavelength'' expansion parameter, and finally $1/N$, where $N$
is the number of scalar fields coupled to gravity.  We have calculated
the ANEC integral to leading order in $1/N$, to quadratic order in
$\varepsilon$, and to the first three nonvanishing terms in $L_p^2 /
{\cal L}^2$.  We restricted our analysis to the case where incoming
gravitational radiation does not dominate the first order metric
perturbation.  Apart from this restriction, we have shown that the
transversely smeared ANEC integral for nearly flat spacetimes is always
strictly positive along every null geodesic in the long wavelength
limit, except in the trivial case of the vacuum state in
Minkowski spacetime, where the ANEC integral vanishes identically.

There were several independent places in our analysis in which, {\it a
priori}, a violation of ANEC could easily have arisen.  In particular,
the ANEC integral for pure states need not have vanished at first
order in $\varepsilon$; the coefficient (\ref{Cans}) need not have
been of a defenite sign; or any one of the extra terms that appear at
second order in $\varepsilon$ in the coefficient $\Delta C$ need not
have been identically vanishing.  Indeed, several times during the
course of this work we believed to have discovered a serious violation
of ANEC, only to find on more careful analysis that this was not the
case.  Therefore, we consider our results to be evidence in favor of
the conjecture that ANEC comes sufficiently close to holding in
general solutions of the semiclassical equations to rule out
macroscopic traversable wormholes.

\nonum
\section{ACKNOWLEDGMENTS}
We are grateful to Bob Geroch for suggesting the
argument used in Appendix \ref{exactanec}, and for bringing to our
attention the relevance of the heat equation (\ref{heat}) to the
discussion in Sec.~\ref{proposal1}.  We also thank Matt Visser for
some useful conversations.  This research was supported by 
NSF grants PHY 92-20644 and PHY 95-14726 to the University of Chicago,
and by an Enrico Fermi fellowship to \'E. Flanagan.

\appendix{Exact solutions of first order perturbation equations}
\label{firstordersolutions}

In this appendix we derive all of the exact solutions to the first
order semiclassical equation (\ref{firstorderexplicit}) whose spatial
Fourier transforms exist.  Some of these exact solutions have
been discussed by Horowitz \cite{Horowitz}, in the special
case of the homogeneous version of the equation, without the source
term (\ref{source}).  Here we generalize the treatment of Horowitz to
allow for first order perturbations to the quantum state.  To solve
Eq.~(\ref{firstorderexplicit}) we can fix the incoming state
perturbation $\omega_{\rm in}^{(1)}$ (this is freely specifiable up
to some regularity conditions discussed in Appendix \ref{FTstress}),
and solve for the metric perturbation $h_{ab}^{(1)}$.  Because the
equation depends on $h^{(1)}$ only through its linearized Einstein tensor
$G^{(1)}$, we can first solve for $G^{(1)}$ and then use this to obtain
$h^{(1)}$, as suggested by Horowitz \cite{Horowitz}.

The exact solutions to Eq.~(\ref{firstorderexplicit}) are closely
analogous to the solutions of the Klein-Gordon equation with negative
mass squared
\begin{equation}
(\Box - m^2 )\Phi(x) = \rho(x),
\label{simpleEq}
\end{equation}
where $\rho(x) = \rho({\bf x},t)$ is a source.  We start by recalling
the nature of the solutions of Eq.~(\ref{simpleEq}).  The general
solution
can be written in terms of the spatial Fourier transform ${\tilde
\rho}({\bf k},t)$ of the source as $\Phi = \Phi_< + \Phi_>$, where
\begin{eqnarray}
\Phi_>({\bf x},t) &=&
 \int_{|{\bf k}|>m} {d^3{\bf k} \over (2 \pi)^3}\ e^{i {\bf k} \cdot
{\bf x}} \, \bigg[ A({\bf k}) e^{i \omega_{\bf k} t} + B({\bf k}) e^{-i
\omega_{\bf k} t} \nonumber \\
\mbox{} && + \int_{-\infty}^\infty dt^\prime \ G_{\rm sym,osc}(t-t^\prime;
\omega_{\bf
k}) {\tilde \rho}({\bf k},t^\prime) \, \bigg],
\label{tachyonsoln}
\end{eqnarray}
and
\begin{eqnarray}
\Phi_<({\bf x},t) &=&
 \int_{|{\bf k}|<m} {d^3{\bf k} \over (2 \pi)^3} \ e^{i {\bf k} \cdot
{\bf x}} \, \bigg[ C({\bf k}) e^{ \kappa_{\bf k} t} + D({\bf k}) e^{-
\kappa_{\bf k} t} \nonumber \\
\mbox{} && + \int_{-\infty}^\infty dt^\prime \ G_{\rm sym,exp}(t-
t^\prime;
\kappa_{\bf k}) {\tilde \rho}({\bf k},t^\prime) \, \bigg].
\label{expsoln}
\end{eqnarray}
Here $\omega_{\bf k} = \sqrt{{\bf k}^2 - m^2}$, $\kappa_{\bf k} =
\sqrt{m^2
- {\bf k}^2}$,
$
G_{\rm sym,osc}(t;\omega) = \sin (\omega |t|) \, / (2 \omega),
$
$
G_{\rm sym,exp}(t;\kappa) = - e^{- \kappa |t|}/(2 \kappa),
$,
and the functions $A({\bf k}), \ldots, D({\bf k})$ are arbitrary
except for the reality conditions $A(-{\bf k}) = A({\bf k})^*$ etc.
Following Horowitz, we will refer to the portion
(\ref{tachyonsoln}) of the solution as the tachyonlike or oscillatory
part, and the portion (\ref{expsoln}) as the exponential part.

We explicitly display these solutions to the negative mass squared Klein
Gordon equation because the solutions to the semiclassical equation
(\ref{firstorderexplicit}) have a very similar character.  In particular
these solutions can be divided into ``oscillatory'' and
``exponential'' pieces.  We will obtain the general solutions by
spacetime Fourier methods.  As background, we start by recalling how
to obtain the solution (\ref{tachyonsoln}) -- (\ref{expsoln}) of
Eq.~(\ref{simpleEq}) by spacetime Fourier transforms (as opposed to
merely spatial Fourier transforms).

It is clear that the ``tachyonlike'' portion $\Phi_>$ of the general
solution can be straightforwardly obtained using Fourier transforms,
when the source $\rho({\bf
x},t)$ is sufficiently well behaved.  The Fourier transform with
respect to time of the Greens function obtained from
Eq.~(\ref{simpleEq}) has two poles on the real $\omega$ axis, and the
choice of ``$i \epsilon$'' regularization prescription is equivalent
to the choice of Greens function; for instance, as is well known,
demanding that ${\tilde G}(\omega)$ be analytic in the upper half
plane yields the retarded Greens function.  The freely specifiable
first term in Eq.~(\ref{tachyonsoln}) can clearly be written down by
inspection using the location of the poles of the Greens function.

The situation is slightly different for the exponentially growing
portion (\ref{expsoln}) of the solution.  In this case the Greens
function in the Fourier domain
\begin{equation}
{\tilde G}(\omega) \propto 1 / (\omega^2 + \kappa^2)
\end{equation}
has no poles on the real axis, and hence Fourier transform methods
produce a unique Greens function in the time domain, which is just the
particular Greens function $G_{\rm sym,exp}(t)$ that we choose to use in
Eq.~(\ref{expsoln}).  Of course, other Greens functions appropriate to
different boundary conditions do exist in the time domain, but $G_{\rm
sym,exp}(t)$ is the only one whose Fourier transform exists.
Hence, solving
Eq.~(\ref{simpleEq}) using Fourier transform methods will reproduce
the last term in Eq.~(\ref{expsoln}).  The freely specifiable first
two terms in Eq.~(\ref{expsoln}) are not directly obtained, but
clearly again can be written down by inspection, using the location of
the poles in ${\tilde G}(\omega)$.

We now turn to a similar analysis of the semiclassical equation
Eq.~(\ref{firstorderexplicit}).  We
can obtain very general solutions to the equation by determining the
analytic structure in $\omega$ of the appropriate Greens function in
the Fourier domain.  See Sec.~\ref{solns1} and Appendix \ref{FTstress}
for a discussion of the existence of the Fourier transform of the
source term (\ref{source}).
Using Eqs.~(\ref{firstorderexplicit}), (\ref{ABdef}) and
(\ref{tildeHlambda}) we find the
following formal expression for the general Fourier transformable
solution
\FL
\begin{eqnarray}
\kappa {\tilde G}_{ab}^{(1)}(k) &=& { {\tilde s}_{ab} \over F_1(k)}
\nonumber \\
\mbox{} && + {F_2(k) \over F_1(k) F_3(k) } (k_a k_b - \eta_{ab} k^2)
{\tilde s}_c^{\ c} ,
\label{sol1}
\end{eqnarray}
where
\begin{equation}
F_1(k) = 1 - 16 \pi a L_p^2 k^2 {\tilde H}_\lambda(k),
\label{F1def}
\end{equation}
\begin{equation}
F_2(k) = (8 \pi) \left[({2\over3}a + 2 b) L_p^2 {\tilde H}_{\lambda}(k)
+
2 \beta L_p^2 \right],
\end{equation}
and
\begin{equation}
F_3(k) = 1 + 6 L_p^2 k^2 [\beta + b {\tilde H}_{\lambda}(k)].
\label{F3def}
\end{equation}

While Eq.~(\ref{sol1}) is not the general solution we are looking for,
it is straightforward to write down the general solution by
inspection, essentially by adding extra terms to Eq.~(\ref{sol1}) that
correspond to poles of the functions $1/F_1$ and $1/(F_1 F_3)$.
Consider first the function $1/F_1$.
{}From Eqs.~(\ref{tildeHlambda}) and (\ref{F1def}) we can write $F_1$ as
the limit of a function analytic on the upper half $\omega$ plane: $F_1(k)
= \lim_{\epsilon \to 0^+} G_1({\bf k},\omega + i
\epsilon)$, where
\begin{equation}
G_1({\bf k},\omega) = 1 + \left[ {\omega_{\bf k}^2 - \omega^2  \over
\omega_c^2} \right]
\, \ln \left[ {\hat \lambda}^2 (\omega_{\bf k}^2 - \omega^2)\right],
\label{G1}
\end{equation}
$\omega_{\bf k}= |{\bf k}|$, $\omega_c^2 = 1/ (32 \pi^2 a L_p^2)$,
${\hat \lambda }=\lambda \exp(\gamma - 1/2)$, and $\gamma$ is Euler's
constant.  The function $G_1$ has
branch cuts on the real axis at $\omega > \omega_{\bf k}$ and $\omega <
-\omega_{\bf k}$.  The poles in the Greens function are just the zeros
of
$G_1$.  These location of these zeros depend on $\lambda$ and
$\omega_{\bf k}$ in the following way (see Fig.~\ref{poles}).  Define
\begin{equation}
\lambda_{\rm crit} = 4 \pi L_p e^{-\gamma} \sqrt{2 a},
\label{lambdacrit}
\end{equation}
which is a fixed length of the order of the Planck length.  Let $z_i$,
$1 \le i \le 4$, denote the four roots of the equation $1 + z \ln
[ z {\hat \lambda}^2 \omega_c^2] =0$.  These complex roots
depend only on the ratio $\lambda/\lambda_{\rm crit}$.  Then the zeros
of the Greens function (\ref{G1}) are
\begin{equation}
\omega_i = \sqrt{\omega_{\bf k}^2 - \omega_c^2 \,z_i[\lambda /
\lambda_{\rm crit}]}, \ \ \ 1\le i\le4.
\label{omegai}
\end{equation}
When $\lambda \ge \lambda_{\rm crit}$,
there are four complex zeros $\pm \omega, \pm \omega^*$, where
$\omega = \alpha_0 + i \beta_0$
lies in the first quadrant.
When $\lambda \le \lambda_{\rm crit}$ there are three separate cases.
When $ \exp(- \omega_c^2 / \omega_{\bf k}^2) / \omega_{\bf k}^2 >
{\hat \lambda}^2$, then there are two real roots $\pm \alpha_0$ and
two imaginary roots $\pm i \beta_0$.  When $\exp(-
\omega_c^2 / \omega_{\bf k}^2) / \omega_{\bf k}^2 < {\hat \lambda}^2$,
then if
$\omega_{\bf k} > \omega_c$ there are four real roots $\pm \alpha_0,
\pm \alpha_0^\prime$, and if $\omega_{\bf k} < \omega_c$ there are
four imaginary roots $\pm i \beta_0, \pm i \beta_0^\prime$.

The analytic structure of the function $1/F_3(k)$ is somewhat simpler.
There exists a unique $k_0 > 0$ depending on
$\lambda$ such that the locus of the zeros of $F_3(k)$ is $k^2 =
\omega_k^2 - w^2 =  k_0^2$.

The general solution to Eq.~(\ref{firstorderexplicit}) can be written
as
\begin{equation}
G_{ab}^{(1)} = G_{ab}^{(1),\,{\rm inhom}} + G_{ab}^{(1),\,{\rm
free}} + G_{ab}^{(1),\,{\rm free,T}}.
\label{generalsoln}
\end{equation}
We discuss these three terms in turn.  The inhomogeneous part of
the solution is given by
\FL
\begin{eqnarray}
\kappa G_{ab}^{(1),\,{\rm inhom}}(x) &=& \int {d^4 k \over (2 \pi)^4} e^{ i k \cdot
x} \bigg[ { {\tilde s}_{ab} \over F_1(k)} \nonumber \\
\mbox{} && + {F_2(k) \over F_1(k) F_3(k) } (k_a k_b - \eta_{ab} k^2)
{\tilde s}_c^{\ c} \bigg].
\label{inhomsym}
\end{eqnarray}
In this expression it is understood that the poles on the real
$\omega$ axis in the functions $1/F_1$ and $1/F_3$ are regulated with
the appropriate $i \epsilon$ prescription to pick out the ``half
retarded plus half advanced'' type contribution from each pole.  [We
choose this particular prescription for convenience, it would clearly
be possible to use instead for example the ``retarded'' type contribution
from each pole.]  As discussed above, poles on the real $\omega$ axis will
occur only for $\lambda < \lambda_{\rm crit}$ for the function $1/F_1$, but
will occur for all values of $\lambda$ for the function $1/F_3$.  Note
also that in the conformally coupled case $\xi=1/6$, the trace of the
source tensor $s_{ab}$ vanishes, and hence the second term in
Eq.~(\ref{inhomsym}) vanishes.  However, as we discuss below, the
freely specifiable piece of the solution associated with the second
term in Eq.~(\ref{inhomsym}) [Eq.~(\ref{sol4}) below] does not vanish even
for conformal coupling.

The second term $G_{ab}^{(1),\,{\rm
free}}$ in Eq.~(\ref{generalsoln}) is the freely specifiable, homogeneous
piece of the general solution associated with
the Greens function $1/F_1(k)$.  In the case $\lambda > \lambda_{\rm
crit}$, it can be written as \cite{caveatdecay}
\FL
\begin{eqnarray}
\kappa G_{ab}^{(1),\,{\rm free}}(x) &=&  \int d^3{\bf k} \ e^{i {\bf k} \cdot
{\bf x}} \,
 \bigg[ C_{ab}({\bf k}) e^{- i \alpha_0 t}
e^{\beta_0 t}  \nonumber \\
\mbox{} && + D_{ab}({\bf k}) e^{+ i \alpha_0 t}
e^{-\beta_0 t}  \bigg] + {\rm c.c.}.
\label{sol3}
\end{eqnarray}
Here $C_{ab}({\bf k})$ and $D_{ab}({\bf k})$ are arbitrary except that
they must be traceless and satisfy $k^a C_{ab} = l^a D_{ab} =0$, where
$k^a = ({\bf k},\alpha_0+i\beta_0)$ and $l^a = ({\bf
k},-\alpha_0-i\beta_0)$.  The quantities $\alpha_0$ and $\beta_0$
depend on the mode frequency $\omega_k = |{\bf k}|$, and on the
lengthscale $\lambda$ as indicated by Eq.~(\ref{omegai}).  This is
purely an exponentially growing/decaying type solution and has no
oscillatory parts.

{\vskip -1.2cm}
{\plotoneNew{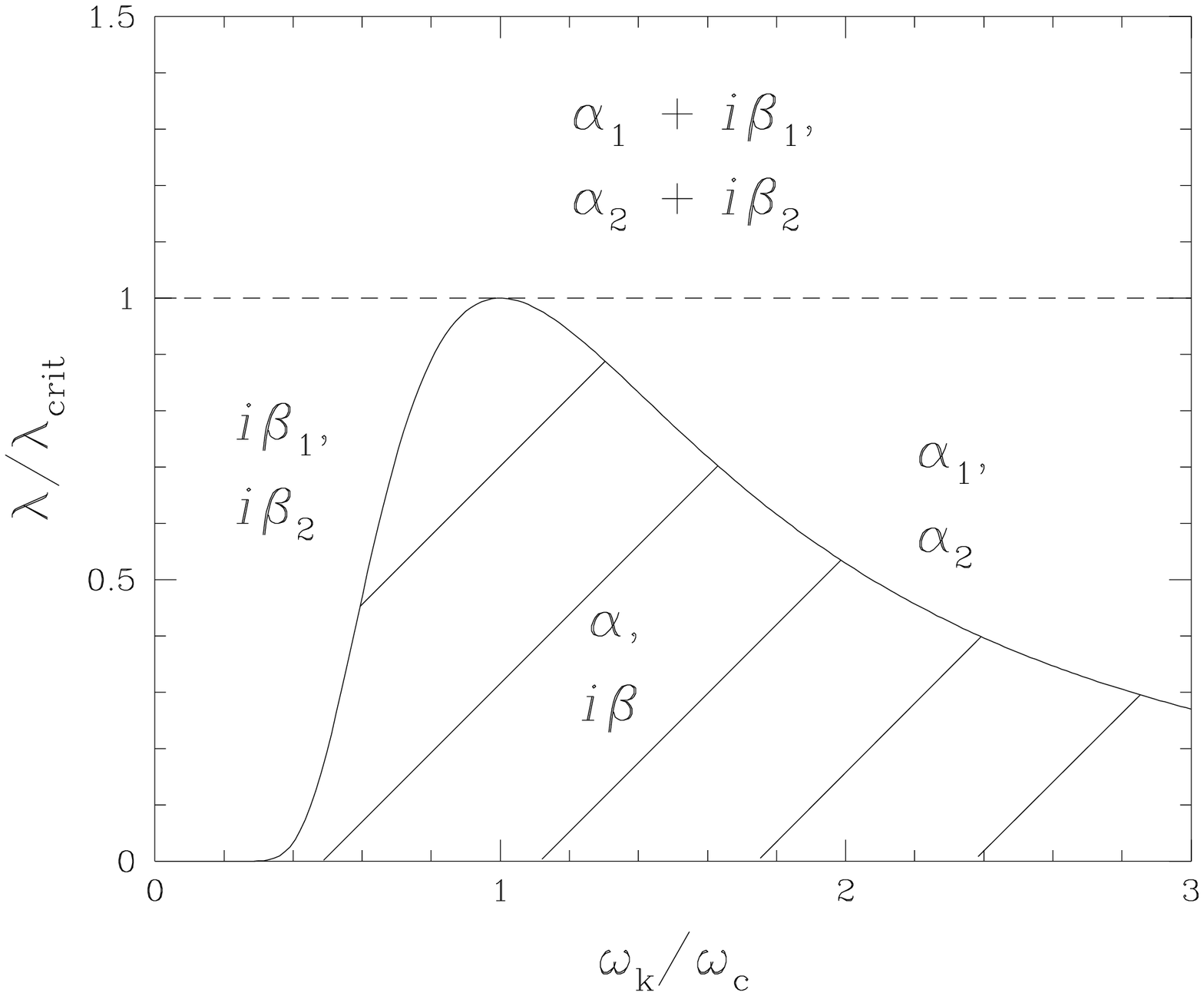}}
{\vskip -0.5truein}
\figure{
An illustration of the dependence of the locations of the poles of the
Greens function $1/F_1(k)$ on the parameters $\omega_{\bf k}$, the
frequency of the plane wave mode in question, and $\lambda$, the
additional free parameter with dimensions of length that appears in
the quantum theory but not in the classical theory.  In general there
are four poles of the form $\pm z_1$, $\pm z_2$, where the locations
of the two poles $z_1$ and $z_2$ are as follows.  In the hatched
region below the curve, there is one real pole and one imaginary pole.
Above the curve and below the dashed line, in the left region both
poles are imaginary, and in the right region both poles are real.
Finally, in the region above the dashed line, both poles are complex.
$\lambda_{\rm crit}$ is a critical length of order the Planck length,
and $\omega_c$ is a critical frequency of order the Planck frequency.
\label{poles}}
{\vskip 0.6cm}

The homogeneous term $G_{ab}^{(1),\,{\rm free}}$
is more complicated in the case $\lambda <
\lambda_{\rm crit}$.  In this case there are again freely
specifiable terms for each of the poles of $1/F_1$ (see Fig.~\ref{poles}).
The poles off the 
real axis will give exponentially growing/decaying terms, and the
poles on the real axis will give ``tachyonlike'' contributions
analogous to Eq.~(\ref{tachyonsoln}).  In
particular, there will be a portion of the general solution which has
the form of an integral over the values of ${\bf
k}$ for which both poles are real.  This expression will
have the same form as $\Phi_>$
of Eq.~(\ref{tachyonsoln}), except that $A$ and $B$ are replaced by
transverse traceless tensors, and the quantity $ m^2$ is replaced by
either of two independent constants depending only on $\lambda$
(i.e.,~there is a sum of two terms).  This portion of the general
solution was obtained by Horowitz \cite{Horowitz}.  It can be written as
\begin{equation}
G_{\rm tachyon}^{(1)}(x)_{ab} = \int_{\cal H} L_{ab}(k) e^{i k \cdot x}.
\label{tachyon}
\end{equation}
Here ${\cal H}$ is the union of two spacelike hyperbola $k^2 =
 k_1^2 >0 $ and $k^2 = k_2^2 > 0$ in
momentum space, where $k_1$ and $k_2$ are constants depending only on
$\lambda$.  The tensor $L_{ab}(k)$ is a freely specifiable transverse
traceless tensor on ${\cal H}$ which falls off sufficiently fast at
infinity.

The third term $G_{ab}^{(1),\,{\rm free,T}}$ in Eq.~(\ref{generalsoln}) 
is a purely transverse, freely specifiable piece of the solution which
is associated with the Greens function $1/F_3$.  It can be written as
\cite{caveatdecay} 
\FL
\begin{eqnarray}
&&  \int_{|{\bf k}|< k_0} d^3{\bf k} \ e^{i {\bf k}
\cdot
{\bf x}} \, \bigg[ E({\bf k}) e^{ \kappa_{\bf k} t} + F({\bf
k}) e^{-
\kappa_{\bf k} t} \bigg] \, (k_a k_b - \eta_{ab} k^2) \nonumber \\
\mbox{} && +  \int_{|{\bf k}|> k_0} d^3{\bf k} \ e^{i {\bf k}
\cdot
{\bf x}} \, \bigg[ I({\bf k}) e^{ i \nu_{\bf k} t} + J({\bf k}) e^{-
i \nu_{\bf k} t} \bigg] \, (k_a k_b - \eta_{ab} k^2) \nonumber \\
\mbox{} && + {\rm c.c.}
\label{sol4}
\end{eqnarray}
Here $\nu_{\bf k} = \sqrt{{\bf k}^2 - k_0^2}$, $\kappa_{\bf
k} = \sqrt{ k_0^2- {\bf k}^2}$, and
the functions $E$, $F$, $I$ and $J$
are freely specifiable functions of ${\bf k}$.  Thus,
there are both tachyonlike and exponentially growing modes of this
type for all
values of $\lambda$.  Also, as remarked above, this freely specifiable
transverse piece of the solution does not vanish in the conformally
coupled case $\xi=1/6$, despite the fact that the 
the analogous transverse contribution to the inhomogeneous
piece of the general solution [the second term in Eq.~(\ref{inhomsym}) above]
does vanish for conformal coupling.


Finally, we remark that exact solutions to the alternative, rescaled
version (\ref{alphadep}) of the first order perturbation equation can
be obtained from the above analysis using the substitutions $h_{ab}^{(1)}
\to
\alpha^2 {\hat h}_{ab}^{(1)}$, $L_p \to L_p / \alpha$, $\lambda \to
\lambda / \alpha$ and $s_{ab} \to {\bar s}_{ab} \equiv \langle
T_{ab}^{(0)},{\bar \omega}_{\rm in}^{(1)} \rangle$.

\appendix{ANEC integral for exact solutions}
\label{exactanec}

In this appendix we consider the specific subclass ${\cal S}$ of the
solutions of the exact, first order, semiclassical equation discussed
in Appendix \ref{firstordersolutions}, given by using half
advanced plus half retarded Greens function to obtain the linearized
Einstein tensor.  In the case of
exponential type modes discussed in Sec.~\ref{proposal2} above, the
use of this Greens function to pick out a class of solutions is
equivalent to throwing away the runaway solutions by hand.  For the
oscillatory type modes, this choice of Greens function yields a
particular subclass of solutions.
We shall show that the solutions in ${\cal S}$ have the property that
their transversely smeared ANEC integral is always non-negative, even
outside of the long wavelength limit, whenever $\lambda > \lambda_{\rm
crit}$.  However, we also shall show that some exact solutions outside
of this subclass do violate ANEC.

As explained in Sec.~\ref{proposal2} above, we see
no reason to view solutions in ${\cal S}$ as being any more physically
meaningful than any other subclass of exact solutions.
However, given a solution in ${\cal S}$, any other exact solution
obtained from the same incoming state will have the same perturbative
expansion in $1/\alpha^2$ and $\ln \alpha / \alpha^2$ [or
equivalently, in $\hbar$ and $\hbar \ln(\hbar)$].  This perturbative
series up to any finite order also should coincide with what would be
obtained by carrying the reduction of order procedure of the
semiclassical equation to the appropriate order in $1/\alpha^2$ and
solving exactly the new, reduced order equation.  Thus, an expansion
in $1/\alpha^2$ and $\ln \alpha / \alpha^2$ of the positivity result
of this appendix provides an alternative proof of the results we
established in Sec.~\ref{firstorderexact} above for the solutions of
the reduced order equation (\ref{firstordermixed3}), at least for
$\lambda > \lambda_{\rm crit}$.  In other words, we can use the analysis
of the solutions in ${\cal S}$ as a mathematical tool to to establish
a positivity result for solutions of the reduced order equations.
The alternative proof that this appendix provides also gives insight
into the otherwise mysterious positivity properties of the
coefficients $A$, $B$ and $C$ (with and without smearing) discussed in
Sec.~\ref{long1} above.

We now turn to a proof of the positivity of the ANEC integral for this
class of solutions.  The analysis of the ANEC integral for the exact
solutions to the unmodified semiclassical equation parallels that
given in Sec.~\ref{solns1} for the solutions of the reduced order
equation, with the only difference being that the functions $S_1$ and
$S_2$ of Eqs.~(\ref{S1def}) and (\ref{S2def}) are replaced by the
expressions $1/F_1$ and $F_2 / F_3$ respectively, where the functions
$F_1$, $F_2$ and $F_3$ were defined in Appendix
\ref{firstordersolutions}.  Correspondingly we again obtain the
formula (\ref{anec6}), except that the function $K({\bf x}_T)$ is
replaced by $K_1({\bf x}_T)$, where
\begin{equation}
{\tilde  K}_1({\bf \Delta k}_T) = \left[1 + {{\bf \Delta k}_T^2 \over
\omega_c^2} \ln \left( {\hat \lambda^2} {\bf \Delta k}_T^2
\right)\right]^{-1}.
\label{tildeK1def}
\end{equation}

Note that if $\lambda > \lambda_{\rm crit}$ then ${\tilde K}_1({\bf
k}_T)$ is a continuous function; its Fourier transform $K_1({\bf x}_T)$
is $L^2$ and is continuous everywhere away from the origin.
The function $K_1({\bf x}_T)$ for $\lambda = 1.2 \lambda_{\rm crit}$
is shown in Fig.~\ref{Kfunction}.  For $\lambda <
\lambda_{\rm crit}$, the expression (\ref{tildeK1def}) blows up
at some finite value of $|{\bf k}_T|$ so that ${\tilde K_1}$ is not
even $L^2$.  In this case an appropriate $i \epsilon$
regularization prescription should be understood to apply to the
formula (\ref{anec6}), the precise prescription being determined by the
fact that we are choosing half retarded plus half advanced solutions.
The regularization prescription yields a well defined distribution
${\tilde K}_1$, with well defined distributional Fourier transform
$K_1({\bf x}_T)$.  We assume from now on that $\lambda > \lambda_{\rm
crit}$; however, it is possible that the results in this appendix
continue to hold for $\lambda \le \lambda_{\rm crit}$.

{\vskip -0.5cm}
{\plotoneNew{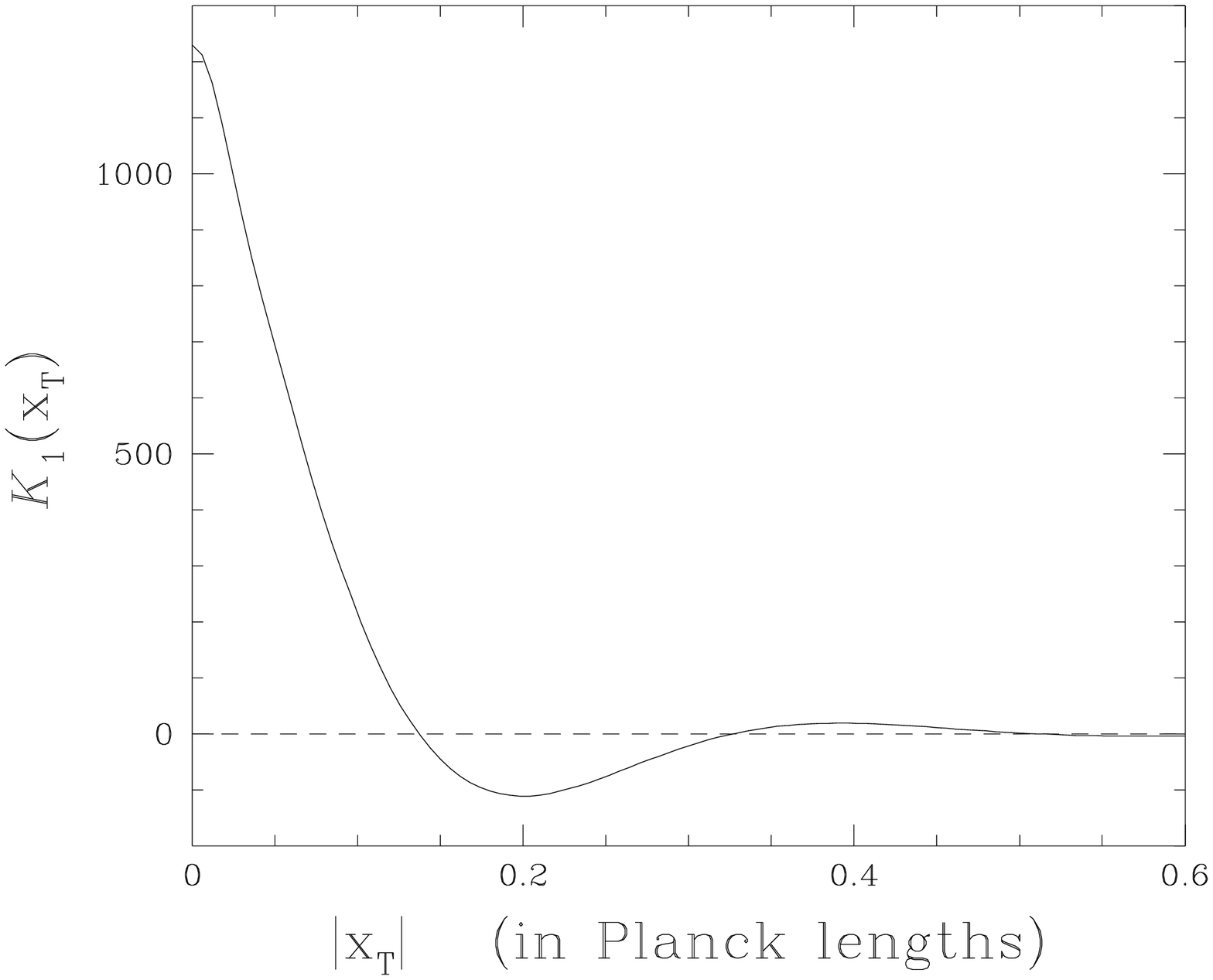}
{\vskip -0.4truein}
\figure{
The ANEC integral for exact, "half advanced plus half retarded"
solutions to the semiclassical equation can be expressed in terms of
the function $I_F({\bf x}_T)$ [which is obtained from the ANEC
integral in Minkowski spacetime evaluated on transversely displaced
geodesics] integrated against a particular function $K_1({\bf x}_T)$
with width of order of the Planck length, see Eq.~(\ref{anec7}) above.
Here we plot the function $K_1$ as a function of the transverse distance
$|{\bf x}_T|$ from the original geodesic, in the case where $\lambda =
1.2 \lambda_{\rm crit}$.  The fact that this function is negative for
some values of its argument implies that there are incoming states
for which the ANEC integral including backreaction is negative.
\label{Kfunction}}
{\vskip 0.6cm}

{}From Eq.~(\ref{anec6}), the smeared ANEC integral
is proportional to
\begin{equation}
\int d^2 {\bf x}_T \ I_F({\bf x}_T) [K_1 \circ S_{\rm dil}]({\bf x}_T),
\label{anec8}
\end{equation}
where $S_{\rm dil}({\bf x}_T) = S({\bf x}_T / \Lambda_T)$ is a
``dilated'' smearing function and $K_1 \circ S_{\rm dil}$ denotes the
convolution of $K_1$ and $S_{\rm dil}$.  Since we know $I_F({\bf
x}_T)$ is positive, to make the integral (\ref{anec8}) positive it
suffices to find smearing functions whose convolutions with $K_1$ are
pointwise positive everywhere.  We now show that for the choice
(\ref{smearingfn}) of smearing function, there exists some
$\Lambda_T >0$ such that the function $S({\bf x}_T / \Lambda_T)$ will
satisfy the positive convolution condition.  In fact, our proof below
can be easily extended to apply to {\it any} positive smearing
function which does not fall off at large ${\bf x}_T$ more rapidly
than $K_1$ does ($\propto |{\bf x}_T|^{-4}$); in other words, any
function which falls off slowly enough will be a suitable smearing
function when sufficiently ``dilated''.

We now outline a proof of that the convolution is pointwise positive.
We would like to show that there exists some $\Lambda$ such that the function
\begin{equation}
I_1(\Lambda,{\bf x}_0)  \, \equiv \Lambda^2 \, \int d^2{\bf x} \,\,K_1(\Lambda
{\bf x}) S({\bf x} + {\bf x}_0)
\label{I1int}
\end{equation}
is positive for all ${\bf x}_0$.  [For the remainder of this
appendix, we drop the subscript $T$ on $\Lambda_T$.] Clearly, for any fixed
${\bf x}_0$, we
can find a $\Lambda$ such that the integral (\ref{I1int}) is positive,
since $\Lambda^2 K_1(\Lambda {\bf x}) \to \delta^2({\bf x})$ as $\Lambda
\to \infty$.  Let $\Lambda_{\rm min}({\bf x}_0)$ denote the smallest
number $\Lambda_0$ such that the integral (\ref{I1int}) is positive
for all $\Lambda > \Lambda_0$.
Then we need just to show that $\Lambda_{\rm min}({\bf x}_0)$ is
bounded above as a function of ${\bf x}_0$, so that there exists some
positive $\Lambda$ which works for all ${\bf x}_0$.

To show this we suppose conversely that there exists some sequence
$(\Lambda_j, {\bf x}_{0,\,j})$ such that $\Lambda_j \to \infty$ and
\begin{equation}
I_1(\Lambda_j,{\bf x}_{0,\,j}) = 0.
\label{I1sequence}
\end{equation}
Now suppose that the sequence ${\bf x}_{0,\,j}$ is bounded.  Then by
passing to a subsequence we can without loss of generality assume that ${\bf
x}_{0,\,j} \to {\hat {\bf x}}_0$ for some ${\hat {\bf x}}_0$, so that
$(\Lambda_j^{-1},{\bf x}_{0,\,j}) \to (0,{\hat {\bf x}}_0)$.  However,
the function $I_1(\Lambda,{\bf x}_0)$ is jointly continuous as a
function of $\Lambda^{-1}$ and ${\bf x}_0$ even at $\Lambda^{-1} =0$,
as can be deduced from the formula
\begin{equation}
I_1 = \int d^2{\bf y} K_1({\bf y}) S({\bf x}_0 + \Lambda^{-1} {\bf y})
\label{I1inta}
\end{equation}
and by using the properties of the functions $S$ and $K_1$.  Hence it
follows that $I_1(\Lambda_j,{\bf x}_{0,\,j}) \to I_1(\infty,{\hat {\bf
x}}_0) = S({\hat {\bf x}}_0) > 0$.  This contradicts
Eq.~(\ref{I1sequence}) above, and so we conclude that the sequence
${\bf x}_{0,\,j}$ is unbounded.

To exclude the possibility of the existence of such a sequence with
${\bf x}_{0,\,j}$ unbounded, we now derive an estimate for the
function $I_1(\Lambda,{\bf x}_0)$ for large ${\bf x}_0$.  First, we
note that $K_1({\bf x}) \sim - |{\bf x}|^{-4}$ at large $|{\bf x}|$,
and in particular that there exists a $k_0 > 0$ and $\epsilon_0>0$ such
that for all $\epsilon < \epsilon_0$,
\begin{equation}
|K_1({\bf x})| \le \epsilon
\label{K1bound}
\end{equation}
whenever $|{\bf x}| \ge k_0 \epsilon^{-1/4}$ \cite{ftnote}.  Moreover,
from the formula (\ref{smearingfn}) for $S({\bf x})$ we can show that
for any $\eta$ with $0 < \eta < 1$, there exists $k_1 >0$ such that
\begin{equation}
(1-\eta) S({\bf x}_0) \le S({\bf x} + {\bf x}_0) \le (1 + \eta) S({\bf
x}_0)
\label{Sbound}
\end{equation}
whenever $|{\bf x}| \le k_1 |{\bf x}_0|$.
Let $K^+$ be the integral of $|K_1|$ over the domain where it is positive,
and similarly define $K^->0$, so that
\begin{equation}
1 = \int d^2{\bf x} K_1({\bf x}) \, = K^+ - K^-.
\end{equation}

Consider now the integral
(\ref{I1int}).  For any $\epsilon$ with $0 < \epsilon < \epsilon_0$ ,
we can split up this integral into three different parts:  There is a
contribution from the region
where $|{\bf x}| > k_0 \epsilon^{-1/4}
/\Lambda$, which is bounded below by $- C_0 \epsilon \Lambda^2$ from
Eq.~(\ref{K1bound}), for some constant $C_0$.
There is a contribution
from the portion of the region $|{\bf x}| < k_0 \epsilon^{-1/4}
/\Lambda$ in which $K_1$ is positive.  Using Eq.~(\ref{Sbound}), this
will be bounded below by  $K^+(1-\eta) S({\bf x}_0) / \Lambda^2$,
if
\begin{equation}
k_0 \epsilon^{-1/4} /\Lambda \le k_1 |{\bf x}_0|.
\label{necccondt}
\end{equation}
We can ensure that this condition holds by choosing $\epsilon = (k_1
|{\bf x}_0| \Lambda / k_0)^{-4}$; this requires that
\begin{equation}
|{\bf x}_0|\, \Lambda \ge (k_0/k_1) \epsilon_0^{-4}.
\label{restriction1}
\end{equation}
Finally, there will be a
contribution from the corresponding region in which $K_1$
is negative, which with the above choice of $\epsilon$ is bounded
below by $- K^- (1 + \eta) S({\bf x}_0) / \Lambda^2$.
Combining these bounds and choosing $\eta = (2 + 4 K^-)^{-1}$ yields
that
\begin{equation}
I_1(\Lambda,{\bf x}_0) \ge {- C_1 \over x_0^4 \Lambda^2} + {1 \over 2(
1 + x_0^4)},
\end{equation}
where $C_1$ is a constant, whenever the condition (\ref{restriction1})
is satisfied.  It clearly
follows that the function $\Lambda_{\rm min}({\bf x}_0)$ is bounded
above for large ${\bf x}_0$.  This contradicts the unboundedness of
the above sequence ${\bf x}_{0,\,j}$, and we conclude that the
function $\Lambda_{\rm min}({\bf x}_0)$ is bounded above for all ${\bf
x}_0$.

We conclude that there is some fixed length $\Lambda_T$ such that for
the smearing function (\ref{smearingfn}), the ANEC integral
(\ref{anec8}) is always non-negative.  It is clear on dimensional
grounds that the critical
value of $\Lambda_T$ is of order of the Planck length.  We remark that
this positivity result would not hold for any smearing function which
falls off more rapidly than $|K_1|$.  Thus, for example, there is no
positivity result for Gaussian transverse smearing.  However, as
mentioned above, our proof could easily be modified to apply to
smearing functions that fall off more slowly than $|K_1|$.

Finally, we consider the exact solutions of
Eq.~(\ref{firstorderexplicit}) outside of the subclass ${\cal S}$ of
half retarded plus half advanced solutions.  We now show that when $\lambda
< \lambda_{\rm crit}$, there exist exact solutions outside of ${\cal
S}$ which violate ANEC, even when transversely smeared.  [It can be
shown that violations of ANEC do not occur when $\lambda >
\lambda_{\rm crit}$, if we we discard the exponentially growing and
decaying pieces of the solutions.]
Using the momentum space coordinates $\beta$, $\gamma$,
${\bf k}_T$, the first order ANEC integral (\ref{firstorderanec}) can
be written as
\begin{equation}
I^{(1)} = \int_{-\infty}^\infty d\gamma \, \int d^2 {\bf k}_T \, {\tilde
G}^{(1)}_{ab}(\beta=0, \gamma, {\bf k}_T) \lambda^a \lambda^b.
\label{firstorderanectachyon}
\end{equation}
For the tachyon type solution (\ref{tachyon}), the linearized Einstein
tensor can be written as
\FL
\begin{equation}
{\tilde G}_{ab}(\beta,\gamma,{\bf k}_T) = \delta(-2 \gamma \beta +
{\bf k}_T^2 - k_0^2) \, {\cal F}_{ab}(\beta,\gamma,{\hat {\bf k}}_T),
\label{tachyon2}
\end{equation}
where ${\cal F}_{ab}$ is transverse and traceless but otherwise freely
specifiable and ${\hat {\bf k}}_T =
{\bf k}_T / |{\bf k}_T|$.  From Eqs.~(\ref{firstorderanectachyon}) and
(\ref{tachyon2}), and writing ${\hat {\bf k}}_T = (\cos \chi,\sin
\chi)$, it follows that
\FL
\begin{equation}
I^{(1)} = \int_{-\infty}^\infty d\gamma \int_0^{2 \pi} d\chi \, {1 \over 2
k_0} {\cal F}_{ab}(0,\gamma,\chi) \lambda^a \lambda^b.
\end{equation}
Since ${\cal F}_{ab}(\beta,\gamma,{\hat {\bf k}}_T)$ is freely
specifiable away from $2 \beta \gamma + k_0^2=0$, it is clear that in general
$I^{(1)}$ is non vanishing and can be of either sign.  Moreover, it is
easy to see that a negative $I^{(1)}$ cannot always be made positive
by transverse smearing in this case.

Note that if $h_{ab}^{(1)}(x;\alpha)$ is a one parameter family of exact
solutions such that ANEC is violated, then it will still be true that
ANEC will be satisfied order by order in $1/\alpha^2$ and $\ln \alpha
/ \alpha^2$.  In other words, the violations of ANEC are
non-perturbative in $1/\alpha^2$, or equivalently in $\hbar$.

\appendix{Existence of spacetime Fourier Transform of expected
Stress-Energy Tensor}
\label{FTstress}

In this paper we have imposed a condition on the Minkowski spacetime
in-states we consider, which requires that the two-point functions
be smooth and have suitable falloff properties at spatial infinity.  In
this
appendix we derive some implications of these falloff assumptions,
which are used in the body of the paper.

We start by describing the class of states we are considering.  Let
the two point bidistribution of the state be
\begin{equation}
G(x,x^\prime) = G_0(x,x^\prime) + F(x,x^\prime),
\label{Gg}
\end{equation}
where $G_0(x,x^\prime)$ is the vacuum two-point bidistribution in
Minkowski spacetime (which was denoted $G_{\rm in,0}$ in the body of
the paper).  Throughout this appendix, we drop the
subscripts ``in'' that appeared in the body of the paper.  Also the
conditions on $F$ that we discuss below apply to both the functions
$F_{\rm
in}^{(1)}$ and $F_{\rm in}^{(2)}$ that appear in the body of the paper, and
our conclusions
about the expected stress tensor in the state (\ref{Gg}) clearly then
will apply to the source term (\ref{source}) and to the first term in
Eq.~(\ref{secondordersource}).

The function $F$ is determined by the restrictions of the four functions
$F$, $\partial F /\partial t$, and $\partial F /
\partial t^\prime$ and $\partial^2 F / (\partial t \partial t')$
to $\Sigma \times \Sigma$, where $\Sigma$ is
the hypersurface $t=0$.  We assume that all these functions lie
in the class ${\cal V}$ defined in Sec.~\ref{firstorderexact} --- that
is, that they are smooth and that all of their spatial
derivatives are $L^1$ on $\Sigma \times \Sigma$.
Note that imposing this condition on
any other surface of constant time would yield the same class of states.
[It would be sufficient for all of our results to assume only that the
all spatial derivatives up to seventeenth order are $L^1$; we
have not investigated what the sharp falloff requirements
are.]

As in the body of the paper, we can express the two point function in
terms of functions $f$ and $g$ via
\begin{eqnarray}
F(x,x^\prime) &=& \int d^3 {\bf k} \, \int d^3 {\bf k}^\prime \,
f({\bf k},{\bf k}^\prime) \, e^{i k \cdot x} \, e^{i k^\prime \cdot
x^\prime} \nonumber \\
\mbox{} && + \int d^3 {\bf k} \, \int d^3 {\bf k}^\prime \,
g({\bf k},{\bf k}^\prime) \, e^{i k \cdot x} \, e^{- i k^\prime \cdot
x^\prime} \nonumber \\
\mbox{} && + {\rm c.c.}
\label{fgdefine}
\end{eqnarray}
where $k = ({\bf k},\omega_k)$, $\omega_k = |{\bf k}|$, and ``c.c.''
means the complex conjugate.
This equation defines $f$ and $g$ to be suitable complex linear
combinations of the spatial Fourier transforms of the above-mentioned
four functions restricted to $\Sigma \times \Sigma$: we have
\begin{equation}
\label{transform}
\left( \begin{array}{c}
f \\
g \\
f^* \\
g^* \end{array} \right)= {1 \over 4}
\left[ \begin{array}{cccc}
1  & 1 & 1 & 1 \\
1 & 1 & -1 & -1 \\
1 & -1 & -1 & 1 \\
1& -1 & 1 & -1 \end{array} \right]
\left( \begin{array}{c}
{\tilde F}({\bf k},{\bf k}')\\
i {\tilde F}_{,t}({\bf k},{\bf k}') / \omega_{\bf k}\\
i {\tilde F}_{,t^\prime}({\bf k},{\bf k}') / \omega_{{\bf k}'}\\
- {\tilde F}_{,t\,t'}({\bf k},{\bf k}') / (\omega_{\bf k} \omega_{{\bf
k}'})
\end{array} \right).
\label{explicitfgdef}
\end{equation}
Our assumptions on $F$ imply that
$f$ and $g$ are continuous
away from ${\bf k} =0 $ and ${\bf k}^\prime=0$ and satisfy, for any
integer $N$,
\begin{eqnarray}
\max \left\{ \, |f({\bf k},{\bf k}^\prime)|, |g({\bf k},{\bf
k}^\prime)| \, \right\} \, &\le& {C_N \over (1 + {\bf k}^2  + {\bf
k}^{\prime \,2})^N} \nonumber \\
\mbox{} &\times& \left(1 + {1 \over \omega_{\bf k}}\right) \,
 \left(1 + {1 \over \omega_{{\bf k}^\prime}}\right)
\label{bound}
\end{eqnarray}
for some constant $C_N$.
[Note that the requirement that the total energy of the
state be finite should imply an inequality of the form (\ref{bound})
but without the $1 + 1/\omega_{\bf k}$ factors and with $N = 5/2 +
\epsilon$.]  Moreover our assumptions imply the functions ${\hat
f} \equiv \omega_{\bf k} \omega_{{\bf k}'} f$ and ${\hat g}
\equiv \omega_{\bf k} \omega_{{\bf k}'} g$ defined before
Eq.~(\ref{omegak}) are at least $C^0$, a fact which is used in the
body of the paper.

Now the expected stress tensor in the state (\ref{Gg}) will be
automatically smooth, from the relation
\begin{equation}
T_{ab}(x) =  {\cal D}_{ab} \,F(x,x')
\end{equation}
and the fact that the initial data for $F$ on $\Sigma \times \Sigma$
is smooth, so that $F$ itself is smooth on $M \times M$.  Therefore
its Fourier transform ${\tilde T}_{ab}(k)$ exists as a distribution.  We
now
investigate the regularity properties of this distribution.  In our
notation below we make use
of delta functions; the steps can be made rigorous by integrating all
equations on both sides against a smooth test tensor field $f^{ab}(x)$
of compact support on $M$.  The expression for ${\cal D}_{ab}$ is
given in Eq.~(\ref{calDdef}) above.  We restrict attention below to the
piece
\begin{equation}
{\tilde S}_{ab}(l_a) =
\int d^4x e^{-i l_a  x^a} \lim_{x' \to x} \nabla_a \nabla_{b'}
F(x,x^\prime)
\end{equation}
of this expression; a similar analysis applies to the other type of
term involving $\nabla_a \nabla_b S$ in Eq.~(\ref{calDdef}).  From
Eqs.~(\ref{Gg}) and (\ref{fgdefine}) we
obtain that
\begin{equation}
{\tilde S}_{ab}(l_a) = F_{ab}(l_a) + F_{ab}(-l_a)^*,
\end{equation}
where $l_a = ({\bf l},\omega)$ and
\begin{eqnarray}
F_{ab}({\bf l},\omega) &=&  \int d^3 {\bf k} \, \omega_{\bf k}
\omega_{{\bf k} - {\bf l}} \,\,
(1,{\bf n})_a \nonumber \\
\mbox{}& & \times \big[\delta(\omega - \omega_{\bf k} + \omega_{{\bf
k} - {\bf l}})
g({\bf k},{\bf k} - {\bf l})
 (1,-{\bf n}^\prime)_b \nonumber \\
\mbox{} && - \delta(\omega - \omega_{\bf k} - \omega_{{\bf k} - {\bf
l}})
f({\bf k}, {\bf l} - {\bf k})
 (1,{\bf n}^\prime)_b \big].
\label{formal}
\end{eqnarray}
Here ${\bf n}$ and ${\bf n}^\prime$ are unit vectors in the directions
of ${\bf k}$ and ${\bf k} - {\bf l}$ respectively.  We write the
integral over ${\bf k}$ as
\begin{equation}
\int_0^\infty dk k^2 \int_{-1}^1 d\mu \int_0^{2 \pi} d\varphi_k,
\end{equation}
where $\mu$ is the cosine of the angle between ${\bf k}$ and ${\bf
l}$.  The $\delta$-function in the first term in Eq.~(\ref{formal})
can be rewritten as
\begin{equation}
\Theta\left[k - (\omega+l)/2\right] \, \Theta(l-|\omega|) \delta(\mu -
\mu_1(k,l,\omega)) {|k-\omega| \over k l},
\end{equation}
and similarly the $\delta$-function in the second term becomes
\begin{equation}
\Theta\left[l/2 - |k - \omega/2|\right] \, \Theta(\omega-l) \delta(\mu -
\mu_2(k,l,\omega)) {|k-\omega| \over k l}.
\label{subst}
\end{equation}
Here $\Theta$ is the step function, and $\mu_1$ and $\mu_2$ are the
appropriate values of $\mu$ that are enforced by the $\delta$-function.
Equations (\ref{formal})-(\ref{subst}) yield
\begin{eqnarray}
F_{ab}({\bf l},\omega) &\propto&
{\Theta(l-|\omega|)  \over l}
\int_{(l+\omega)/2}^\infty dk  \int d\varphi_k \,k^2
(k-\omega)^2 \nonumber \\
\mbox{} & & \times \left[ (1,{\bf n})_a (1,{\bf n}^\prime)_b \,\,
g({\bf k},{\bf k} - {\bf l}) \,\right]
(\mu_1) \nonumber \\
\mbox{} &-&  {\Theta(\omega-l) \over l}
\int_{(\omega-l)/2}^{(\omega+l)/2} dk  \int d\varphi_k \, k^2
(k-\omega)^2 \nonumber \\
\mbox{} && \times \left[ (1,{\bf n})_a (1,-{\bf n}^\prime)_b \,\,
f({\bf k},{\bf l} - {\bf k}) \right](\mu_2).
\label{formal1}
\end{eqnarray}
{}From the properties of the functions $f$ and $g$ and the form of
Eq.~(\ref{formal1}), it is clear that $F_{ab}$ and hence also ${\tilde
S}_{ab}({\bf l},\omega)$ is continuous everywhere away from the
lightcone, and that $|{\bf l}|F_{ab}({\bf l},\omega)$ is bounded in a
neighborhood of ${\bf l}=0$.  Similar conclusions apply to the entire
stress tensor ${\tilde T}_{ab}(k)$.

Finally, we note that the stress tensor ${\tilde T}_{ab}(k)$ is $L^2$,
from which it follows that $T_{ab}(x)$ is $L^2$ on Minkowski spacetime.
We can prove this using the formula (\ref{formal1}) as follows.
Let $e^{ab}$ be any constant tensor, and let $F_{ab}^I$
denote the first term in Eq.~(\ref{formal1}).  Then from
Eq.~(\ref{bound}) and using the fact that the $\delta$ functions in
Eq.~(\ref{formal}) enforce $|{\bf k} - {\bf l}| = |\omega - k|$, it
follows that
\begin{eqnarray}
&&\int | e^{ab} F_{ab}^I({\bf l},\omega) |^2\, d^3 {\bf l} \,d
\omega \,\, \le C_0 \int_0^\infty d l \int_{-l}^l d \omega
\nonumber \\ &&
 \left\{
\int_{(l+\omega)/2}^\infty d k \, \, {k^2 (k-\omega)^2 \over \left[1 +
k^2 + (k-w)^2 \right]^N} \right\}^2,
\end{eqnarray}
where $C_0$ is a constant, which is finite for $N \ge 5$.  A similar
analysis applies to the second term in Eq.~(\ref{formal1}), and to the
other types of term involving $\nabla_a \nabla_b S$ in
Eq.~(\ref{calDdef}).

\appendix{Fourier transform of the function ${\bf k_T}^2 \log[ {\bf
k}_T^2]$}
\label{technical}

In this appendix we prove the following result which is used
in the body of the paper.

{\it Lemma \, } \ \ Let h({\bf x}) be a non-negative, smooth, $L^1$
function on ${\bf R}^2$ such that
\begin{equation}
h({\bf 0}) = h_{,i}({\bf 0}) = h_{,ij}({\bf 0}) = h_{,ijk}({\bf 0}) =
0,
\label{cc5a}
\end{equation}
where the commas denote partial derivatives.  Then
\begin{equation}
\int d^2 {\bf k} \left[ \int d^2 {\bf x} e^{i {\bf k} \cdot {\bf x}}
h({\bf
x}) \right] {\bf k}^2 \log({\bf k}^2) \ \ge 0,
\label{lemmaint}
\end{equation}
with equality iff $h \equiv 0$.

{\it Proof: \, } \ The essential idea is that the Fourier transform of
${\bf k}^2 \log({\bf k}^2)$ consists of a smooth positive function
away from the origin, plus distributional contributions at the origin
whose effects are unimportant because of the condition (\ref{cc5a}).

Using polar coordinates $k,\varphi,x,\chi$ defined by ${\bf x} = (x
\cos \chi, x \sin \chi)$, ${\bf k} = (k \cos(\chi + \varphi), k
\sin(\chi + \varphi))$, we can write the integral (\ref{lemmaint}) as
\begin{equation}
\int_0^{2 \pi} d \chi \lim_{K \to \infty} \, \int_0^\infty d x \,
h(x,\chi) \, {\cal I}(K,x),
\label{IK0}
\end{equation}
where
\begin{equation}
{\cal I}(K,x) \equiv \int_0^K d k\, \int_0^{2 \pi} d \varphi \,
k^3 \log(k^2) e^{i k x \mu},
\label{IK}
\end{equation}
and $\mu \equiv \cos \varphi$.  By evaluating the integral over $k$
in Eq.~(\ref{IK}) and using the identities
\begin{equation}
\int d\varphi \, {f(\mu) \over \mu^2} =
\int d\varphi \, f^\prime(\mu) \, {1 - \mu^2 \over \mu}
\end{equation}
and
\begin{equation}
\int d\varphi \, {f(\mu) \over \mu^4} =
\int d\varphi \, f^\prime(\mu) \, {(2 \mu^2 + 1) (1 - \mu^2) \over 3
\mu^3},
\end{equation}
it can be shown that
\begin{equation}
{\cal I}(K,x) = 12 \pi / x^3 + \int_0^{2 \pi} d\varphi \,\Delta {\cal
I} (K,x,\mu).
\label{IK1}
\end{equation}
Here $\Delta {\cal I}(K,x,\mu)$ consists of linear combinations of terms
of the form
\begin{equation}
{ (\log K)^q K^p \mu^w \over x^n} \cos(K \mu x)
\label{costerms}
\end{equation}
or
\begin{equation}
{ (\log K)^q K^p \mu^w \over x^n} \sin(K \mu x),
\label{sinterms}
\end{equation}
where $q$,$p$,$w$ and $n$ are integers with $0 \le q \le 1$, $0 \le p
\le 5$, $0 \le w \le 8$ and $0 \le n \le 3$.

After integrating over $x$, the contribution from such terms becomes
proportional to
\begin{equation}
{\tilde g}_{\chi,n}(K \mu) (\log K)^q K^p \mu^w,
\label{IK2}
\end{equation}
where $g_{\chi,n}(x) \equiv h(x,\chi)/x^n$ is the restriction of the
smooth function $h({\bf x})/x^n$ to the line at angle $\chi$ in the
${\bf x}$
plane, and ${\tilde g}_{\chi,n}$ is the (one dimensional) Cosine or Sine
transform of this function.  The function $g_{\chi,n}$ is smooth
because of the condition (\ref{cc5a}) since $0 \le n \le 3$.
It follows that ${\tilde g}_{\chi,n}(K)$ will fall off
at large $K$ faster than any power of $K$, and hence the absolute
value of the expression (\ref{IK2}) is bounded above by
\begin{equation}
{ |\log K|^q K^p \mu^w C_p \over 1 + (\mu K)^{p+1}},
\label{IK3}
\end{equation}
where we have chosen the power to be $p+1$ and $C_p$ is a constant.
Carrying out the integral over $\varphi$ yields an expression which is
bounded above at large $K$ by
\begin{equation}
{ |\log K|^q K^p  C^\prime \over C^{\prime\prime} +  K^{p+1}},
\label{IK4}
\end{equation}
where $C^\prime$ and $C^{\prime\prime}$ are additional
constants that depend on $p$ and $w$.  Now it can be seen that these
terms give a
vanishing contribution in the limit $K \to \infty$ in Eq.~(\ref{IK0}).
Thus, from Eq.~(\ref{IK1}) the original integral (\ref{lemmaint}) can
be written as
\begin{equation}
\int_0^{2 \pi} d \chi  \, \int_0^\infty d x \,
h(x,\chi) \, \left[{12 \pi \over x^3}\right],
\label{IK5}
\end{equation}
which is manifestly positive.

\appendix{Consistency of our results with examples of negative
(unsmeared) ANEC integrals in selfconsistent solutions}
\label{visserexample}

As mentioned in the Introduction, in a very recent paper Visser
\cite{Visser96} has shown that the expected stress tensor of a scalar
field in the Boulware vacuum outside a Schwarzschild black hole
violates ANEC.  Visser argued that this result suggests that similar
violations would occur in self-consistent solutions with backreaction.
Moreover, these violations of ANEC would not be confined to a
Planck-scale tube surrounding a particular null geodesic, but would
instead occur over a macroscopic region.  In this appendix we review
Vissers argument and show that --- at least in the context of
perturbation theory off of Minkowski spacetime --- his class of
examples is consistent with our positivity results.  We also present a
simple explicit model which illustrates this point, and discuss
implications for the existence of traversable wormholes.

A key element in Visser's argument is a method for obtaining
approximate self consistent solutions to the semiclassical equations,
starting from solutions of the classical Einstein equation.
Essentially, the idea is an extension of our long-wavelength or small
$\hbar$ expansion beyond the context of perturbation theory about flat
spacetime.  We now describe this approximation scheme in a language
similar to that we have used in the body of the paper.  Let $\Psi_{\rm
cl}$ be some classical field with stress tensor $T_{ab}^{\rm
cl}\left[\Psi_{\rm cl},g_{cd}\right]$, and let $g_{ab}^{(0)}$,
$\Psi^{{\rm cl},(0)}$ be a solution of the classical Einstein equation
\begin{equation}
\kappa G_{ab}[g_{cd}^{(0)}] = T_{ab}^{\rm cl}\left[\Psi^{{\rm
cl},(0)},g_{cd}^{(0)}\right].
\label{classicalsoln}
\end{equation}
Now consider an additional quantum field ${\hat \Phi}$, and let us
seek solutions of the semiclassical equation
\begin{equation}
\kappa G_{ab}[g_{cd}] = T_{ab}^{\rm cl}\left[\Psi_{\rm
cl},g_{cd}\right] + \langle {\hat T}_{ab} \rangle[g_{cd}].
\label{fulltheory}
\end{equation}
Now if the state of the quantum field is the incoming vacuum state,
then the second term on the right hand side above
will be in order of magnitude $\sim L_p^2 / {\cal L}^4$, whereas the
first term should be $\sim 1 / {\cal L}^2$, where ${\cal L}$ is the
lengthscale determined by the classical background solution
$g_{ab}^{(0)}$.  Therefore if ${\cal L} \gg L_p$, the quantum stress
tensor can be treated as a small perturbation, and a leading order
approximation to the self consistent solution will be given by
\begin{eqnarray}
g_{ab} &=& g_{ab}^{(0)} + g_{ab}^{(1)} \nonumber \\
\mbox{} \, \Psi^{\rm cl} &=& \Psi^{{\rm cl},(0)} + \Psi^{{\rm cl},(1)},
\end{eqnarray}
where $g_{ab}^{(1)}$ and $\Psi^{{\rm cl},(1)}$ are calculated from the
linearized version of Eq.~(\ref{fulltheory}) with source
$\langle T_{ab} \rangle [g_{ab}^{(0)}]$ as well as the linearized
field equation for $\Psi^{\rm cl}$.  At leading order, in regions
of the spacetime where $\Psi^{{\rm cl},(0)}$ and $\Psi^{{\rm cl},(1)}$
vanish, the Einstein tensor for this self consistent solution will be
just the test field quantum stress tensor on the classical background, as
claimed by Visser \cite{Visser96}.

Consider now the application of this scheme to the Boulware vacuum
outside a Schwarschild black hole.  This state is unphysical because
the expected stress tensor diverges on the horizon.  However, Visser
argues that the expected stress tensor far from the horizon will
likely be approximately the same as the stress tensor for the static
vacuum state outside a spherical star with radius close to its
Schwarschild radius.  [The stress tensors will not be exactly the same
because the expected stress tensor in the static vacuum state does
have a non-local dependence on the spacetime geometry.]  Therefore,
the approximate self consistent solution obtained from the above
scheme starting from a compact star and a test field in the static
vacuum state should violate ANEC over macroscopically large regions
far from the star.

It might appear that this violation of ANEC is qualitatively different
from the cases treated in our analysis in the body of the paper, and,
thus, that new possibilities are open for severe violations of ANEC.
However, we shall now argue that this is not the case by showing that
examples of this type also exist in the context of perturbation theory
off of Minkowski spacetime, and that these examples
satisfy our smeared version of ANEC provided only that stress-energy
of the classical matter itself satisfies ANEC.  To see this, consider
perturbing Minkowski spacetime by adding a classical stress tensor
$T_{ab}^{(1), \, {\rm cl}}$.  For example this could be the linearized
stress tensor for a static spherical star.  The first order metric
perturbation in the semiclassical theory will then satisfy a modified
version of Eq.~(\ref{firstorder}) wherein the term $T_{ab}^{(1), \,
{\rm cl}}$ is added to the right hand side.  When one goes to the long
wavelength limit and performs a reduction of order, one finds that, to
lowest order, the metric perturbation satisfies
Eq.~(\ref{firstordermixed3}) with $\hbar s_{ab}$ replaced by
$T_{ab}^{(1), \, {\rm cl}}$.  [Note that the resulting equation simply
says that $\kappa$ times the Einstein tensor of the semiclassical
solution is just the sum of the classical stress tensor
$T_{ab}^{(1),\,{\rm cl}}$ and the vacuum polarization stress tensor in
the spacetime $\eta_{ab} + h_{ab}^{\rm cl}$, where
$G^{(1)}_{ab}[h^{\rm cl}] = T_{ab}^{(1),\,{\rm cl}}$, in agreement
with our discussion above.]  However, our proof of
the positivity of the smeared ANEC integral given in
Sec.~\ref{mixedsec} above used only the positivity of the ordinary
Minkowski ANEC integral for $s_{ab}$.  Thus, if $T_{ab}^{\rm(1),~cl}$
satisfies ANEC in Minkowski spacetime, our analysis shows that the
smeared ANEC integral in the semiclassical spacetime will be positive
to first order.

A simple example will illustrate these points.
Consider a point mass $m$ moving along the geodesic
$x^a(\tau) = \tau u^a$ in Minkowski spacetime.  Let $h_{ab}^{\rm cl}$
be the linearized gravitational field of this point mass, which is
given by
\begin{equation}
{\tilde h}^{\rm cl}_{ab}({\vec k}) = {2 \pi \over \kappa} \, {m u_a u_b \over {\vec k}^2 + i
\varepsilon} \, \delta ({\vec k} \cdot {\vec u}).
\end{equation}
Next, using Eq.~(\ref{linearvacpol}) we can calculate the expected
stress tensor in the incoming vacuum state on the spacetime with
metric $\eta_{ab} + h_{ab}^{\rm cl}$, which is given by
\begin{eqnarray}
{\tilde T}_{ab}({\vec k}) &=& {m {\tilde H}_\lambda(k) \over (2 \pi)^3
\kappa } \delta({\vec k}
\cdot {\vec u}) \bigg[ 2 a {\vec k}^2 u_a u_b \nonumber \\
&&  - (2 a/3 + 2 b) (k_a k_b - \eta_{ab} k^c k_c)\bigg].
\label{lvp}
\end{eqnarray}
From this stress tensor we can obtain a second linearized metric
perturbation $h_{ab}^{\rm quantum}$.  Finally, we calculate the ANEC
integral along the geodesic 
$x^a(\lambda) = \Delta^a + \lambda \lambda^a$ of the Einstein tensor
of the spacetime $\eta_{ab} + h_{ab}^{\rm cl} + h_{ab}^{\rm quantum}$, 
which is given by
\begin{equation}
I = {1 \over (2 \pi)^3 \kappa} \int d^4 k \,{\tilde G}_{ab}(k) \lambda^a
\lambda^b \,e^{i {\bf k}_T \cdot {\bf \Delta} } \, \delta( {\vec k}
\cdot {\vec \lambda}).
\end{equation}
Using Eqs.~(\ref{tildeHlambda}), (\ref{IK5}) and (\ref{lvp}) we
obtain for $\Delta_T \ne 0$ 
\begin{equation}
I = - { 6 m a \over (2 \pi)^4 \kappa} |{\vec u} \cdot {\vec \lambda}| \,
{1 \over \Delta_T^4},
\label{neg1}
\end{equation}
where $\Delta_T$ is the length of the component of $\Delta^a$
perpendicular to $\lambda^a$ and $u^a$, i.e., the impact parameter of
the null geodesic with respect to the point mass.  Equation
(\ref{neg1}) is essentially the large $r$, weak gravity limit of
Visser's counterexample to ANEC involving a static star.

Thus, the ANEC integral is negative along all geodesics away from the
point mass in the self consistent solution.  Nevertheless, this is
consistent with our result of positivity of smeared ANEC, since in
calculating the smeared ANEC integral for a geodesic a distance
$\Delta_T$ from the point mass, the negative contribution (\ref{neg1})
will be compensated by the positive contribution from the point masses
stress tensor itself, which also scales like $\Delta_T^{-4}$ because
of our smearing function (\ref{smear2}) \cite{lastnote}.

This example illustrates that our results are consistent with having
negative ANEC integrals 
over a macroscopic region.  The price one must pay is that the amount
of ANEC violation is restricted to be very small compared to distant
mass scales.  We now give a crude argument which suggests that the
restriction is easily sufficient to prevent the existence of
macroscopic traversable wormholes.  Let us characterize the
region with negative ANEC integral by the quantity with dimensions of mass 
\begin{equation}
M_- = \int d^2 {\bf x}_T \int d\lambda \,T_{ab} \lambda^a \lambda^b,
\end{equation}
and suppose that there is a region of positive ANEC integral a
distance $\sim \Delta$ away characterized by the mass $M_+$.  Then, in order of
magnitude, our result implies that
\begin{equation}
M_- + \left({L_p \over \Delta}\right)^4 M_+ \ge 0.
\end{equation}
Consider now a static macroscopic wormhole, and suppose that the
wormhole can be characterized by one lengthscale ${\cal L}$.  On
dimensional grounds, the energy density required to hold it open
should be be of order $\sim - 1 / {\cal L}^2$.  [This is confirmed by
explicit calculations in specific examples by Ford and Roman
\cite{Ford-Roman95c}.]  Consequently we have $M_- \sim - {\cal L}$,
and therefore
$$
M_+ \agt \left( {\Delta \over L_p}\right)^4 {\cal L}.
$$
Since the distance $\Delta$ to the positive mass region should be
$\agt {\cal L}$, we obtain $M_+ \agt 10^{130} M_\odot ({\cal L} / 1 \,
{\rm cm})^5$, which is a ridiculuously large mass.  Moreover, the
natural requirement that the $\Delta$ be larger than the gravitational
radius $M_+$ of the positive mass yields the restriction ${\cal L}
\alt (L_p / \Delta)^3 L_p \le L_p$.

\newpage
\onecolumn

\begin{table}
\caption{
A summary of our results for the ANEC integral in the different cases.
``Pure to first order'' indicates that the two point function of the
scalar field is pure to first order in $\varepsilon$.  ``Always $>0$''
means that the smeared ANEC integral is always strictly positive for
all solutions of the equations except for the trivial, flat
spacetime/vacuum solution.
\label{table1}}
\begin{tabular}{lllll}
&\mbox{\ \ \ \ \ \ \ \ \ \ \ \ \ \ Pure to first order}&&\mbox{\ \ \ \
\ \ \ \ \ \ \ \ \ \ Mixed at first order} &\mbox{}\\
\tableline
&\mbox{No\ smearing}&\mbox{Smearing}&\mbox{No\ smearing}&\mbox{Smearing}
\\
\tableline
\mbox{First order in }$\varepsilon$& $=0$ & $=0$ &Can be $<0$ &Always
$>0$ \\
&&&&in long wavelength limit\\

\mbox{\ }&  & & & \\
\mbox{Second order in }$\varepsilon$&Can be $<0$&Always $>0$&N/A&N/A \\
&&in long wavelength limit&&\\
\end{tabular}
\end{table}

\newpage
\onecolumn

\begin{table}
\caption{In this table, for the aid of the reader, we list in
alphabetical order some of the symbols that appear in the paper.
We do not list symbols whose meaning is very conventional, or symbols
which are used only in the immediate vicinity of where they are
introduced.  For each item listed, we give a brief description, and
also a reference to the equation in the text where the symbol first
appears, or after which the symbol is first introduced.  Except in
special cases, we do not
list separately the following variants of symbols:  symbols with
tildes or symbols with the superscripts $(0)$, $(1)$ or $(2)$.  The
former
always denote a Fourier transform, and the latter superscripts always
denote expansion coefficients in an expansion of a quantity in powers
of $\varepsilon$.
\label{table2}}
\begin{tabular}{llllll}
&\mbox{Symbol}&\mbox{\ \ Meaning}&\mbox{\ }&\mbox{Equation in
which}&\mbox{Other relevant}\\
&\mbox{\ }&\mbox{\ }&\mbox{\ }&\mbox{first appears}&\mbox{Equations}\\
\tableline

&
$a$
&\mbox
{Coefficient that controls the anomalous scaling
}&&(\ref
{logscalegeneral}
)&(\ref
{abdef}
)\\

&
$A$
&\mbox
{Expansion coefficient in long wavelength expansion
}&&(\ref
{anecexpand0}
)&\mbox{\ }\\

&
$A_{ab}$
&\mbox
{Fourth rank local curvature tensor
}&&(\ref
{Adefgeneral}
)&(\ref
{logscalegeneral}
)\\

&
$A_{ab}^{(1)}$
&\mbox
{Linearized local curvature tensor for metric perturbation $h_{ab}^{(1)}$
}&&(\ref
{ABdef}
)&(\ref
{linearvacpol}
)\\

&
$A_{ab}^{(1)}[\cdot]$
&\mbox
{Operator that acts on metric perturbations to yield a local curvature
tensor
}&&(\ref
{alphadep}
)&(\ref
{Zabdef}
)\\

&
${\cal A}_{ab}$
&\mbox
{Local curvature tensor at zeroth order in $1/\alpha^2$ expansion
}&&(\ref
{firstordermixed3}
)&(\ref
{Zabdef}
)\\

&
$b$
&\mbox
{Coefficient that controls the anomalous scaling
}&&(\ref
{logscalegeneral}
)&(\ref
{abdef}
)\\

&
$B$
&\mbox
{Expansion coefficient in long wavelength expansion
}&&(\ref
{anecexpand0}
)&\mbox{\ }\\

&
$B_{ab}$
&\mbox
{Fourth rank local curvature tensor
}&&(\ref
{Bdefgeneral}
)&(\ref
{logscalegeneral}
)\\

&
$B_{ab}^{(1)}$
&\mbox
{Linearized local curvature tensor for metric perturbation $h_{ab}^{(1)}$
}&&(\ref
{ABdef}
)&(\ref
{linearvacpol}
)\\

&
$B_{ab}^{(1)}[\cdot]$
&\mbox
{Operator that acts on metric perturbations to yield a local curvature
tensor
}&&(\ref
{alphadep}
)&(\ref
{Zabdef}
)\\

&
${\cal B}_{ab}$
&\mbox
{Local curvature tensor at zeroth order in $1/\alpha^2$ expansion
}&&(\ref
{firstordermixed3}
)&(\ref
{Zabdef}
)\\


&
$C$
&\mbox
{Expansion coefficient in long wavelength expansion
}&&(\ref
{anecexpand0}
)&\mbox{\ }\\

&
${\cal D} \ ({\cal D}_x, {\cal D}_y)$
&\mbox
{D'Alembertian type wave operator (wrt $x$ or $y$)
}&&(\ref
{diffop}
)&\mbox{\ }\\

&$
{\cal D}_{ab}
$&\mbox
{Operator arising in point splitting prescription, consisting of
}&&(\ref
{Tabformal}
)&(\ref
{calDdef}
)\\
&&\mbox{\ \ \  differential operator followed by (implicit)
coincidence limit}&&&\\

&$
{\cal D}_{ab}^{(1)}
$&\mbox
{First order change in ${\cal D}_{ab}$ due to metric perturbation
$h_{ab}^{(1)}$
}&&(\ref
{expandall}
)&\mbox{\ }\\

&$
{\cal D}_{ab}^{(1)}[\cdot]
$&\mbox
{Operator that acts on metric perturbations giving first order change
in ${\cal D}_{ab}$
}&&(\ref
{expandall}
)&\mbox{\ }\\

&$
{\cal E}[s_x,s_y]
$&\mbox
{Operator taking sources $s_x$ and $s_y$ to bisolution of wave
equation
}&&(\ref
{deltaF}
)&\mbox{\ }\\

&$
F_{\rm in}
$&\mbox
{Regularized two point function of incoming state $\omega_{\rm in}$.
}&&(\ref
{newFin}
)&\mbox{\ }\\

&$
{\bar F}_{\rm in}
$&\mbox
{Regularized two point function of rescaled incoming state ${\bar
\omega}_{\rm in}$.
}&&(\ref
{Finalpha}
)&(\ref
{rhoexpand1}
)\\


&$
F_1(k)$, $F_2(k)$, $F_3(k)$
&\mbox
{Functions arising in exact solutions
}&&(\ref{F1def}) - (\ref{F3def})
&(\ref
{sol1}
)\\

&$
f
$&\mbox
{``Pure frequency'' part of two point function, in momentum space
}&&(\ref
{F1explicit}
)&(\ref
{fgdefine}
)\\

&$
{\hat f}
$&\mbox
{Same as $f^{(1)}$ but multiplied by $|{\bf k}| \, |{\bf k}'|$
}&&(\ref
{omegak}
)&(\ref
{F1explicit1}
)\\

&$
G_{ab}^{(1)}
$&\mbox
{Linearized Einstein tensor for metric perturbation $h_{ab}^{(1)}$.
}&&(\ref
{G12def}
)&\mbox{\ }\\

&$
G_{ab}^{(1)}[\cdot]
$&\mbox
{Operator that acts on metric perturbation to give linearized Einstein
tensor
}&&(\ref
{G12def}
)&\mbox{\ }\\

&$
G_{ab}^{(2)}
$&\mbox
{Second order part of Einstein tensor, depending on $h_{ab}^{(1)}$ and
$h_{ab}^{(2)}$.
}&&(\ref
{Gsdef}
)&\mbox{\ }\\

&$
{\bar G}_{ab}^{(2)}
$&\mbox
{Rescaled version of second order part of Einstein tensor
}&&(\ref
{Gsdef}
)&\mbox{\ }\\

&$
G_{ab}^{(2)}[\cdot,\cdot]
$&\mbox
{Operator that acts on pairs of metric perturbations
}&&(\ref
{G12def}
)&\mbox{\ }\\

&$
g
$&\mbox
{``Mixed frequency'' part of two point function, in momentum space
}&&(\ref
{F1explicit}
)&(\ref
{fgdefine}
)\\

&$
{\hat g}
$&\mbox
{Same as $g^{(1)}$ but multiplied by $|{\bf k}| \, |{\bf k}'|$
}&&(\ref
{omegak}
)&(\ref
{F1explicit1}
)\\

&$G
$&\mbox
{Two point bidistribution of the state $\omega$ on $(M,g_{ab})$.
}&&(\ref
{Gdef}
)&\mbox{\ }\\

&$G_0
$&\mbox
{Two point bidistribution of the in vacuum state $\omega_0$ on
$(M,g_{ab})$.
}&&(\ref
{newF}
)&\mbox{\ }\\

&$G_{\rm in}
$&\mbox
{Two point bidistribution of the in state $\omega_{\rm in}$ on
$(M,\eta_{ab})$.
}&&(\ref
{newF}
)&\mbox{\ }\\

&$G_{\rm in,0}
$&\mbox
{Two point bidistribution of the vacuum state $\omega_{\rm in,0}$ on
$(M,\eta_{ab})$.
}&&(\ref
{newF}
)&\mbox{\ }\\

&$
h_{ab}^{(1)}$, $h_{ab}^{(2)}
$&\mbox
{Metric perturbations
}&&(\ref
{expandall}
)&\mbox{\ }\\

&$
{\hat h}_{ab}^{(1)}$, ${\hat h}_{ab}^{(2)}
$&\mbox
{Rescaled metric perturbations in treatment of long-wavelength limit
}&&(\ref
{salpha}
)&(\ref
{hhatdefs}
)\\

&$
H_\lambda
$&\mbox
{Horowitz distribution that enters the expression for vacuum
polarization
}&&(\ref
{Hlambda}
)&(\ref
{linearvacpol}
)\\


&$
{\cal H}
$&\mbox
{Usual Fock space of states on Minkowski spacetime
}&&(\ref
{secondorder}
)&\mbox{\ }\\

&$
I$, $I^{(1)}$, $I^{(2)}
$&\mbox
{ANEC integral and its expansion coefficients
}&&(\ref
{anecbasic}
)&\mbox{\ }\\

&$
I_s$, $I_s^{(1)}$, $I_s^{(2)}
$&\mbox
{Generalized ANEC integral and its expansion coefficients
}&&(\ref
{generalanec}
)&\mbox{\ }\\

&$
{\bar I}_s$, ${\bar I}_s^{(1)}$, ${\bar I}_s^{(2)}
$&\mbox
{Limiting form of generalized ANEC integral, with only transverse
smearing
}&&(\ref
{generalaneclimit1}
)&\mbox{\ }\\

&$
I_F({\bf x}_T)
$&\mbox
{ANEC integral without backreaction as a function of transverse
displacement
}&&(\ref
{IFdef}
)&(\ref
{IFformula}
)\\

&$
{\bar I}_F({\bf x}_T)
$&\mbox
{Same as $I_F$ but for rescaled incoming state ${\bar \omega}_{\rm in}$
}&&(\ref
{IFalpha}
)&\mbox{\ }\\

&$
J(k,k^\prime)
$&\mbox
{Function that arises in solutions of reduced order semiclassical
equations
}&&(\ref
{Jfn}
)&(\ref
{I1full}
)\\

&$
k, k^a, {\bf k}_T
$&\mbox
{Coordinates on momentum space
}&&(\ref
{kspace}
)&\mbox{\ }\\

&$
K({\bf x}_T)$, ${\tilde K}({\bf k}_T)$
&\mbox
{Distribution that describes the effect of backreaction on the ANEC
integral
}&&(\ref
{tildeKdef}
)&(\ref
{anec6}
)\\
&&\mbox{\ \ \  for solutions of reduced order equations}&&&\\

&$
K_1({\bf x}_T)$, ${\tilde K}_1({\bf k}_T)$
&\mbox
{Function that describes the effect of backreaction on the ANEC integral
}&&(\ref
{tildeK1def}
)&(\ref
{anec8}
)\\
&&\mbox{\ \ \  for solutions of original semiclassical equations}&&&\\

&$
k_0, k_1, k_2
$&\mbox
{Fixed, Planck scale frequencies controlling tachyon type solutions
}&&(\ref
{tachyon}
)&\ref
{sol4}
)\\

&$
{\cal L}
$&\mbox
{Lengthscale of incoming state or of semiclassical solution
}&&(\ref
{basic}
)&\mbox{\ }\\

&$
N
$&\mbox
{Number of scalar fields in the $1/N$ expansion
}&&(\ref
{basic}
)&\mbox{}\\

&$
s_{ab}
$&\mbox
{Source tensor in linearized semiclassical equation
}&&(\ref
{source0}
)&(\ref
{status}
)\\

&$
S({\bf x}_T)
$&\mbox
{Transverse smearing function
}&&(\ref
{geodesicfunction}
)&(\ref
{generalanec}
)\\

&$
S_{ab}$, $S
$&\mbox
{Coincidence limits of derivatives of two point function
}&&(\ref{Sdef}),(\ref{Sabdef})
&\mbox{\ }\\

&$
T_{ab}[g_{cd}]
$&\mbox
{Linear map on states on $(M,\eta_{ab})$ returning conserved stress
tensor on $(M,g_{ab})$
}&&(\ref
{notationdefine}
)&\mbox{\ }\\

&$
T_{ab}^{(0)}
$&\mbox
{Usual Minkowski spacetime stress tensor (linear map on states)
}&&(\ref
{STexpand}
)&\mbox{\ }\\

&$
T_{ab}^{(1)}[\cdot]$, $T_{ab}^{(2)}[\cdot,\cdot]
$&\mbox
{Expansion coefficients of $T_{ab}[g_{cd}(\varepsilon)]$ that act on
metric perturbations
}&&(\ref
{STexpand}
)&\mbox{\ }\\
&&\mbox{\ \ \  or pairs of metric perturbations and return linear maps
on Minkowski}&&&\\
&&\mbox{\ \ \  spacetime states}&&&\\

&$
{\bf x}_T
$&\mbox
{Two dimensional transverse coordinate on Minkowski spacetime
}&&(\ref
{geodesicfunction}
)&(\ref{kspace})\\

&
$Z_{ab}$
&\mbox
{Combination of fourth rank local curvature tensors that enters into
}&&(\ref
{Zabdef1}
)&(\ref
{logscalegeneral}
)\\
&&\mbox{\ \ \  anomalous scaling of stress tensor}&&&\\

&
$Z_{ab}^{(1)}$
&\mbox
{Linearized form of $Z_{ab}$ for metric perturbation $h_{ab}^{(1)}$
}&&(\ref
{Zabdef}
)&(\ref
{linearvacpol}), (\ref{Hscaling})\\

&
$Z_{ab}^{(1)}[\cdot]$
&\mbox
{Operator that acts on metric perturbations to yield anomalous scaling
tensor
}&&(\ref
{Zabdef}
)&\mbox{\ }\\

&$
\alpha
$&\mbox
{Parameter used to characterize long wavelength limit
}&&(\ref
{Finalpha}
)&\mbox{\ }\\

&$
\alpha
$&\mbox
{Unknown numerical coefficient of local curvature term in stress tensor
}&&(\ref
{STgeneral}
)&(\ref{linearvacpol})\\

&$
\alpha_0
$&\mbox
{Real part of complex frequency in oscillatory modes of exact solutions
}&&(\ref
{omegai}
)&\mbox{\ }\\

&$
\beta
$&\mbox
{Coordinate on momentum space
}&&(\ref
{kspace}
)&\mbox{\ }\\

&$
\beta
$&\mbox
{Unknown numerical coefficient of local curvature term in stress tensor
}&&(\ref
{STgeneral}
)&(\ref{linearvacpol})\\

&$
\beta_0
$&\mbox
{Imaginary part of complex frequency in oscillatory modes of exact
solutions
}&&(\ref
{omegai}
)&\mbox{\ }\\

&$
\gamma
$&\mbox
{Null geodesic in spacetime $(M,g_{ab})$, also zeroth order geodesic
in $(M,\eta_{ab})$
}&&(\ref
{generalanec}
)&\mbox{\ }\\

&$
\gamma
$&\mbox
{Coordinate on momentum space
}&&(\ref
{kspace}
)&\mbox{\ }\\

&$
\varepsilon
$&\mbox
{Basic expansion parameter of perturbation expansion
}&&(\ref
{metricexpand}
)&(\ref{rhoexpand0})\\

&$
\epsilon
$&\mbox
{Generic small parameter in discussion of higher order time derivative
}&&Sec.~\ref
{proposal2}
&\mbox{\ }\\
&&\mbox{\ \ \  equations of motion}&&&\\

&$
\zeta
$&\mbox
{Null coordinate on Minkowski spacetime, also a Fermi-Walker
coordinate
}&&(\ref
{geodesicfunction}
)&(\ref{firstorderanec1})\\
&&\mbox{\ \ \   on $(M,g_{ab})$ in a neighborhood of the null
geodesic}&&&\\

&$
\Theta
$&\mbox
{Step function
}&&(\ref
{tildeHlambda}
)&\mbox{\ }\\

&$
\Theta_\gamma
$&\mbox
{Function entering definition of generalized ANEC integral
}&&(\ref
{geodesicfunction}
)&(\ref{firstorderanec1})\\

&$
\kappa
$&\mbox
{Inverse of Newtons constant
}&&(\ref
{lagrangian}
)&\mbox{\ }\\

&$
\lambda
$&\mbox
{Affine parameter along geodesic $\gamma$, also Fermi-Walker
coordinate on $(M,g_{ab})$
}&&(\ref
{geodesicfunction}
)&(\ref{kspace})\\
&&\mbox{\ \ \ in a neighborhood of $\gamma$, also null coordinate on
Minkowski spacetime}&&&\\

&$
\lambda
$&\mbox
{Undetermined lengthscale appearing in expression for linearized stress
tensor
}&&(\ref
{linearvacpol}
)&(\ref{tildeHlambda})\\

&$
{\hat \lambda}
$&\mbox
{Rescaled version of above lengthscale $\lambda$
}&&(\ref
{tildeKdef}
)&(\ref{G1})\\

&$
\lambda_{\rm crit}
$&\mbox
{Critical value of lengthscale $\lambda$
}&&(\ref
{lambdacrit}
)&\mbox{\ }\\

&$
\lambda^a
$&\mbox
{Tangent vector to geodesic $\gamma$, also vector field on Minkowski
spacetime
}&&(\ref
{geodesicfunction}
)&(\ref{kspace})\\

&$
\Lambda$, $\Lambda_T$, $\Lambda_L
$&\mbox
{Lengthscales entering definition of generalized ANEC integral
}&&(\ref
{geodesicfunction}
)&(\ref{generalanec})\\

&$
\Lambda
$&\mbox
{Used in Appendix \ref{exactanec} instead of $\Lambda_T$.
}&&(\ref
{I1int}
)&\mbox{\ }\\

&$
\Lambda_{\rm T,crit}
$&\mbox
{Critical value of transverse smearing lengthscale
}&&(\ref
{Lambdacrit}
)&\mbox{\ }\\

&$
\nu_0
$&\mbox
{Numerical coefficient in expansion of Fourier space smearing function
}&&(\ref
{sfnexpand}
)&\mbox{\ }\\

&$
\xi
$&\mbox
{Curvature coupling coefficient
}&&(\ref
{lagrangian}
)&\mbox{\ }\\

&$
{\hat \rho}^{(1)}$, ${\hat \rho}^{(2)}
$&\mbox
{Expansion coefficients of incoming state for Fock space states
}&&(\ref
{rhoexpand}
)&\mbox{\ }\\

&$
{\bar {\hat \rho}}^{(1)}$, ${\bar {\hat \rho}}^{(2)}
$&\mbox
{Expansion coefficients of rescaled incoming state for Fock space states
}&&(\ref
{rho1pure2}),(\ref{rho2mixed2})
&\mbox{\ }\\

&$
\tau
$&\mbox
{Timescale characterizing radiation reaction effects
}&&(\ref
{tau}
)&\mbox{\ }\\

&$
\tau
$&\mbox
{Generic small parameter in discussion of reduction of order
}&&(\ref
{redord}
)&\mbox{\ }\\

&$
\tau^*
$&\mbox
{Evolution timescale for solutions of radiation reaction equation
}&&(\ref
{restriction}
)&\mbox{\ }\\

&$
{\hat \Phi}_+$, ${\hat \Phi}_-
$&\mbox
{Positive and negative frequency pieces of field operator
}&&(\ref
{posneg}
)&\mbox{\ }\\

&$
\chi^{(1)}_{ab}$, $\chi^{(2)}_{ab}
$&\mbox
{Rescaled versions of metric perturbations $h^{(1)}_{ab}$,
$h^{(2)}_{ab}$
}&&(\ref
{ansatz}
)&(\ref{hhatdefs})\\

&$
\chi^{(1,0)}_{ab}$, $\chi^{(2,0)}_{ab}
$&\mbox
{Leading order approximations to $\chi^{(1)}_{ab}$, $\chi^{(2)}_{ab}$
in long wavelength expansion
}&&(\ref
{chiexpand}
)&(\ref{chi10eqn}),(\ref{chi20eqn})\\

&$
\omega_{\rm in,0}
$&\mbox
{Vacuum state on Minkowski spacetime
}&&(\ref
{rhoexpand0}
)&\mbox{\ }\\

&$
\omega_{\rm in}^{(1)}$, $\omega_{\rm in}^{(2)}
$&\mbox
{Expansion coefficients of incoming, Minkowski spacetime state
}&&(\ref
{rhoexpand0}
)&\mbox{\ }\\

&$
{\bar \omega}_{\rm in}^{(1)}$, ${\bar \omega}_{\rm in}^{(2)}
$&\mbox
{Expansion coefficients of rescaled, incoming, Minkowski spacetime state
}&&(\ref
{rhoexpand1}
)&\mbox{\ }\\

&$
\omega_c
$&\mbox
{Critical frequency of the order of the Planck length
}&&(\ref
{tildeKdef}
)&(\ref{G1})\\

&$
\langle \cdot, \cdot \rangle
$&\mbox
{Product notation for stress tensors acting on Minkowski spacetime
states
}&&(\ref
{notationdefine}
)&\mbox{\ }\\

\end{tabular}
\end{table}
\newpage


\begin{references}
\bibitem{particleproduction}
If a stress tensor $T_{ab}$ initially vanishes, is conserved, and
satisfies the dominant energy condition, then it will vanish for all
time \cite{Hawking-Ellis}.  Therefore a violation of the dominant
energy condition is essential in any particle production situation where
the stress tensor is initially vanishing.  See Ref.~\cite{Zeldovich} above.
\bibitem{suppress}
In the large $N$ limit which we discuss later in Sec.~\ref{status},
the fluctuations in the stress tensor will however be suppressed
relative to  
its expected value by a factor of $\sqrt{N}$.  Hence this objection
to the applicability of the semiclassical equation does not apply in
the large $N$ context.
\bibitem{Kipsmodel}
See, for example, the suggested model in
Ref.~\cite{Morris-Thorne-Yurtsever} of a wormhole held open by the
Casimir energy between two concentric metal spheres.  As explained in
Ref.~\cite{Morris-Thorne-Yurtsever}, the model is probably unrealistic
due to finite skin-depth effects.
\bibitem{others}
Formal derivations of the equations have also been given starting from
the Wheeler-deWitt equation \cite{kiefer}, and also from from a
formal, decoherent histories formulation of quantum gravity
\cite{Hartle}.  However, for the purposes of our discussion it is
sufficient to concentrate on the two most conventional methods of
derivation.
\bibitem{vacuumnote}
One also could expand about a non-vacuum solution, but then the metric
and scalar field perturbations would be coupled at linear order, and
it would not be mathematically consistent to keep the matter loop
contributions but ignore the graviton loop contributions.
\bibitem{ambiguity}
Strictly speaking the axioms do allow higher order local curvature
terms, an example of which is
$$
\lambda_1^2 {1 \over \sqrt{-g}} {\delta \over \delta g^{ab}} \int d^4
x \sqrt{-g} \, R^3,
$$
but only at the expense of introducing additional explicit
lengthscales into the theory ($\lambda_1$ in the above example).  Such
terms are excluded by the assumption that the semiclassical theory
does not determine any preferred lengthscale not present in the
classical theory other than the lengthscale $\lambda_0$ that enters
into the logarithmic scaling (\ref{logscalegeneral}), and also that
the expected stress tensor depends on $\lambda_0$ only through the
logarithmic scaling.  This assumption is supported by the results
of explicit one-loop calculations in perturbation theory about flat
space \cite{oneloop}, as, in particular, standard regularization
prescriptions do not introduce such additional lengthscales.
Note also that the above expression with
$\lambda_1 \sim L_p$ is one of the types of correction terms that
would be expected at two loops.
\bibitem{note3}
At linear order in perturbation theory about flat space (but only in
this context), an assumption that is equivalent to demanding that
there be no higher-order
local curvature tensors is that the stress tensor should not depend
on derivatives of the metric of sixth order or higher.  This is the
assumption used in the derivation of the stress tensor in
Ref.~\cite{Horowitz}.  The reason for the equivalence is the
following.  For any local conserved curvature tensor, we can define
the order $n_{\rm derivative}$ of the highest order derivative of the
metric that the tensor contains, and also the dimension $n_{\rm
dimension}$ such that the physical dimension of the tensor is $({\rm
length})^{-n_{\rm dimension}}$.  For linearized local curvature
tensors $n_{\rm dimension} = n_{\rm derivative}$, but in general it is
only true that $n_{\rm dimension} \ge n_{\rm derivative}$.
\bibitem{localHs}
The tensor $B_{ab}$ is often denoted by ${}^{(1)}H_{ab}$
\cite{Birrell-Davies,Simon}, while the linear combination $A_{ab}/2 +
B_{ab}/3$ which is the variational derivative of the scalar $R_{cd}
R^{cd}$, is often denoted by ${}^{(2)}H_{ab}$.
\bibitem{scalingnote}
Indeed, in the classical theory, the scaling transformation $g_{ab}
\to \mu^2 g_{ab}$, $\Phi \to \Phi$, maps solutions into solutions [see
Eqs.~(\ref{Einsteineq}) and (\ref{waveeqn}) above].  The existence of
this scaling invariance may be viewed as a consequence of the absence
of any preferred lengthscale in the classical theory. On the other
hand, the natural scaling operation applied to metrics and states in
semiclassical gravity will fail to map solutions of Eq.~(\ref{basic})
to solutions.  This failure of scale invariance can be attributed to
the fact that there is a preferred length present in the semiclassical
theory, namely the Planck length $L_p$.
\bibitem{finiteenergy}
The incoming state perturbations must be of finite energy for the
solutions for the metric perturbations to exist.  Finiteness of energy
follows from the regularity properties on the states which we assume;
see the discussion after Eq.~(\ref{bound}).
\bibitem{caveat3b}
More generally we could allow $\varphi_0$ to be an arbitrary fixed
diffeomorphism.  However, this portion of the gauge freedom acts
trivially on the perturbation equations, so we do not consider it in
what follows.
\bibitem{dimensions}
We adopt a convention in which the metric tensor is dimensionless and
the spacetime coordinates have dimensions of length.  The scalar field
has dimensions of $({\rm length})^{-1}$.
\bibitem{caveatqg}
Note that this freedom would not be present in a full quantum theory
of gravity, if we assume that the gravitational initial data at
${\cal J}^-$ is a graviton state for the standard quantization of
linearized vacuum general relativity (i.e.~that the higher time
derivative effects, which are presumably artifacts of the
semiclassical approach, are not present), and if we also
demand that the ``particle content'' of the incoming
graviton state be preserved while the wavelengths are dilated.  This
requirement would dictate the particular scaling $h_{ab}^{(1)\, ,{\rm
in}}(x,\alpha) = \alpha^{-1} {\bar h}_{ab}^{(1)\, ,{\rm
in}}(x/\alpha)$ near ${\cal J}^-$ for some fixed ${\bar h}_{ab}^{(1)\,
,{\rm in}}$.
\bibitem{preacceleration}
This feature of the solutions is not as unphysical as it might appear,
because the magnitude of the pre-acceleration effects should always be
smaller than the ambiguity in the particles trajectory due to finite
size of the particle; see Ref.~\cite{Landau}.
\bibitem{noteS}
This is apparently the reason why the various proposals in
Refs.~\cite{Simon1,Simon2,Simon3} --- that is, accepting only
approximate perturbative solutions, accepting only perturbatively
expandable (exact) solutions, and reducing the order of the equations
--- are not clearly distinguished.  These proposals are all equivalent
to first order in $\tau$ or $\hbar$.  Once it is realized that
first order solutions are not sufficient, it becomes necessary to
discuss (as we do) these ideas as separate, independent
proposals.
\bibitem{note2}
See, e.g., Ref.~\cite{Kip} for an analysis of the effects of
correction terms of this type in an analogous situation involving
gravitational radiation reaction.  Corrections to the shape of the
inspiral orbit will be of relative magnitude $\sim \tau / \tau_p$ in
the case of non-vanishing eccentricity; for example, the total phase
taken to shrink the orbit by a factor of two will will have absolute
corrections of order unity but fractional corrections of order $\sim
\tau/\tau_p$.
\bibitem{notecc}
Note that for the perturbative semiclassical equations in the
special case of conformal coupling with $\lambda > \lambda_{\rm
crit}$, there are no extra degrees of freedom of the oscillatory type.
[All the poles of the relevant Greens functions are off the real axis;
see Appendix \ref{firstordersolutions}.]  Thus, in this case, one could
simply throw out the exponential-type extra solutions [the second term
in Eq.~(\ref{sol3})] by hand.
\bibitem{notcommute}
The issue is not that the process of reduction of order does not
commute with making the change of variables.  Rather, it is that the
process of reducing the order and then solving the resulting equation
for some finite value of $\tau$ does not commute with making the
change of variables.  This issue is not particular to the reduction of
order proceedure; it arises whenever one has an equation that is known
only up to first order in some small parameter.
\bibitem{caveat5}
More precisely, the finite size corrections will be
larger than this by the factor $\sim {\cal L} / (c \tau) \agt 1$,
c.f., Eq.~(\ref{radiationreactionforce}) above and associated
discussion.
\bibitem{note8}
This seemingly somewhat convoluted method of specifying the gauge is
necessary to translate the specification of the gauge in the
``passive'' sense (i.e., the choice of coordinates) to a specification
of the gauge in the ``active'' sense which is coordinate independent.
\bibitem{ftnote}
This can be derived from the formula
$$
\int_0^\infty dx \, J_0(k x) {x \over (1 + x^2)^2} = k K_1(k) /2,
$$
where $K_1$ is a modified Bessel function ({\it not} the function
$K_1$ discussed in the body of the paper), and by using the expansion
$$
k K_1(k) = 1 + k^2 \log(k)/2 + \left[(2 \gamma -1)/4 -
\log(2)/2\right]  k^2  + O(k^3).
$$
\bibitem{caveatdecay}
Strictly speaking, although the exponentially growing pieces of
Eq.~(\ref{sol3}) satisfy the equation of motion
(\ref{firstorderexplicit}), the exponentially decaying pieces do not
because for these pieces the integral over the past light cone in the
non-local term in the equation of motion does not converge.
However, from the time reversal invariance of the semiclassical theory
it is apparant that the exponentially growing and exponentially
decaying solutions are equally physical.  If we had chosen to write
the equations of motion in terms of ``out states'' at ${\cal J}^+$
instead of ``in states'' at ${\cal J}^-$, then we would have obtained
a first order semiclassical equation which posseses exponentially
decaying solutions but no exponentially growing solutions.
\end{references}
\end{document}